\documentclass{midl} 

\usepackage{caption}
\usepackage{hyperref}
\usepackage{tablefootnote}
\usepackage{bm}
\usepackage{bbm}
\usepackage{multirow}
\usepackage{makecell}
\usepackage[export]{adjustbox}
\usepackage{paralist}

\DeclareMathOperator*{\argmin}{arg\,min}

\usepackage{mwe} 
\jmlrvolume{-- Under Review}
\jmlryear{2022}
\jmlrworkshop{Full Paper -- MIDL 2022 submission}

\title[On learning adaptive multi-coil MRI acquisition policies]{On learning adaptive acquisition policies for undersampled multi-coil MRI reconstruction}

\midlauthor{\Name{Tim Bakker\midljointauthortext{Majority of work was done while interning at Facebook AI Research.}\nametag{$^{1}$}} \Email{t.b.bakker@uva.nl}\\
 \addr $^{1}$ University of Amsterdam \\
 \AND
 \Name{Matthew Muckley\nametag{$^{2}$}} \Email{mmuckley@fb.com}\\
 \Name{Adriana Romero-Soriano\nametag{$^{2}$}} \Email{adrianars@fb.com}\\
 \Name{Michal Drozdzal\midljointauthortext{Contributed equally.}\nametag{$^{2}$}} \Email{mdrozdzal@fb.com}\\
 \Name{Luis Pineda\nametag{$^{\dagger2}$}} \Email{lep@fb.com}\\ 
 \addr $^{2}$ Facebook AI Research}

\def \fastMRI {{fastMRI}\xspace}
\definecolor{luiscolor}{RGB}{50,135,168}

\newcommand{\eg}{\textit{e}.\textit{g}.,\ }
\newcommand{\ie}{\textit{i}.\textit{e}.,\ }

\begin{document}

\maketitle

\begin{abstract}
Most current approaches to undersampled multi-coil MRI reconstruction focus on learning the reconstruction model for a fixed, equidistant acquisition trajectory. In this paper, we study the problem of joint learning of the reconstruction model together with acquisition policies. To this end, we extend the End-to-End Variational Network with learnable acquisition policies that can adapt to different data points. We validate our model on a coil-compressed version of the large scale undersampled multi-coil \fastMRI dataset using two undersampling factors: $4\times$ and $8\times$. Our experiments show on-par performance with the learnable non-adaptive and handcrafted equidistant strategies at $4\times$, and an observed improvement of more than $2\%$ in SSIM at $8\times$ acceleration, suggesting that potentially-adaptive $k$-space acquisition trajectories can improve reconstructed image quality for larger acceleration factors. However, and perhaps surprisingly, our best performing policies learn to be explicitly non-adaptive.
\end{abstract}


\section{Introduction}
Magnetic resonance imaging (MRI) is one of the best non-invasive methods for assessing soft-tissue structure in the clinic. However, widespread MRI adoption is limited due to its long acquisition times. Almost since its inception, substantial research has been done to reduce these acquisition times, yielding a variety of undersampled MRI reconstruction techniques such as parallel imaging~\cite{sodickson1997simultaneous, pruessmann1999sense,griswold2002generalized}, compressed sensing~\cite{cs} and deep learning (DL)~\cite{schlemper2017deep,hammernik2018learning,aggarwal2018modl}. Although the DL-based approaches have been shown to achieve the strongest results, they tend to use either fixed or random $k$-space sampling patterns that do not adapt to the data, which may be suboptimal and lead to underestimation of the maximum possible acceleration rates.

There is a substantial literature - going back decades - on designing sampling trajectories for MRI (e.g. \citet{cao1993, gao2000, seeger2009, haldar2019}). Since the introduction of deep learning, researchers have attempted to further improve DL-based MRI reconstruction by learning $k$-space sampling patterns from the data. These learning-based approaches result in either \emph{non-adaptive} or \emph{adaptive} sampling patterns -- often referred to as policies. Non-adaptive policies learn a dataset specific acquisition trajectory -- \eg each data point in the dataset is reconstructed following the same learnt trajectory --, while adaptive policies are conditioned on the initial $k$-space measurements and, as a result, have the potential to produce different sampling trajectories per data point. Non-adaptive policies have been shown to outperform handcrafted sampling patterns for both single-coil~\cite{bahadir, weiss, huijben2020learning} and multi-coil~\cite{zhang2020extending, wang2021b, zibetti2021alternating} acquisition settings. However, adaptive policies have only been devised in the single-coil setting~\cite{zhang2019reducing, jin19, sanchez3, pineda2020active, pg, codesign,van2021active}, showing promising results which in some cases outperform non-adaptive ones. Learning adaptive policies for deep multi-coil MRI remains, to the best of our knowledge, largely unexplored.

In this work, we are the first to devise such a model for joint learning of 2D deep learning MRI reconstruction together with adaptive $k$-space acquisition trajectories for the more clinically relevant \textit{multi-coil} setting. In particular, we extend recent work that learns adaptive acquisition trajectories~\cite{codesign} to the multi-coil scenario and enhance the End-to-End Variational Network (E2E VarNet)~\cite{varnet}, a standard model for multi-coil reconstruction, with the ability to learn dataset-specific as well as potentially adaptive $k$-space sampling strategies. We perform extensive evaluation on Cartesian sampling for 2D MRI using the multi-coil fastMRI knee dataset~\cite{knoll2020fastmriknee} on $4\times$ and $8\times$  acceleration. We hope our effort provides the community a point of departure for further research into adaptive multi-coil acquisition. Our experiments\footnote{We've made our code and pre-trained models publicly available as part of the: \href{https://github.com/facebookresearch/fastMRI}{\fastMRI repository}.} show that: 

\begin{itemize}
    \item On the $8\times$ setup, our learned policies improve $\sim2\%$ in SSIM over the strongest baseline, highlighting the ability of potentially-adaptive $k$-space acquisition to improve MRI reconstruction under high acceleration factors.
    \item On the $4\times$ setup, the gain due to $k$-space trajectory optimisation reduces, with our policies performing on-par with the strongest competing method.
    \item Interestingly, our top performing policies learn to be explicitly non-adaptive, suggesting that adaptivity of the $k$-space acquisition trajectories may come at the expense of over-regularising the reconstruction model.
\end{itemize}



\section{Preliminaries}


\subsection{Background}

We consider a dataset $\mathcal{D}$ of $k$-space measurements $\bm{y} \in \mathbb{C}^{N \times M}$ from which we can reconstruct MR images. In the \emph{single-coil} setting, the reconstructed MR images can be obtained by the inverse Fourier transform $\mathcal{F}^{-1}$ as $\bm{x} = \mathcal{F}^{-1}(\bm{y})$. However, modern scanners accelerate the acquisition of the $k$-space by using multiple receiver coils that are each sensitive to different regions of the anatomy, thus exploiting redundancies in $k$-space measurements~\cite{sodickson1997simultaneous,pruessmann1999sense,griswold2002generalized}. Hence, in the \emph{multi-coil} setting, we define $\bm{y} \in \mathbb{C}^{N \times M \times K}$, where $K$ is the number of coils and where $\bm{y}_i \in \mathbb{C}^{N \times M}$ represents the output of a measurement by the $i$-th coil. The reconstructed MR images can then be obtained as
\begin{equation}
\label{eq:imfromk}
    \bm{x} = \sum_{i=1}^K \bar S_i\odot\mathcal{F}^{-1} (\bm{y_i}),
\end{equation}
where $\odot$ denotes element-wise multiplication, and $\bar S_i$ is the complex-conjugate of the \linebreak complex-valued sensitivity map associated with each receiver coil $i$, normalised such that $\sum_{i=1}^K \bar S_i S_i = 1$. These sensitivity maps encode how sensitive each coil is to each region in the anatomy, and can be estimated in an auto-calibrating fashion by fully sampling the center of the $k$-space, also known as the auto-calibration signal (ACS) region, with each coil. The acquisition of $k$-space measurements can be further accelerated by collecting fewer measurements and reconstructing the images using a partially observed $k$-space. To simulate partial observations of the $k$-space, we introduce a Cartesian binary sampling mask $\mathbf{M}$ that selects $B<M$ measurements, including the ACS measurements, and define the \emph{undersampled} $k$-space as $\bm{\tilde y}_i = \mathbf{M} \odot \bm{y}_i$, where $\odot$ denotes element-wise multiplication. Note that the same mask is applied to the measurements from all coils. The reconstructed MR images can then be obtained by leveraging the undersampled $k$-space as 
\begin{equation}
\label{eq:imfrompartialk}
  \bm{\tilde x} = \sum_{i=1}^K \bar S_i\odot\mathcal{F}^{-1} (\bm{\tilde y_i}).  
\end{equation}
This however results in blur or aliasing effects in the reconstructed images, which can be mitigated through the use of recent deep learning models, such as the End-to-End Variational Network (E2E VarNet) \cite{varnet}. In particular, the E2E VarNet takes as input the partially observed $k$-space $\bm{\tilde y}$ along with the sampling mask $\mathbf{M}$ decomposed into the mask of ACS measurements $\mathbf{M}_{\textrm{ACS}}$ and the mask of the non-ACS measurements $\mathbf{M}'$, such that $\mathbf{M} = \mathbf{M}_{\textrm{ACS}} + \mathbf{M}'$. The model estimates the sensitivity maps from the ACS region, and uses a cascaded neural network, $g$, to produce a high fidelity image reconstruction, $\bm{\hat x} = g(\bm{\tilde y}, \mathbf{M}_{\textrm{ACS}}, \mathbf{M}'; \phi)$, where $\phi$ are learnable parameters. Note that $\mathbf{M}'$ is commonly handcrafted to select equidistant measurements.

\subsection{Problem formulation}
\label{ssec:problem}
Our goal is to adapt the sampling of measurements in $\mathbf{M}'$ to each MR slice (image). To that end, we seek a policy that, given an initial undersampled $k$-space (\eg the ACS $k$-space, $\bm{\tilde y}_{\textrm{ACS}})$, predicts which measurements to acquire next. More precisely, we aim to learn an acquisition policy $\pi(\bm{\tilde y}_{\textrm{ACS}}, \mathbf{M}_{\textrm{ACS}}; \theta) \rightarrow \mathbf{M}'$, parameterised by $\theta$, that selects the measurements to acquire in order to improve the image reconstruction process defined by $g$:\looseness-1

\begin{equation}
\phi^*, \theta^* = \argmin_{\phi,\,\theta} \sum_j \mathcal{L}\left ( \bm{\hat x}_j, \bm{x}_j \right ),
\end{equation}

\noindent where $\phi^*$ and $\theta^*$ are the optimised reconstruction and policy parameters respectively, $j$ indexes the dataset, and $\mathcal{L}$ is a loss function measuring the discrepancy between the model prediction $\bm{\hat x}$ and the image reconstructed from the fully sampled $k$-space $\bm{x}$.

\section{Method}

This section outlines our Policy network and details its integration with the E2E VarNet. An overview of the proposed system is depicted in Figure~\ref{fig:adaptvar}. From now on, we will use the capitalised `Policy' to refer to our proposed model, while continuing to use `policy' as general descriptor of `subsampling strategy'.

\begin{figure}[ht!]
\floatconts
  {fig:adaptvar}
  {\caption{Overview of our system. The E2E VarNet computes sensitivity maps from the ACS, which are passed to the cascaded reconstruction model together with a subsampled $k$-space. The final reconstruction is reduced to a real-valued MR image by a root-sum-of-squares (RSS) operation. The subsampled $k$-space consists of the ACS and acquisitions suggested by a sampling strategy, which may be adaptive or non-adaptive.\looseness-1}}
  {\includegraphics[width=0.95\linewidth]{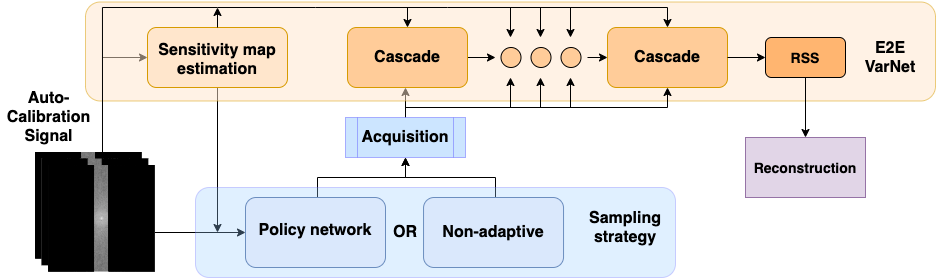}}
\end{figure}

\subsection{Policy network} \label{sec:policynet}

Our Policy is a neural network that takes as input the ACS $k$-space $\bm{\tilde y}{_\textrm{ACS}}$ and the mask of ACS measurements $\mathbf{M}{_\textrm{ACS}}$, and outputs a sampling probability for each measurement in $k$-space. From $\bm{\tilde y}{_\textrm{ACS}}$, we first estimate the sensitivity maps, and reconstruct a complex image $\bm{\tilde x}{_\textrm{ACS}}$ following Eq.~\ref{eq:imfrompartialk}. Using this image as input reduces the Policy network's size while maintaining relevant information, and allows for taking inspiration from convolutional architectures that were successfully used in the single-coil adaptive MRI literature. In particular, we employ the Policy network of~\citet{pg} and extend it to handle complex-valued inputs. The network is composed of a convolutional feature extractor, followed by a fully-connected block that outputs a heatmap encoding the relative salience of each potential $k$-space measurement in the acquisition step: see Appendix~\ref{app:polarch} for details. It remains to normalise these values and then sample $k$-space measurements. To this end, \citet{codesign} have shown that straight-through estimation outperforms reinforcement learning based approaches for backpropagation through discrete sampling in the single-coil setting. We thus employ their formulation, which is as follows: a non-linearity (e.g. a softplus as in~\citet{codesign} or a sigmoid as in~\citet{bahadir, zhang2020extending}) is first applied to ensure non-negative values. To prevent already observed $k$-space measurements in the ACS region from being sampled again, we set their corresponding probabilities to $0$. The resulting vector is normalised to obtain $M$ independent realisations of a Bernoulli distribution from which to sample the measurements to be acquired. We employ rejection sampling to obtain exactly $B$ measurements on each forward pass. As a result of the sampling, we obtain the binary mask of measurements to be acquired, $\mathbf{M}'$. The aforementioned straight-through estimation - employed during backpropagation - is realised by treating the non-differentiable sampling discretisation as a sigmoid function with slope $10$. In following~\citet{codesign}, we additionally enable a fairer comparison to the multi-coil baseline of~\citet{zhang2020extending} (see Section~\ref{sec:baselines}), which employs such straight-through estimation as well.

\subsection{Integrating policy network and E2E VarNet} \label{sec:pol_varnet_integration}

The original E2E VarNet takes as input a partially observed $k$-space $\bm{\tilde y}$, a binary sampling mask of ACS measurements $\mathbf{M}_{\textrm{ACS}}$, and a \emph{predefined} binary sampling mask of non-ACS measurements $\mathbf{M}'$. In the adaptive acquisition setup $\mathbf{M}'$ is instead \emph{predicted} by the policy network, so we further decompose $\bm{\tilde y}$ into the ACS measurements $\bm{\tilde y}_{\textrm{ACS}}$ and the policy prediction $\bm{\tilde y}'$. As a result, our E2E VarNet takes as input $\bm{\tilde y}_{\textrm{ACS}}$, $\bm{\tilde y}'$, $\mathbf{M}_{\textrm{ACS}}$, and $\mathbf{M}'$. Following the original work, we first estimate sensitivity maps from the ACS measurements $\bm{\tilde y}_{\textrm{ACS}}$ by means of a U-Net~\cite{ronneberger2015u}. The weights of the U-Net that predict the sensitivity maps are tied between the E2E VarNet and the Policy network, such that the sensitivity maps are re-used by both networks. Then, the output of the Policy network $\mathbf{M}'$ is used to acquire $\bm{\tilde y}'$. As a result, we obtain a mask of observed measurements $\mathbf{M} = \mathbf{M}_{\textrm{ACS}} + \mathbf{M}'$ and a partially observed $k$-space $\bm{\tilde y} = \bm{\tilde y}_{\textrm{ACS}} + \bm{\tilde y}'$ that are used inside the E2E VarNet to produce a high fidelity MR image reconstruction. The original E2E VarNet is composed of cascaded modules that apply soft data consistency (DC) layers and refinement operations simultaneously. DC layers ensure that $k$-space predictions stay close to the observed $k$-space, while the refinement operations refine the $k$-space predictions by applying a U-Net to the corresponding complex-valued image-space representation. The cascaded and data consistency structure of the E2E VarNet offers several potential interface points with the policy model. After experimentation we chose to re-purpose the DC layers to input acquisitions made by the policy into the E2E VarNet pipeline. More precisely, we introduce hard DC layers into the E2E VarNet~\cite{schlemper2017deep}, which directly replace phantasised measurements in the $k$-space predictions with the observed measurements. To ensure that each cascade possesses all relevant information, we apply DC and refinement operations sequentially, rather than simultaneously. An E2E VarNet cascade now first applies a hard DC layer -- which inputs acquired measurements, and restores changes to observed measurements due to a previous cascade --, followed by a refinement operation on the subsampled $k$-space $\bm{\tilde y}$. \\

\section{Experiments}

\subsection{Data}
We use the \fastMRI multi-coil knee dataset~\cite{fastmri} for all experiments, which contains 973 train volumes and 199 validation volumes of fully sampled $k$-space data. The test volumes are not fully sampled, and therefore cannot be used for our purposes. Our models treat every slice in a volume independently, resulting in 34,742 train slices and 7,135 validation slices to use as our dataset. For ease of experimentation, we reduce the number of coils by taking a Singular Value Decomposition and using the $K = 4$ coils with the largest singular values~\cite{coilcomp}. We further crop the MR slices to the center $(128 \times 128)$ region of $k$-space, see Appendix~\ref{app:crop}. To simulate clinical conditions more closely, we create the ground truth target image from the \emph{uncompressed} $k$-space by applying a coil-wise inverse Fourier transform followed by a root-sum-of-squares (RSS) on the resulting multi-coil image representation. 

\subsection{Baselines} \label{sec:baselines}
We compare our method to two baselines: Equispaced (or Equisdistant) and LOUPE. Equispaced masks were shown by~\citet{hammernik2018learning} to be a strong hand-designed subsampling strategy for deep learning-reconstructed multi-coil MRI. Such masks have been long-used in the parallel imaging literature and are a current standard for clinical 2D imaging. At the same time, they depart from random patterns favoured by compressed sensing, which is applied for single-coil reconstruction. LOUPE is a dataset-specific (\ie non-adaptive) learned strategy that aims to optimise a single subsampling mask for the entire dataset, without conditioning on initial $k$-space measurements \cite{bahadir}. Recently,~\citet{zhang2020extending} extended LOUPE to the multi-coil setting, and we use their method as implemented by~\citet{codesign}, which employs the same normalisation and straight-through estimation used by our Policy network.


\subsection{Training details} \label{sec:training_details}
All models are trained to optimise SSIM~\cite{ssim} using Adam~\cite{adam} for $50$ epochs with a learning rate of $0.001$, decaying it by a factor $10$ on epoch $40$ --- the default \fastMRI E2E VarNet training setting. We initialise the $4\times$ acceleration experiments with the $10$ lowest frequency measurements $k$-space, and acquire $22$ more measurements with our models. The $8\times$ acceleration experiments are initialised with the $4$ lowest frequency measurements, and we acquire $12$ more, instead. This initialisation corresponds, to the ACS used in the \fastMRI E2E VarNet implementation to estimate sensitivity maps\footnote{\href{https://github.com/facebookresearch/fastMRI/commit/ada72932ba5b0486c6f11e13534ea91cb165b789}{As of July 30th, 2021.}}, and is thus a natural starting point for acquisition. See Appendix~\ref{app:implementation} for additional details. We empirically search over several hyperparameter settings for both the reconstruction model and the policy network. For the reconstructor, we consider either $5$ or $7$ cascades, and either $18$ or $36$ channels in the first layer of the refinement modules. We also explore the heatmap non-linearity mentioned in Section~\ref{sec:policynet} and run experiments for both the sigmoid and softplus non-linearities, using $\text{slope} \in \{1, 5, 10\}$ and $\beta \in \{0.5, 1, 5\}$, respectively. Since LOUPE employs the same normalisation and straight-through estimation as our Policy network, we also explore these non-linearities for LOUPE. Unless otherwise specified, we report the best (averaged over seeds) run under the explored hyperparameters. 

\subsection{Results} \label{sec:results}

\begin{figure}
\begin{minipage}[t]{0.49\textwidth}
\centering
\captionof{table}{\small Results on the undersampled multi-coil dataset. Policy performs on-par with the best competing model at $4\times$ and dominates performance at $8\times$.\looseness-1}
\resizebox{1.\textwidth}{!}{
\begin{tabular}{cccc}
  & \bfseries Policy & \bfseries LOUPE & \bfseries Equispaced \\
  $4\times$ & $\textbf{95.63} \pm 0.27$ & $\textbf{95.61} \pm 0.55$ & $95.38 \pm 0.03$ \\
  $8\times$ & $\textbf{93.26} \pm 0.20$ & $91.37 \pm 0.67$ & $91.30 \pm 0.06$ \\
\end{tabular}}
\vspace{0.6cm}

\label{tab:main}
\end{minipage}
\hfill
\begin{minipage}{0.49\textwidth}
     \centering
     \includegraphics[width=1.\textwidth,valign=B]{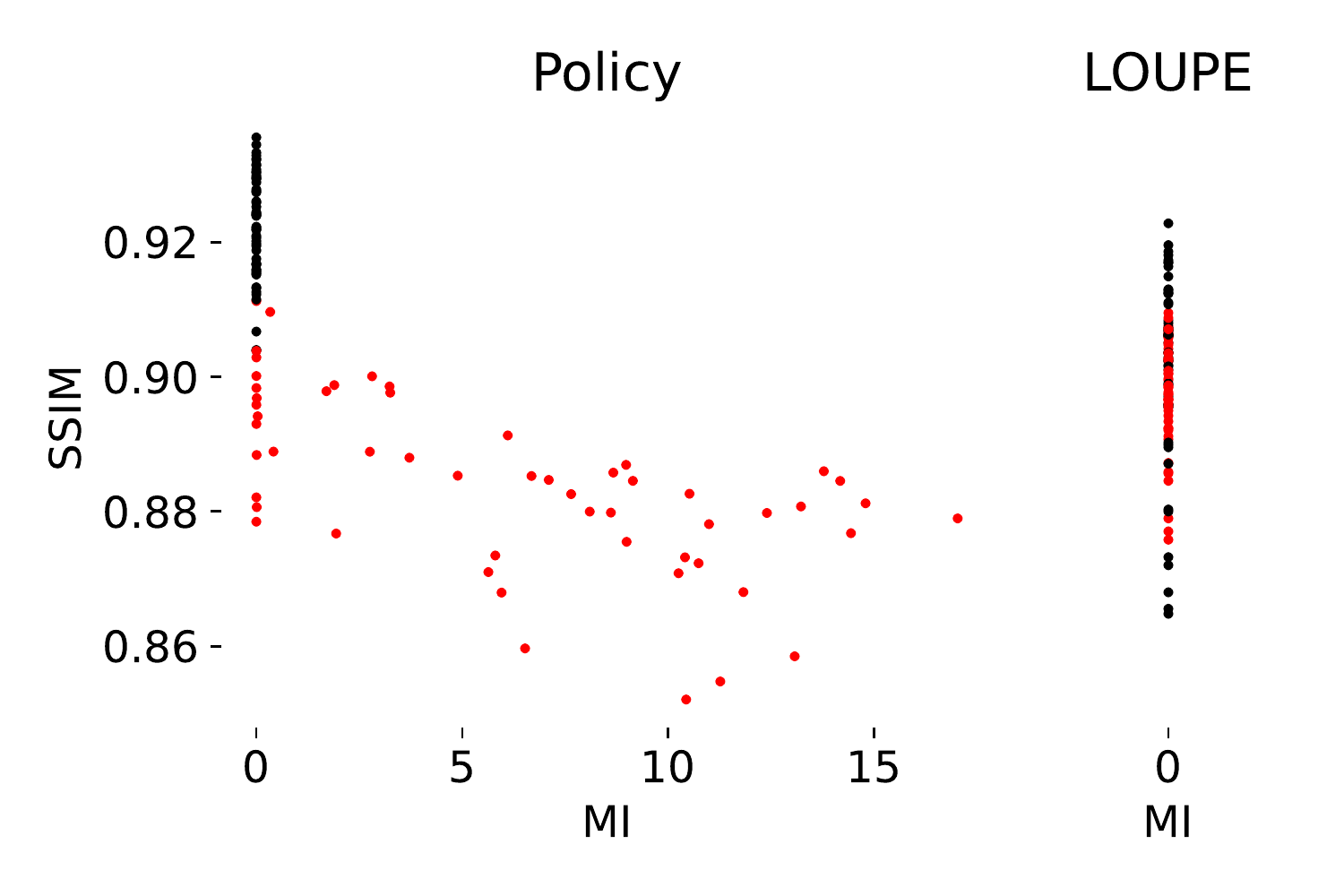}
    \captionof{figure}{\small SSIM as a function of policy mutual information (MI), for the $8\times$ setting. Each dot is a single model. Red: softplus; black: sigmoid.}
    \label{fig:adaptivity_vs_quality}
\end{minipage}
\end{figure}
\begin{figure}[t]
\small 
     \centering
     \subfigure[Target]{\includegraphics[width=0.19\textwidth]{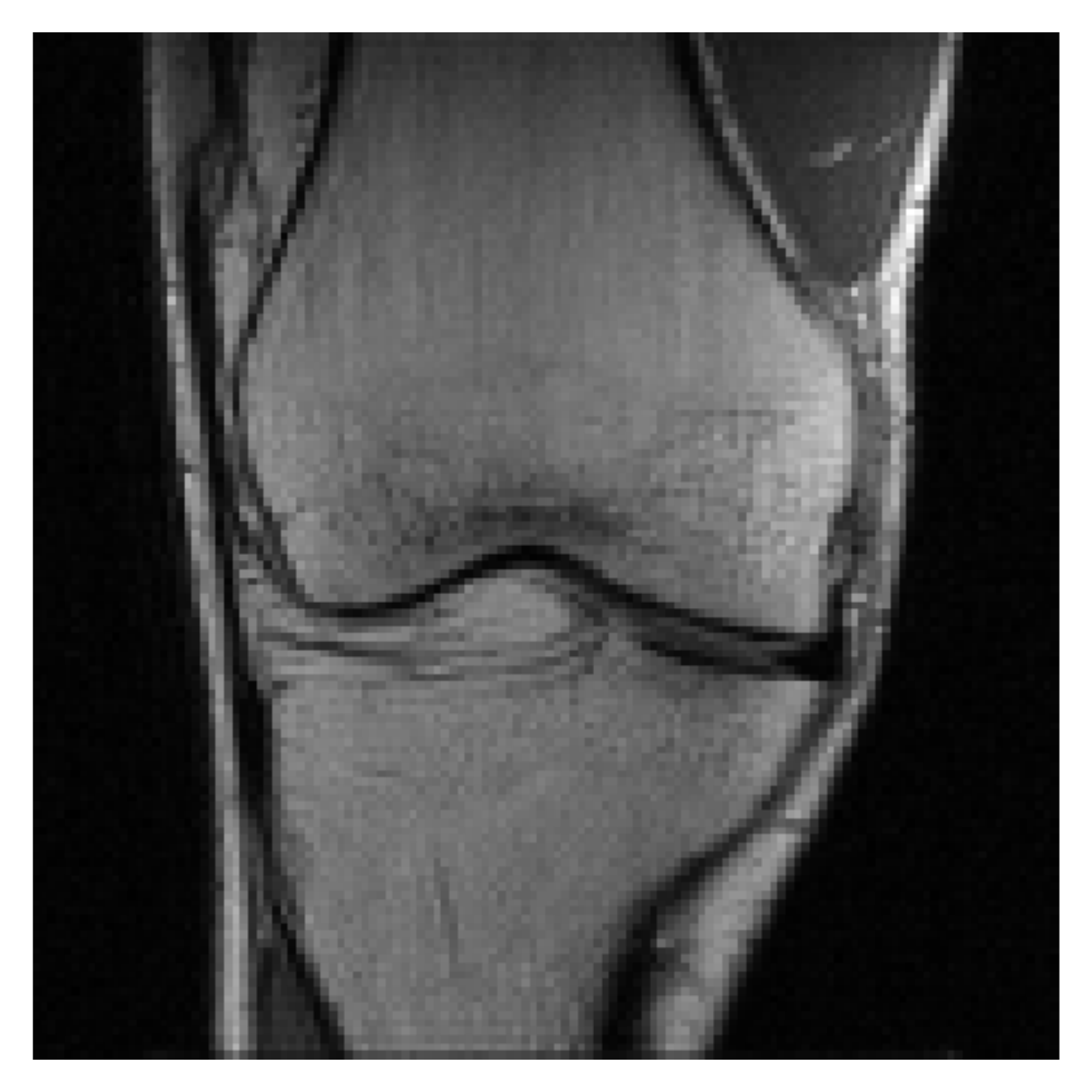}} 
     \subfigure[Equispaced]{\includegraphics[width=0.19\textwidth]{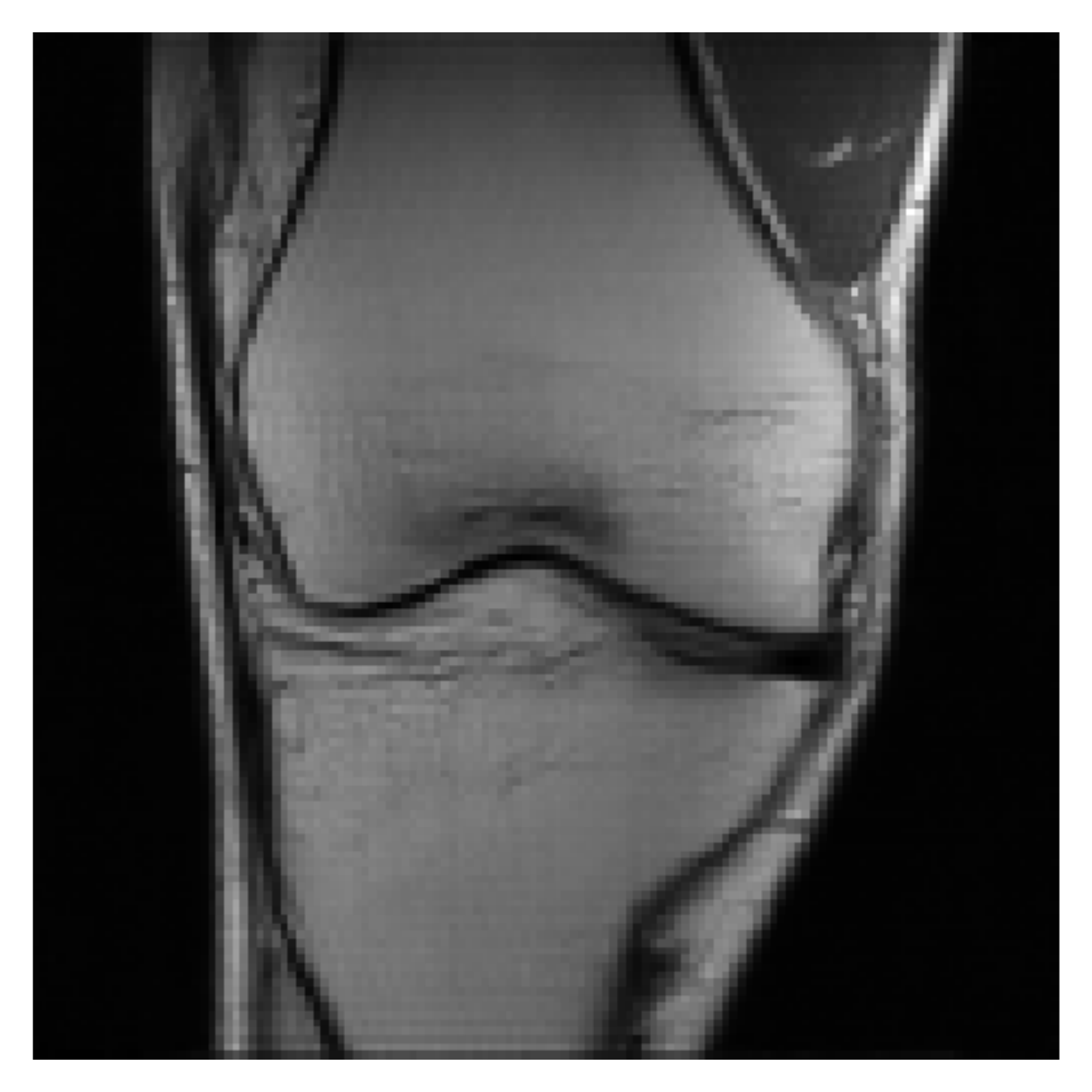}} 
     \subfigure[LOUPE]{\includegraphics[width=0.19\textwidth]{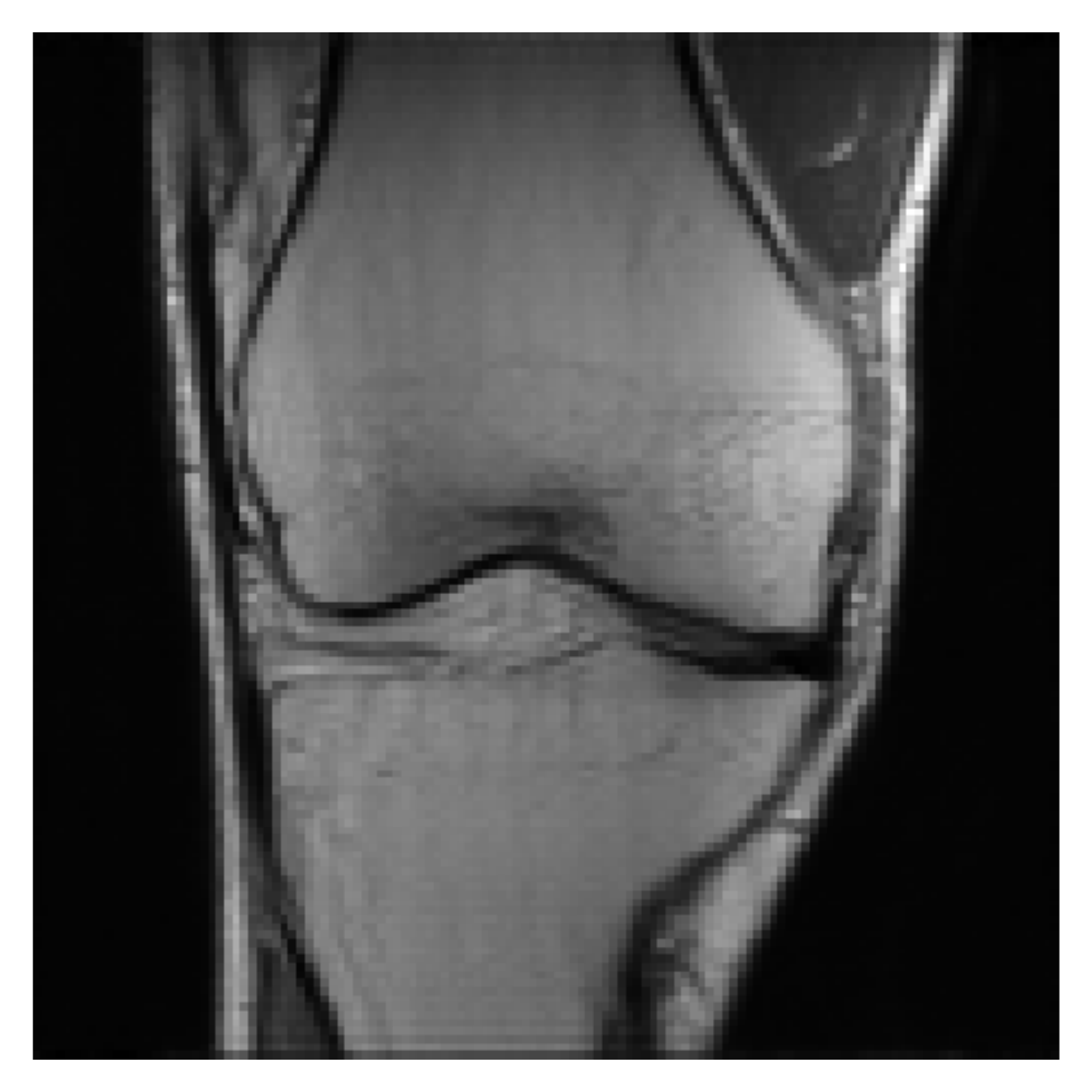}} 
     \subfigure[Sigmoid policy]{\includegraphics[width=0.19\textwidth]{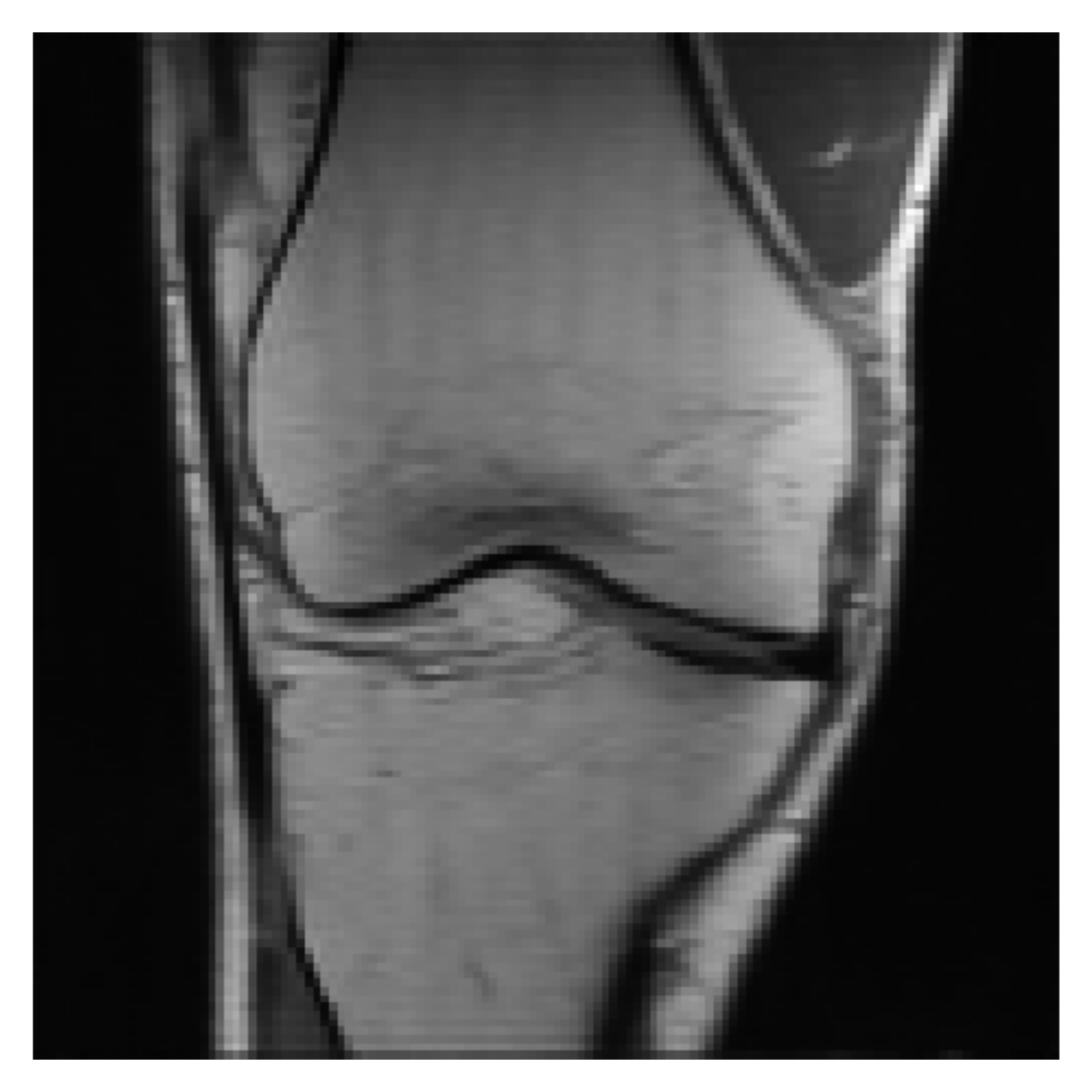}}
     \subfigure[Softplus policy]{\includegraphics[width=0.19\textwidth]{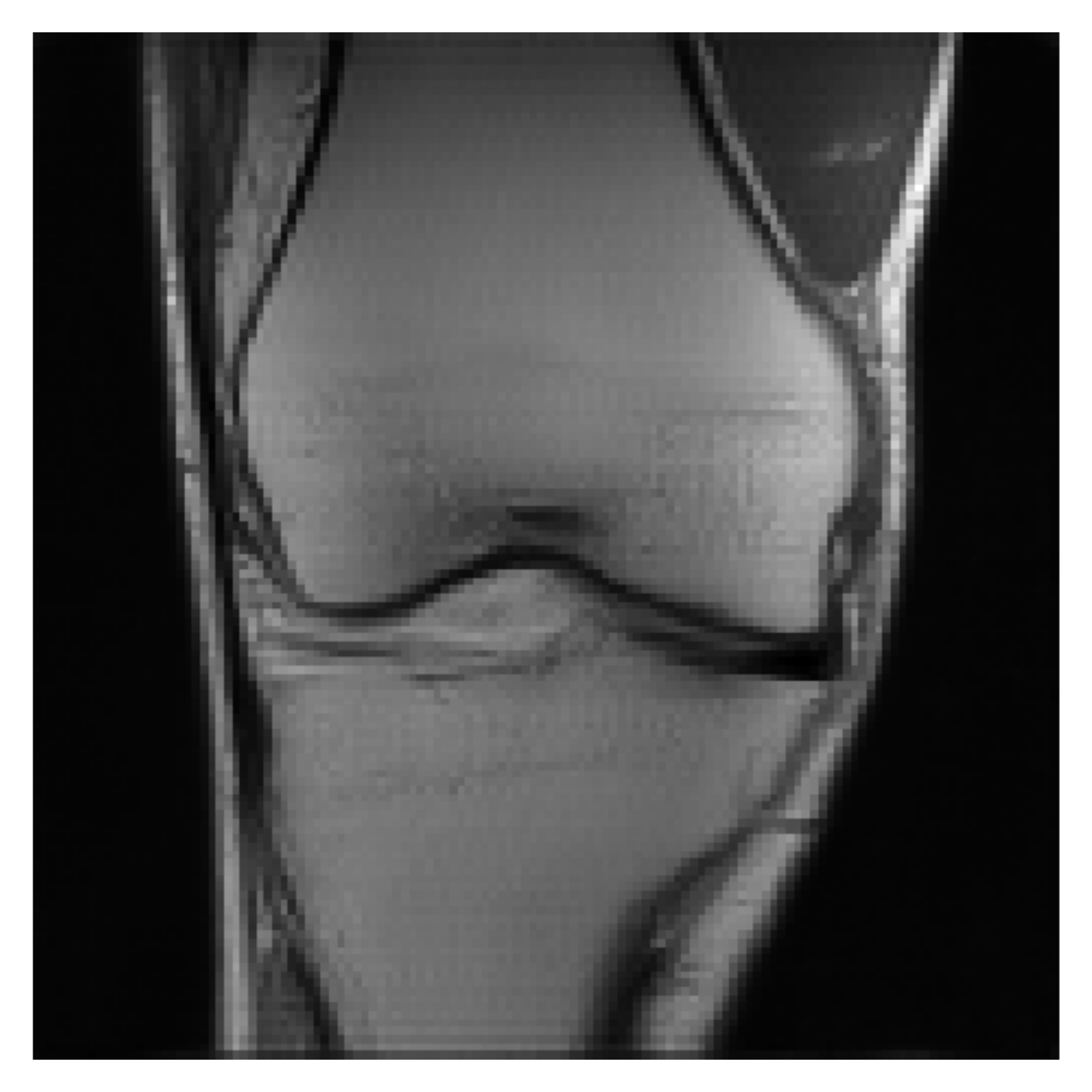}}
     \subfigure[Equispaced]{\includegraphics[width=0.22\textwidth]{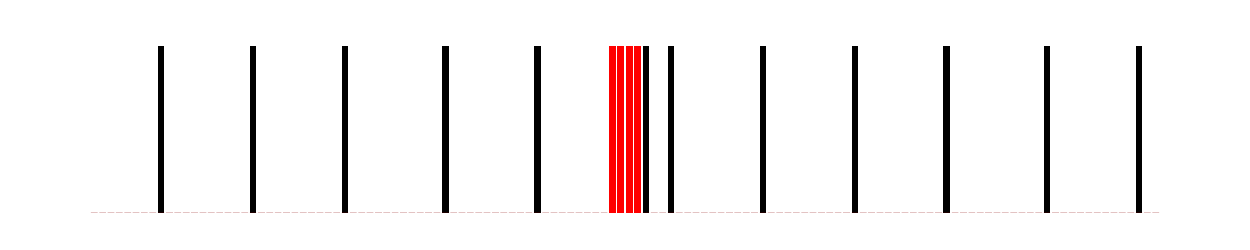}}
     \subfigure[LOUPE]{\includegraphics[width=0.22\textwidth]{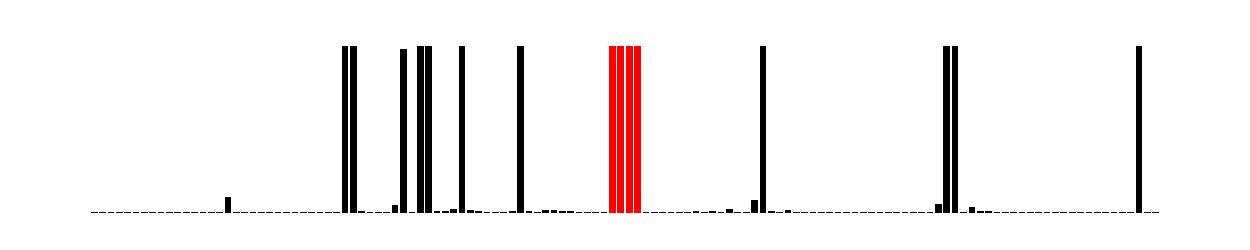}}
     \subfigure[Sigmoid policy]{\includegraphics[width=0.22\textwidth]{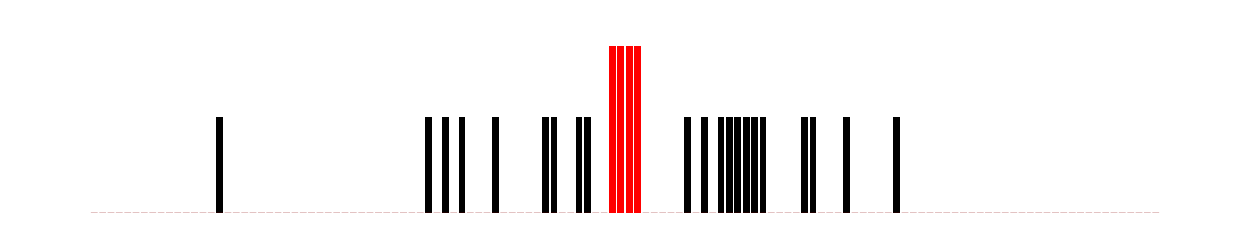}}
     \subfigure[Softplus policy]{\includegraphics[width=0.22\textwidth]{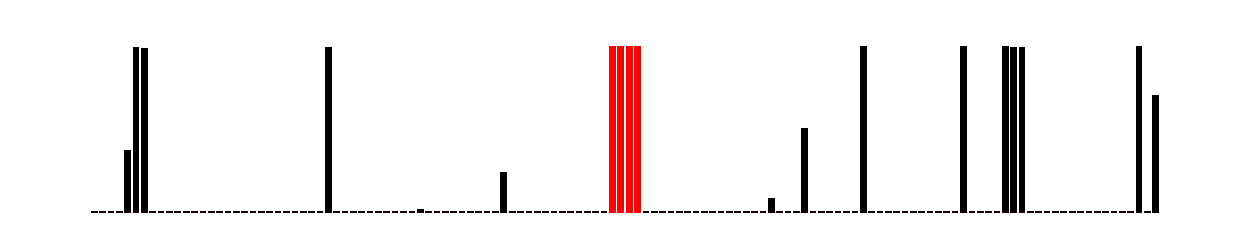}}
    \caption{Qualitative results for $8\times$: (a) Ground truth, (b-e) reconstructions, and (f-g) visualisations of subsampling policies. Each policy depicts 128 probabilities; one per potential $k$-space measurement. The ACS region (red) has probability 1.}
    \label{fig:main_qualitatives}
\end{figure}

We report our main results in Table~\ref{tab:main}, where we present validation SSIM for the best performing hyperparameter setting of each model. While the Policy network performs on-par with LOUPE at $4\times$ acceleration, it outperforms the best competing method at $8\times$ accelerations by $1.89$ SSIM points. To further understand the gains obtained by the Policy model, we inspect the adaptivity of the learned policies by plotting the SSIM score as a function of the mutual information (MI) between probability masks and images of all learned policies (Policy and LOUPE) --- see Figure~\ref{fig:adaptivity_vs_quality} for the 8$\times$ acceleration results. We plot the policies that use the sigmoid non-linearity in black, and policies that use the softplus non-linearity in red. Surprisingly, we observe that the best performing Policy models exhibit zero mutual information, denoting no adaptivity --- \ie the distribution of actions is constant for all data points in our dataset. We observe that early in the learning process all policies are adaptive and that some of them \emph{learn to be non-adaptive}. Moreover, we observe that all sigmoid-based Policies end up being non-adaptive while the majority of the softplus-based Polices converge to adaptive strategies, suggesting that this non-linearity plays a crucial role in learning adaptivity. We hypothesise that it may be easier for Policies using the sigmoid non-linearity to learn non-adaptive strategies -- which requires learning to ignore their input -- given the saturation of the function for both very large positive and negative values. In contrast, the softplus non-linearity only saturates for large negative values, while the model is simultaneously encouraged to assign positive values to at least $B$ actions when sampling from the distribution. Figure~\ref{fig:main_qualitatives} displays some examples of image reconstructions and learnt subsamplings for the $8\times$ acceleration. The sigmoid-based Policy chooses a subset of actions to sample from with equal probability, whereas LOUPE appears to yield very-nearly sparse probabilities; exhibiting probability close to $1$ for a limited set of actions, while most actions end up with a probability near $0$. The more adaptive softplus-based Policy assigns less regular probability values, and seems to favour the center region less than both its sigmoid-based counterpart and LOUPE. Appendix~\ref{app:qual} contains additional qualitative results.

\subsubsection{Discussion} \label{sec:discussion}
In this subsection, we outline three hypotheses as to why the best performing policies learn to be non-adaptive. The first one is model \emph{generalisation}: the model may learn a strong adaptive Policy on the training set that does not generalise to the validation set. However, we examine the Policy adaptivity on the training set in Appendix~\ref{app:mi}, and find no evidence that the lack of adaptivity is caused by overfitting the train data. The second hypothesis is connected to \emph{amortisation} of the parameters of the reconstruction model over the large training dataset. Adaptivity can act as a regulariser on the reconstruction model, since a single reconstruction model needs to reconstruct MR images following multiple different $k$-space sampling patterns. In Appendix~\ref{app:rec_capa_nonlin}, we report results for systems trained with reconstruction models of increasing capacities, and assess whether leveraging higher capacity reconstructors can help achieve Policy adaptivity. Although the higher capacity models lead to slight improvements in terms of SSIM, the best models are still not adaptive. Finally, we hypothesise that the estimation of sensitivity maps -- a significant difference between single-coil and multi-coil reconstruction pipelines -- may affect the adaptivity of the joint model. In the E2E VarNet, sensitivity maps are estimated independently per slice, and this may enable a form of adaptivity beyond the mask selection that we have explored here. However, the interaction between the sensitivity maps and the acquisition policies requires further investigation and is left as future work (see Appendix~\ref{app:future}). We display some learned sensitivity maps for all models in Appendix~\ref{app:sense}.   


\section{Conclusion}
We have explored the problem of jointly optimising an adaptive sampling strategy and a deep reconstruction model for multi-coil 2D MRI. To this end, we have proposed the first method for integrating a policy network with the E2E VarNet reconstruction model, and evaluated it on the large-scale, multi-coil \fastMRI knee dataset. Our Policy networks outperform previous learning based non-adaptive approaches as well as the handcrafted equispaced masks at $8\times$ accelerations. Interestingly however, the Policies learn to be explicitly non-adaptive. The main limitation of our work is the analysis on coil-compressed $(128 \times 128)$ $k$-space acquisitions, rather than the full data. The relevance of our work for the original, un-cropped data should be empirically verified by future research.



\midlacknowledgments{T. Bakker's PhD work is partially supported by the ‘Efficient Deep Learning’ (EDL, \url{https://efficientdeeplearning.nl}) research programme, which is financed by the Dutch Research Council (NWO) domain Applied and Engineering Sciences (TTW).

We are grateful to the Weights\&Biases team \cite{wandb} for providing their experiment tracking software. We would also like to thank Edward Smith for useful discussions.}

\bibliography{references}

\clearpage

\section*{Supplementary material to ``On learning adaptive acquisition policies for undersampled multi-coil MRI reconstruction".}
\appendix

\section{Related work}

Our literature review focuses on online sampling design algorithms for gridded trajectories. As such, we do not cover other substantial fields of MRI acceleration work, such as on non-Cartesian trajectories, dynamic imaging, and 3D sampling~\cite{reviewfastpulse, reviewcompressed, reviewnoncartesian, reviewfancyrecon, reviewdlmri}.

Following the discovery of Fourier encoding~\cite{mansfield1977multi}, Cartesian MRI sampling patterns would collect the entire Fourier transform of the object prior to reconstruction. The advent of parallel imaging~\cite{sodickson1997simultaneous,pruessmann1999sense,griswold2002generalized} enabled to collect less than the full Fourier transform for reconstruction. Sampling patterns with parallel imaging are regular (\ie collecting every other line) due to the beneficial noise properties of regular undersampling that arise from the smoothness of the coil sensitivity patterns. The introduction of compressed sensing~\cite{cs} altered this paradigm, as random patterns are beneficial for compressed sensing. Traditionally in MRI, this divergence has been settled with patterns designed based on heuristics, such as Poisson-disc sampling~\cite{vasanawala2011practical}. Motivation to improve over heuristics has led to a wealth of research on sampling pattern design, which can be broadly classified based on 1) their adaptivity and 2) whether they jointly optimise the reconstructor.

\textbf{Non-adaptive sampling / Non-joint training:} Several works have proposed optimising sampling patterns for sparse reconstruction. Some works considered incoherence between Fourier operators and sparse bases~\cite{chauffert2014variable, adcock2017breaking}. Optimising sampling patterns while considering sensitivity coils is more difficult because the coils are patient-specific, but some works have considered this case with on-the-fly methods~\cite{levine2017fly}.

\textbf{Non-adaptive sampling / Joint training:} In the CS-based literature, the work by~\citet{ravishankar2011adaptive} investigated the use of alternate optimisation to design a fixed sampling pattern for specific reconstruction methods. More recently,~\citet{bahadir} introduced the LOUPE algorithm, which co-jointly trains a non-adaptive probabilistic sampling pattern and deep learning-based reconstruction model. Other variants of this joint training idea have been proposed, exploring different ways to model the sampling pattern and differentiating through the sampling operation~\cite{weiss, huijben2020learning, wang2021b}. Recent work has also studied the use of alternate/iterative optimisation techniques~\cite{razumov2021optimal} in the context of deep learning-based reconstructions, optimising the samples for downstream tasks rather than for reconstruction only. However, in general, the majority of these work has focused on the single-coil setup. 

\textbf{Adaptive sampling / Non-joint training:} A recent line of work has focused on adaptive methods for a pre-trained reconstructor. One approach formalises the MRI acquisition problem as a partially observable Markov decision process (POMDP), and proposed the use of reinforcement learning~\cite{sutton2018reinforcement} to solve it. In particular, \citet{pg} considered the use of policy gradients~\cite{gpomdp}, while \citet{pineda2020active} proposed using the DDQN algorithm~\cite{van2016deep}. An alternative approach considers combining Bayesian inversion with posterior estimation to select $k$-space points~\citet{sanchez3}. The non-joint nature of these models requires a pre-trained reconstruction model, which is usually effectively trained \emph{off-policy} with respect to the subsampling strategy (as it is learned only after reconstruction training concludes). This may be detrimental to overall system performance due to 1) the reconstruction model expending capacity on images resulting from subsampling patterns that are unlikely under the learned policies, and 2) the policies learning under the constraint that its proposed subsampling masks are within distribution for the reconstruction model. These difficulties suggest the need for a jointly optimised system.

\textbf{Adaptive sampling / joint training:} Following the previous discussion, another recent trend involves jointly training the sampler and the reconstructor. \citet{jin19} trained the sampler to mimic Monte Carlo tree search~\cite{silver2017mastering}, while the reconstructor was simultaneously trained on the patterns generated by the sampler, but without any direct gradient path from one to the other. \citet{zhang2019reducing} proposed an uncertainty criterion for action selection, using a GAN-inspired training objective between reconstructor and the sampler. \citet{codesign,van2021active} are the closest to our work, in that they co-design both the sampler and the reconstructor in a fully-differentiable framework, with \citet{codesign} specifically focusing on the question of joint vs non-joint optimisation and finding the former advantageous. However, in contrast to our work, all these approaches are evaluated only in the single-coil MRI setting.

\section{Future work} \label{app:future}
The optimal degree of adaptivity is unknown for both single- and multi-coil MRI subsampling policies, but especially the multi-coil setting remains under-explored. We hope that our results, especially the discussion in Section~\ref{sec:discussion} of the main text, open up promising avenues for future research.

We suggest that one such avenue may be to explore the potential over-regularisation effect of subsampling adaptivity, by employing reconstruction models better equipped to distinguish inputs resulting from varying $k$-space masks. This may, for instance, be achieved by conditioning the reconstruction model on the subsampling mask explicitly, either by adding the subsampling mask as input to the refinement module, or possibly through the use of a hypernetwork~\cite{hypernet} that predicts refinement module parameters conditional on $\mathbf{M}$.

Another potentially fruitful research direction is to explore to what degree the per-slice sensitivity maps of the E2E VarNet confer an intrinsic adaptivity to the joint model. Sensitivity maps represent the primary conceptual difference between the single- and multi-coil settings, and may present a unique challenge in designing adaptive strategies for multi-coil MRI reconstruction.

Finally, beyond adaptive policies, a promising area of research may be to explore \emph{active} multi-coil policies, \ie strategies that dynamically adapt to acquired measurements during the scanning process. While in our experiments we apply a single Policy network to the ACS input, our utilisation of hard DC layers sequentially with refinement allows for placing the Policy in-between cascades of the E2E VarNet as well. We briefly experimented with this, and found no significant differences in performance applying the Policy to the reconstruction resulting from the first cascade instead. However, we did not deeply explore the possibility of interspersing multiple Policies among the cascades of the E2E VarNet: by conditioning the Policy on outputs of previous Policies in such a way, the model may be able to learn active subsampling strategies

\paragraph{Limitations of our work:} We performed our analysis on coil-compressed $(128 \times 128)$ $k$-space acquisitions, rather than the full data. The relevance of our work for the original, un-cropped data should be empirically verified by future research.

\section{Implementation details} \label{app:implementation}
In this section we provide additional implementation details.

\subsection{Cropping} \label{app:crop}
To properly crop the \fastMRI knee $k$-space, we need to account for its oversampling in the frequency encoding direction, as well as preserve the correct aspect-ratio. To this end, we first transform the full MR $k$-space to the image domain, and crop to the \texttt{recon\_size} attribute stored in the \texttt{.h5} file metadata. The result is then transformed to $k$-space, where we finally crop to $(128 \times 128)$. The target is constructed from this cropped $k$-space.

\subsection{Data range}
Computing SSIM and PSNR values requires specifying a data range (or dynamic range) that denote the maximum value a pixel can take. Our coil-compressed implementation employs the maximum of the slice as the data range during training, but validation numbers presented in this paper are computed using the maximum of the volume as data range, as is standard in the MRI literature.

\subsection{Early stopping}
The default fastMRI E2E VarNet implementation employs a form of early stopping, which we also follow: models are trained for the full 50 epochs mentioned in Section~\ref{sec:training_details} of the main text, but only the model that performs best on validation SSIM is saved as a checkpoint and used to report final results.

\subsection{Policy network architecture} \label{app:polarch}
The Policy network starts with an initial $(1\times1)$ convolution with $2$ (real and imaginary) input channels and $16$ output channels, followed by instance normalisation~\cite{instancenorm} and ReLU~\cite{relu} activation. We further employ four convolutional blocks, each consisting of a zero-padded $(3\times3)$ convolution layer that doubles the number of channels, followed by an instance normalisation, ReLU activation, and $(2\times2)$ max-pooling layer. Note that this architecture is reminiscent of the first half of the default fastMRI U-Net~\cite{fastmri}. The resulting tensor is flattened and fed through two dense layers each composed of $256$ output units, followed by a leaky-ReLU~\cite{leakyrelu} activation with slope $0.01$, and one final dense layer with number of output units equal to the number of measurements available in the full $k$-space $M$. The Policy has $4,685,712$ total parameters.

\subsection{Sparse and multiplicative DC} \label{app:sparsemultidc}
As mentioned in Section~\ref{sec:pol_varnet_integration} of the main text, our implementation of a cascade differs slightly from the original E2E VarNet. Specifically, within a cascade we employ a hard DC layer sequentially with the refinement module, rather than a soft DC layer simultaneously with the refinement module. When combined with learned sampling strategies, this change opens up an interesting choice in the DC layer implementation. The original DC layer is implemented using a sparse operation, namely \texttt{torch.where(mask, observed\_kspace - phantasised\_kspace, 0)} on the difference between the observed and phantasised $k$-space. While this is no different from a non-sparse operation (\eg multiplying the difference by the mask values directly) when the observed $k$-space is not part of training, as soon as the observed $k$-space requires gradients (such as when training the Policy network or LOUPE) this sparse operation has the effect of removing gradients for unobserved $k$-space measurements. 

We have explored the effect of this choice by repeating all Policy and LOUPE experiments for both types of DC layer. All results presented in this paper include a search over DC layer type, as well as all the other hyperparameters mentioned. However, while we include this choice in our hyperparameter search, we find empirically that the choice makes no significant difference to the performance of any of our models.

\section{Additional results}

In this section we present and discuss additional quantitative (Appendix~\ref{app:quant}) and qualitative results (Appendix~\ref{app:qual}).

\subsection{Quantitative results} \label{app:quant}

Here we present some additional quantitative results. In Appendix~\ref{app:psnr_nmse} we report our main results table extended with the PSNR and NMSE metrics. Appendix~\ref{app:mi} contains additional mutual information results, as well as a discussion on the hypothesis that generalisation may be preventing the system from learning strong adaptive Policies: we conclude that this is unlikely. Finally, in Appendix~\ref{app:rec_capa_nonlin} we present results for systems trained with reconstruction models of increasing capacities, and further discuss the amortisation hypothesis: we conclude that future research is needed.

\subsubsection{PSNR and NMSE of main results} \label{app:psnr_nmse}

In Table~\ref{tab:extended_main} we report our main results extended with PSNR and NMSE values. SSIM is multiplied by 100, and NMSE by 1000 for readability. Note that SSIM ranking correlates well with PSNR and NMSE ranking. We also split results by non-linearity (sigmoid or softplus). While sigmoid-based strategies slightly outperform softplus-based strategies in all settings, the performance gap is much larger in specifically the 8$\times$ Policy setting.

\begin{table}[htbp]
\small
\floatconts
  {tab:extended_main}%
  {\caption{Extension of primary results. Reported are validation SSIM, PSNR and NMSE of the best performing models in each category, averaged over three seeds. The sigmoid strategies consistently outperform the softplus strategies.}}%
  {\begin{tabular}{ccccccc}
  & \multicolumn{6}{c}{\bfseries Policy} \\
  & \multicolumn{3}{c}{Sigmoid} & \multicolumn{3}{c}{Softplus} \\
  \bfseries Accel & SSIM & PSNR & NMSE & SSIM & PSNR & NMSE \\
  $4\times$ & $\textbf{95.63} \pm 0.27$ & $37.95 \pm 0.49$ & $6.040 \pm 0.60$ & $94.98 \pm 0.30$ & $37.05 \pm 0.35$ & $7.433 \pm 0.58$ \\
  $8\times$ & $\textbf{93.26} \pm 0.20$ & $35.17 \pm 0.12$ & $11.21 \pm 0.28$ & $90.40 \pm 0.55$ & $32.66 \pm 0.49$ & $20.22 \pm 2.11$ \\
  \\
  & \multicolumn{6}{c}{\bfseries LOUPE} \\
  & \multicolumn{3}{c}{Sigmoid} & \multicolumn{3}{c}{Softplus} \\
  \bfseries Accel & SSIM & PSNR & NMSE & SSIM & PSNR & NMSE \\
  $4\times$ & $\textbf{95.61} \pm 0.55$ & $37.73 \pm 0.92$ & $6.387 \pm 1.24$ & $94.94 \pm 0.14$ & $37.02 \pm 0.13$ & $7.457 \pm 0.23$ \\
  $8\times$ & $91.37 \pm 0.67$ & $33.34 \pm 0.45$ & $17.32 \pm 1.84$ & $90.76 \pm 0.17$ & $32.86 \pm 0.09$ & $19.21 \pm 0.40$ \\
  \\
  & \multicolumn{6}{c}{\bfseries Equispaced} \\
  \bfseries Accel & \multicolumn{2}{c}{SSIM} & \multicolumn{2}{c}{PSNR} & \multicolumn{2}{c}{NMSE} \\
  $4\times$ & \multicolumn{2}{c}{$95.38 \pm 0.03$} & \multicolumn{2}{c}{$37.80 \pm 0.06$} & \multicolumn{2}{c}{$6.307 \pm 0.11$} \\
  $8\times$ & \multicolumn{2}{c}{$91.30 \pm 0.06$} & \multicolumn{2}{c}{$33.70 \pm 0.07$} & \multicolumn{2}{c}{$15.77 \pm 0.22$} \\
  \end{tabular}}
\end{table}

\subsubsection{Mutual information} \label{app:mi}

In this section we present further results on Policy adaptivity through analysis of the mutual information of the learned subsampling strategies, following~\citet{pg}. In our case, the MI is computed under the assumption that the policy outputs $M$ independent Bernoulli distributions; one for each potential measurement site. Due to the rejection sampling step -- which guarantees the correct number of measurements $B$ is acquired -- the actual sampled measurements are not fully independent; thus, the independence assumption is an approximation. Note however that, due to the data-processing inequality, our MI approximation provides an upper bound on the MI of the joint acquisition distribution.

\begin{figure}[t]
\centering
\begin{minipage}{.48\textwidth}
  \centering
  \includegraphics[width=.9\textwidth]{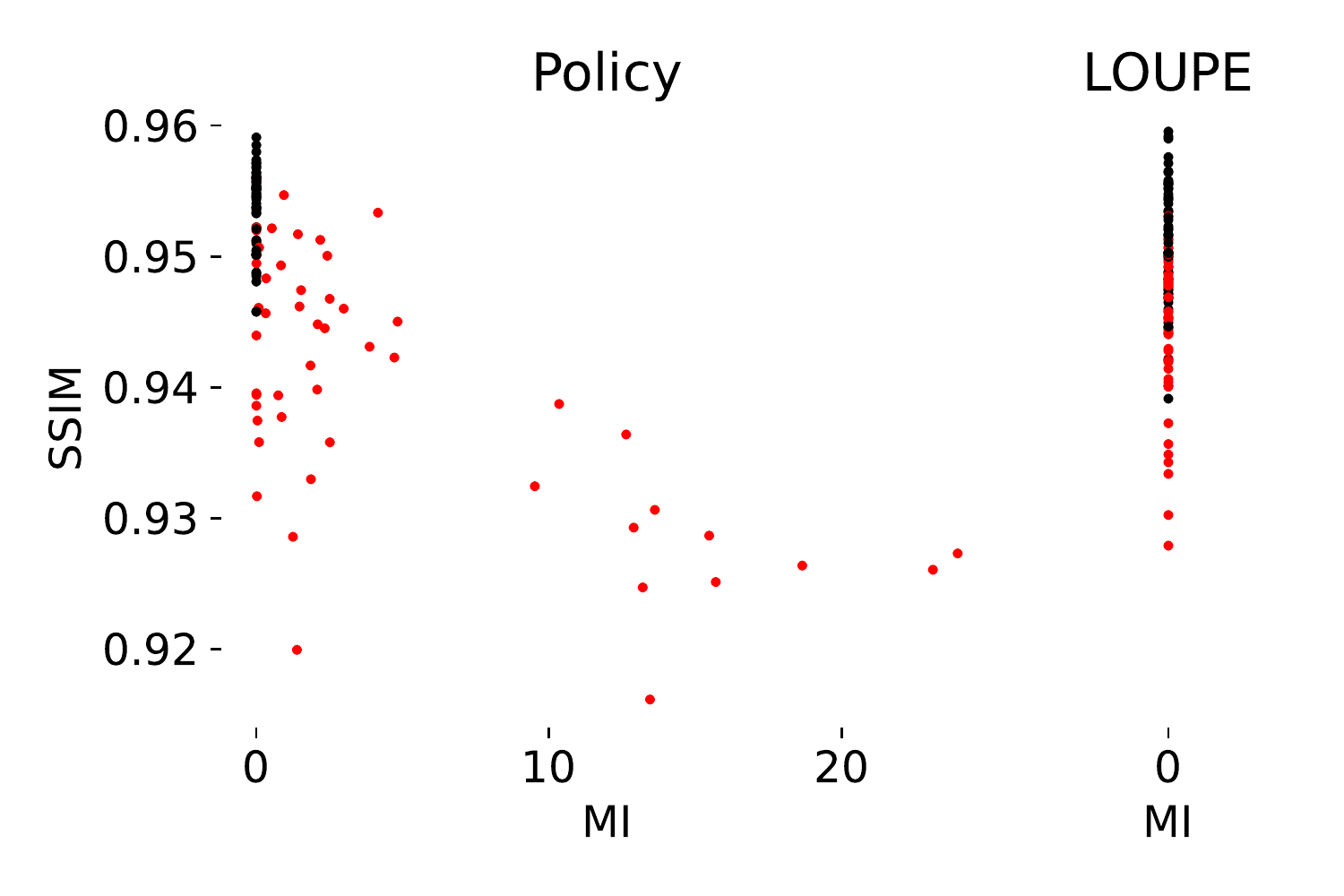}
  \captionof{figure}{\small Validation SSIM as a function of the policy adaptivity measured as the mutual information (MI) for the 4$\times$ acceleration factor. Each dot is a single model. Red: softplus; black: sigmoid.}
  \label{fig:SSIMvsMI_4x}
\end{minipage}\hfill
\begin{minipage}{.48\textwidth}
  \centering
  \includegraphics[width=.9\textwidth]{figures/MI_SSIM_8x_val.pdf}
  \captionof{figure}{\small Validation SSIM as a function of the policy adaptivity measured as the mutual information (MI) for the 8$\times$ acceleration factor. Each dot is a single model. Red: softplus; black: sigmoid.}
  \label{fig:SSIMvsMI_8x}
\end{minipage}
\end{figure}

Figure~\ref{fig:SSIMvsMI_4x} is the $4\times$ counterpart of the $8\times$ Figure~\ref{fig:adaptivity_vs_quality} from the main text, here re-presented as Figure~\ref{fig:SSIMvsMI_8x}. The trends for the Policy models are similar: the highest performing Policies employ the sigmoid non-linearity and are non-adaptive, although the effect is less exaggerated than in the $8\times$ setting. For LOUPE we note a qualitative difference in relative performance of the sigmoid and softplus models: for $4\times$ sigmoid models are clearly favoured while in the $8\times$ setting the difference is smaller.

\begin{figure}[t]
\centering
\begin{minipage}{.48\textwidth}
  \centering
  \includegraphics[width=.9\textwidth]{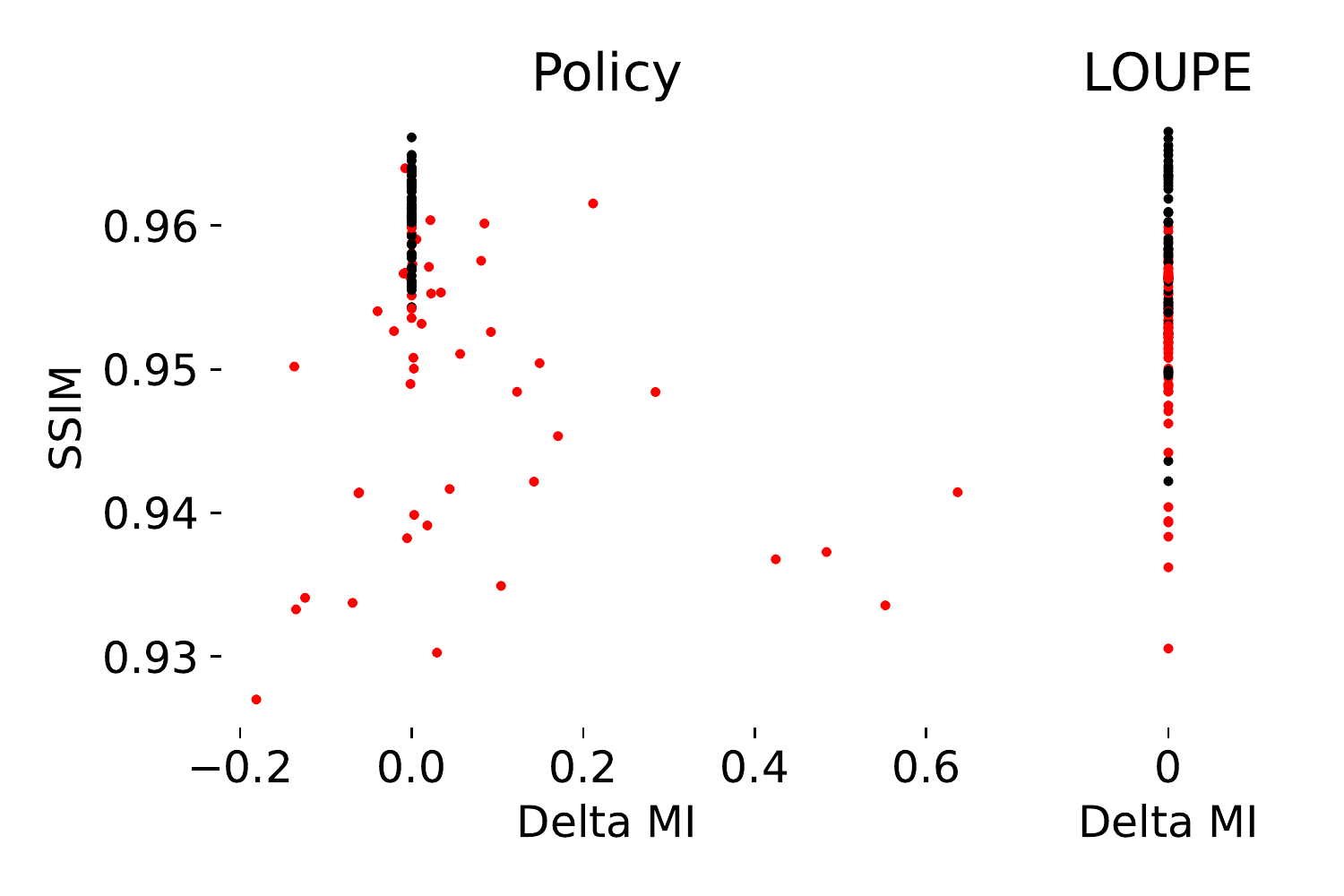}
  \captionof{figure}{\small Train SSIM as a function of the difference (train MI - validation MI), for the 4$\times$ acceleration factor. Each dot is a single model. Red: softplus; black: sigmoid.}
  \label{fig:SSIMvsMI_4x_delta_only_mi}
\end{minipage}\hfill
\begin{minipage}{.48\textwidth}
  \centering
  \includegraphics[width=.9\textwidth]{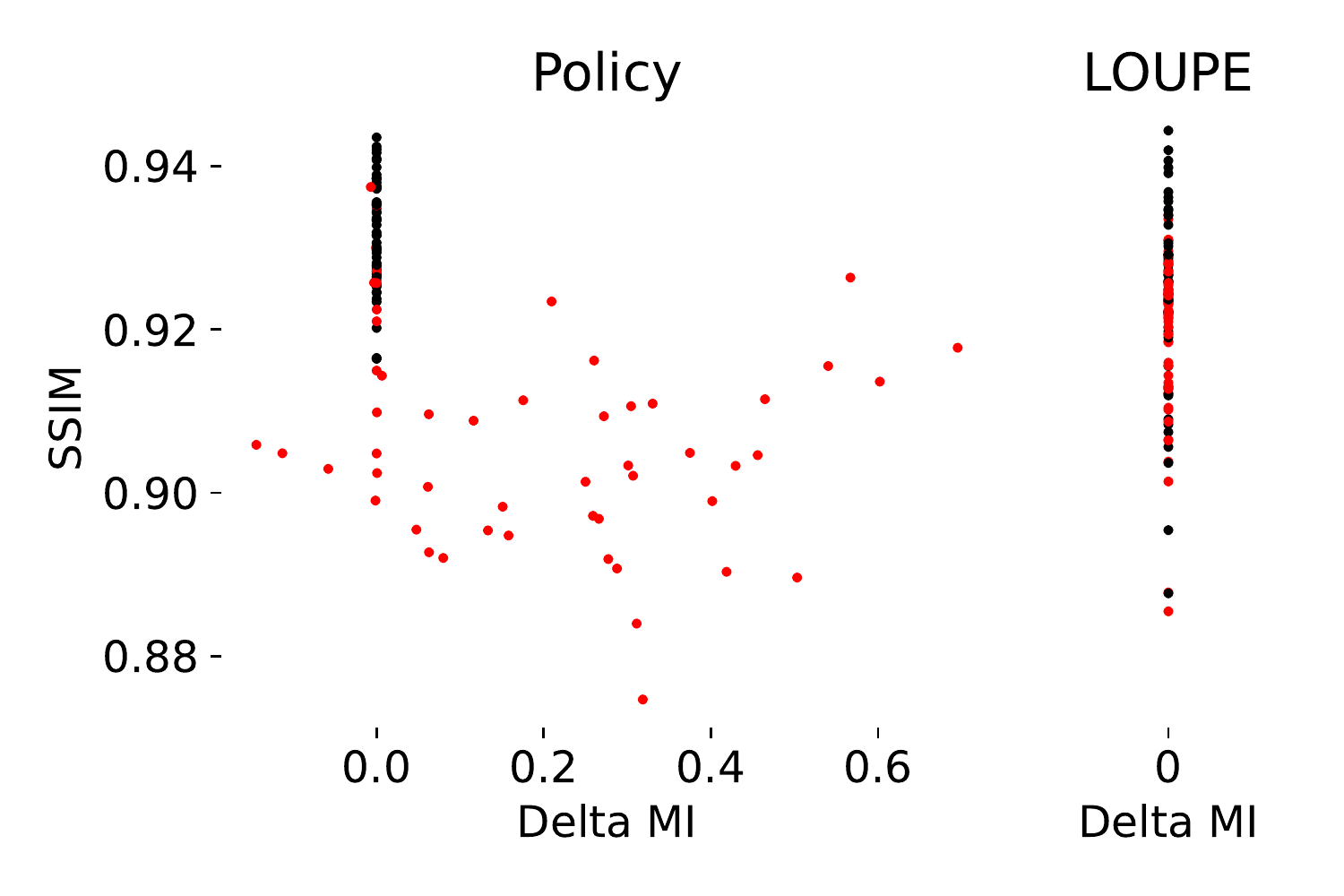}
  \captionof{figure}{\small Train SSIM as a function of the difference (train MI - validation MI), for the 8$\times$ acceleration factor. Each dot is a single model. Red: softplus; black: sigmoid.}
  \label{fig:SSIMvsMI_8x_delta_only_mi}
\end{minipage}
\end{figure}

In Figures~\ref{fig:SSIMvsMI_4x_delta_only_mi} and~\ref{fig:SSIMvsMI_8x_delta_only_mi} we present similar plots, where now \emph{train} SSIM is placed against the difference between train and validation MI (labeled as `Delta MI'). We observe that all the best performing models are no more adaptive on the train data than on the validation data: given that these models are non-adaptive on the validation data, we conclude that they are non-adaptive on the train data as well. This observation provides evidence against the hypothesis that the lack of high-performing adaptive Policies is due to generalisation issues. In Section~\ref{sec:results} we noted that non-adaptivity requires the Policy to learn to ignore the data: indeed, if the Policy has learned to ignore validation data, it must have learned to do so on the training set: this additionally suggests that the generalisation hypothesis is unlikely.

\begin{figure}[t]
\centering
\begin{minipage}{.48\textwidth}
  \centering
  \includegraphics[width=.9\textwidth]{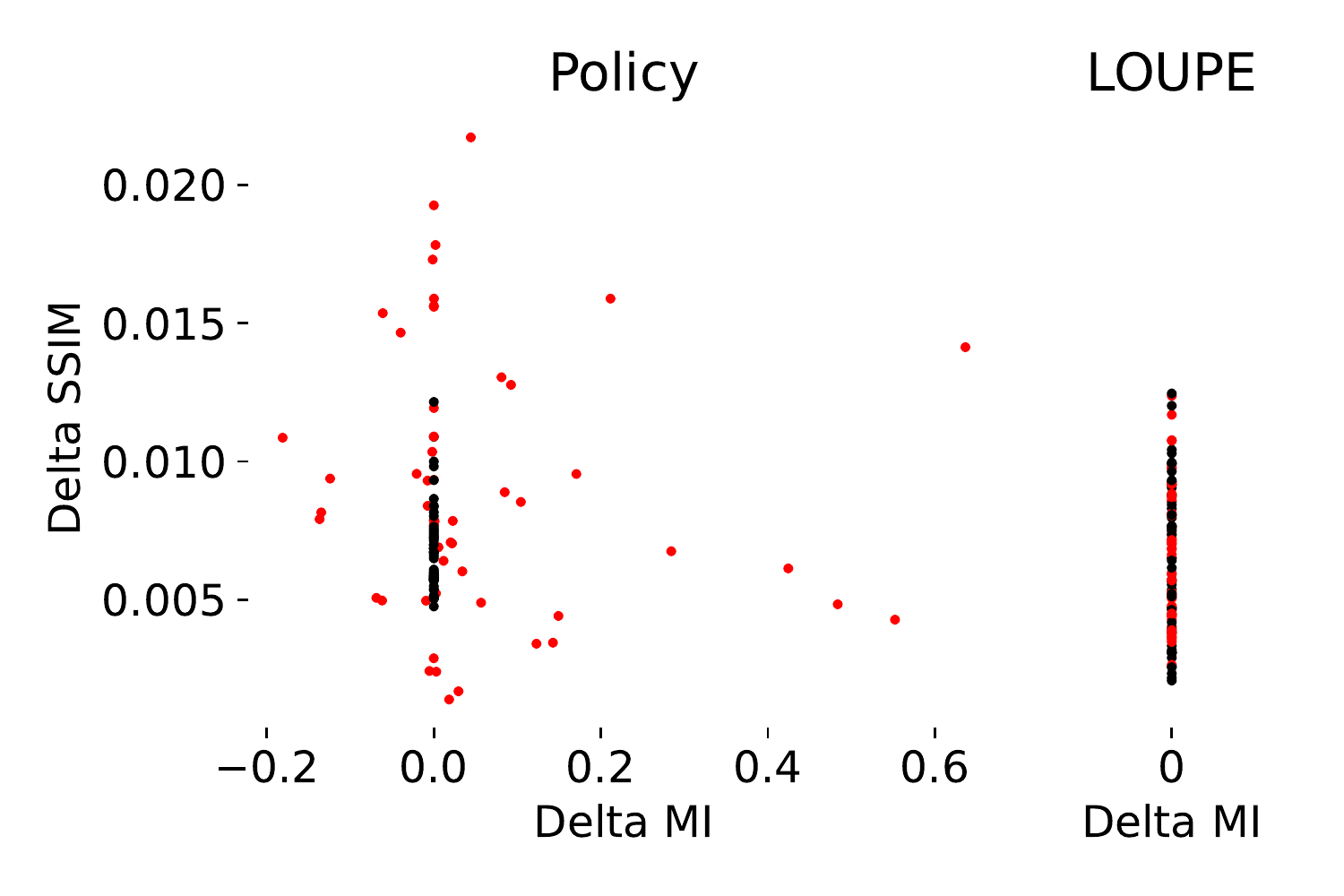}
  \captionof{figure}{\small The difference (train SSIM - validation SSIM) as a function of the difference (train MI - validation MI), for the 4$\times$ acceleration factor. Each dot is a single model. Red: softplus; black: sigmoid.}
  \label{fig:SSIMvsMI_4x_delta}
\end{minipage}\hfill
\begin{minipage}{.48\textwidth}
  \centering
  \includegraphics[width=.9\textwidth]{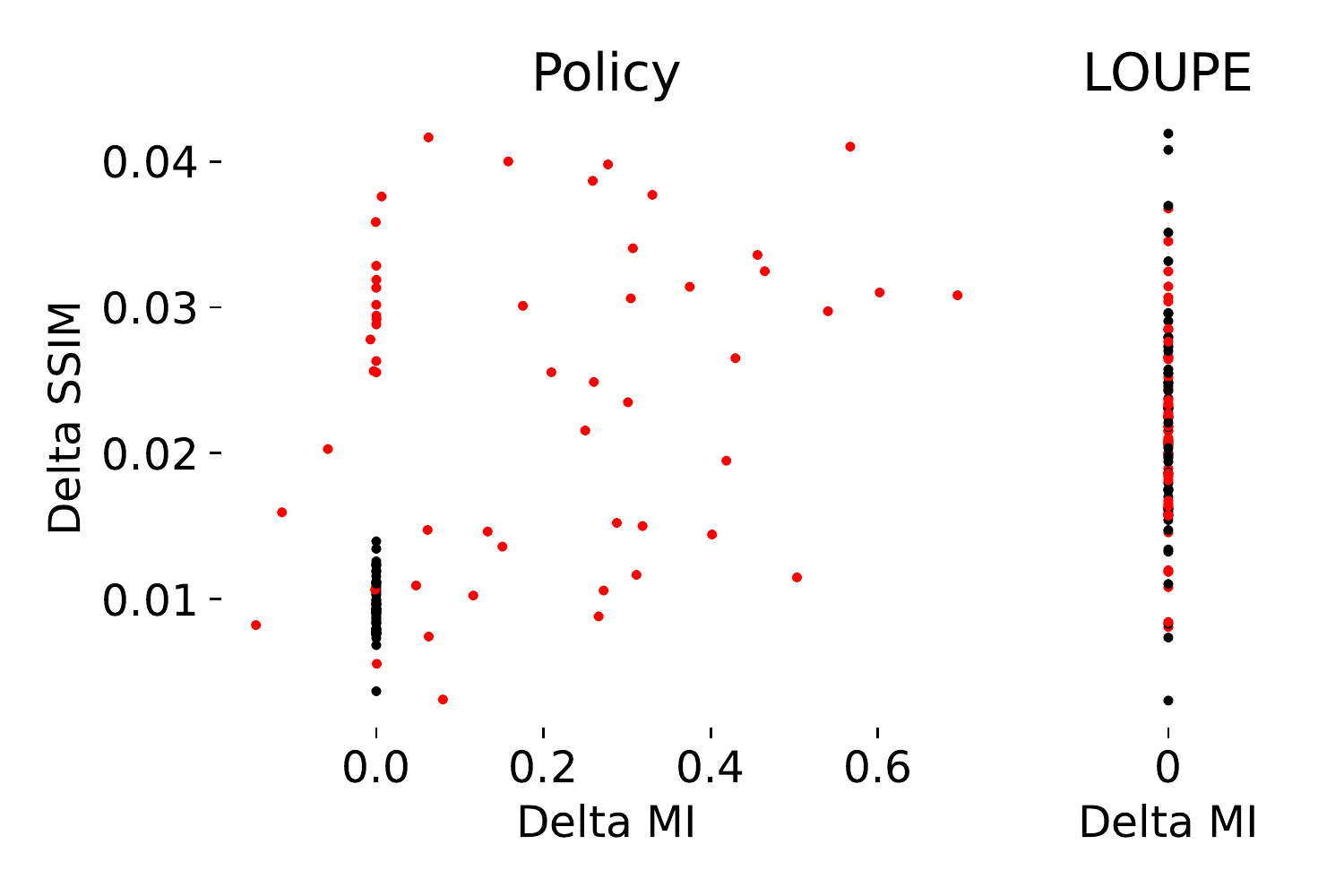}
  \captionof{figure}{\small The difference (train SSIM - validation SSIM) as a function of the difference (train MI - validation MI), for the 8$\times$ acceleration factor. Each dot is a single model. Red: softplus; black: sigmoid.}
  \label{fig:SSIMvsMI_8x_delta}
\end{minipage}
\end{figure}

In Figures~\ref{fig:SSIMvsMI_4x_delta} and~\ref{fig:SSIMvsMI_8x_delta} we have repeated these figures with now the difference between train SSIM and validation SSIM on the y-axis (labeled as `Delta SSIM'). This visualisation shows that the softplus-based policies are more prone to overfitting than the sigmoid-based policies, especially at $8\times$ accelerations, suggesting that overfitting may play a role in the overall lower performance of softplus-based policies. However, note that this effect is not just mediated through adaptivity, as a number of policies that show no significant change in MI between train and validation data overfit quite strongly.

\subsubsection{Reconstruction model capacity} \label{app:rec_capa_nonlin}

In Table~\ref{tab:rec_capa_nonlin} we present results for systems trained with reconstruction models of increasing capacities, as mentioned in Section~\ref{sec:discussion} of the main text. Although the higher capacity models lead to slight improvements in terms of Policy method SSIM, we note that sigmoid models still dominate softplus models, especially in the 8$\times$ Policy setting.

These higher capacity models were already included in the adaptivity visualisations of Figures~\ref{fig:adaptivity_vs_quality} and~\ref{fig:SSIMvsMI_4x}. Note that the best performing Policy models all show no adaptivity. If amortisation is an issue as hypothesised, these results suggest that increased capacity alone may not overcome it. Indeed, as the refinement module does not directly receive information about the subsampling mask as input, the model may be unable to distinguish inputs resulting from various masks, preventing it from using its increased capacity effectively. Further research is needed to explore the degree to which amortisation over the dataset poses a challenge for learning strong adaptive subsampling strategies. We suggest some directions specifically aimed at exploring the amortisation hypothesis in Appendix~\ref{app:future}.

\begin{table}[t]
\small
\floatconts
  {tab:rec_capa_nonlin}%
  {\caption{Effect of hyperparameters on validation SSIM for Policy and LOUPE, averaged over three seeds. $5$c/$18$ch: reconstructor with $5$ cascades and $18$ refinement channels. The sigmoid strategies consistently outperform the softplus strategies.}}%
  {\begin{tabular}{cccccc}
  & & \multicolumn{2}{c}{\bfseries Policy} & \multicolumn{2}{c}{\bfseries LOUPE} \\
  \bfseries Capacity & \bfseries Accel & Sigmoid & Softplus & Sigmoid & Softplus \\
  5c/18ch & $4\times$ & $\textbf{95.42} \pm 0.19$ & $94.31 \pm 0.90$ & $\textbf{95.37} \pm 0.19$ & $94.21 \pm 1.05$ \\
  & $8\times$ & $\textbf{92.55} \pm 0.79$ & $89.28 \pm 0.93$ & $90.46 \pm 0.80$ & $89.76 \pm 0.45$ \\
  7c/18ch & $4\times$ & $\textbf{95.51} \pm 0.13$ & $94.59 \pm 0.58$ & $\textbf{95.61} \pm 0.55$ & $94.94 \pm 0.14$ \\
  & $8\times$ & $\textbf{92.91} \pm 0.69$ & $89.20 \pm 1.89$ & $91.37 \pm 0.67$ & $90.76 \pm 0.17$ \\
  5c/36ch & $4\times$ & $\textbf{95.63} \pm 0.27$ & $94.98 \pm 0.30$ & $\textbf{95.55} \pm 0.22$ & $94.74 \pm 0.43$ \\
  & $8\times$ & $\textbf{93.26} \pm 0.20$ & $90.40 \pm 0.55$ & $91.28 \pm 0.21$ & $90.15 \pm 0.47$ \\
  \end{tabular}}
\end{table}

\subsection{Qualitative results} \label{app:qual}

In this section we provide visualisations of our model outputs. Specifically, we visualise some of the learned subsampling strategies in Appendix~\ref{app:masks}, reconstructions in~\ref{app:recon}, and sensitivity maps in~\ref{app:sense}.

\subsubsection{Subsampling visualisation} \label{app:masks}

\begin{figure}[t]
     \centering
     \subfigure[Target]{\includegraphics[width=0.1\textwidth]{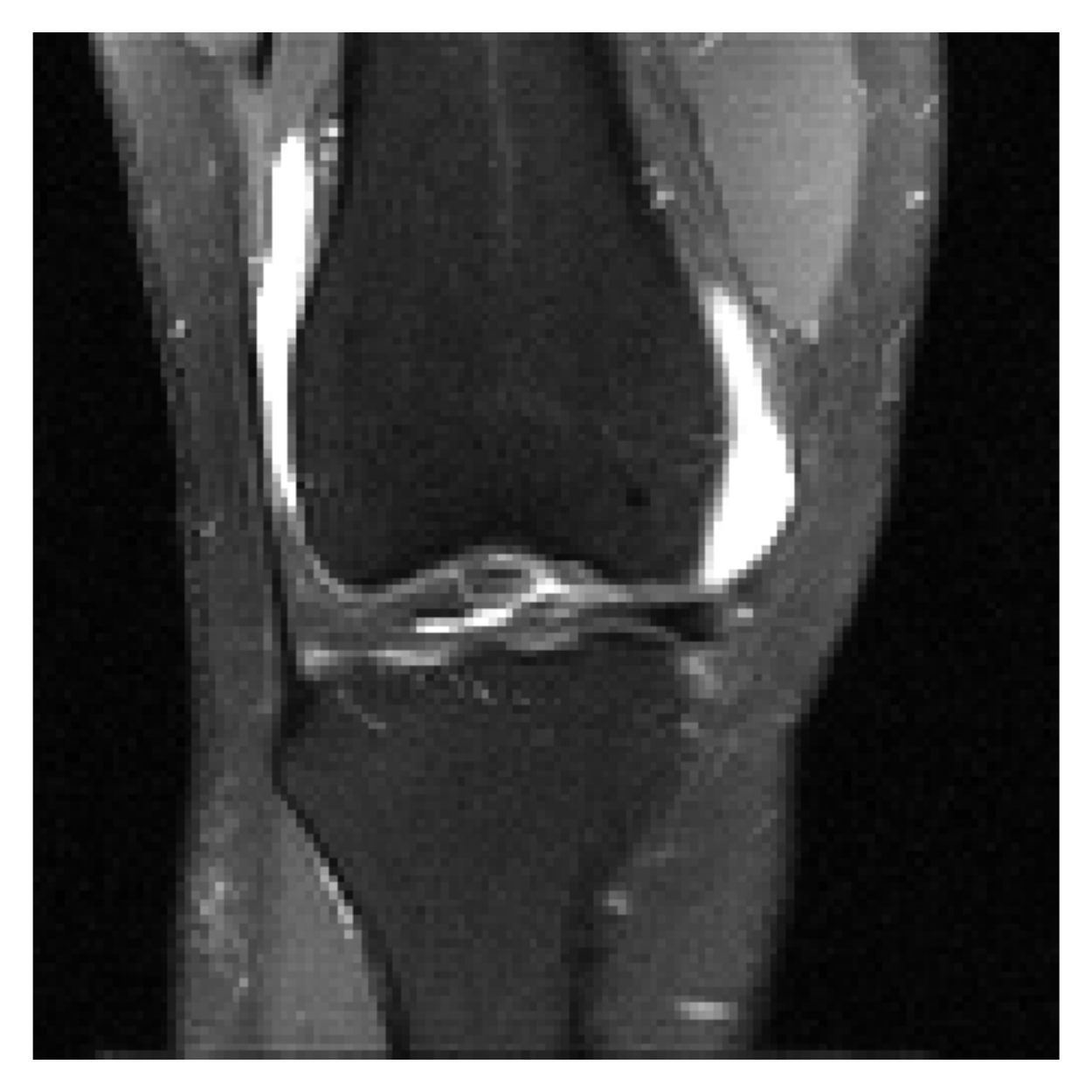}} 
     \subfigure[Equispaced]{\includegraphics[width=0.21\textwidth]{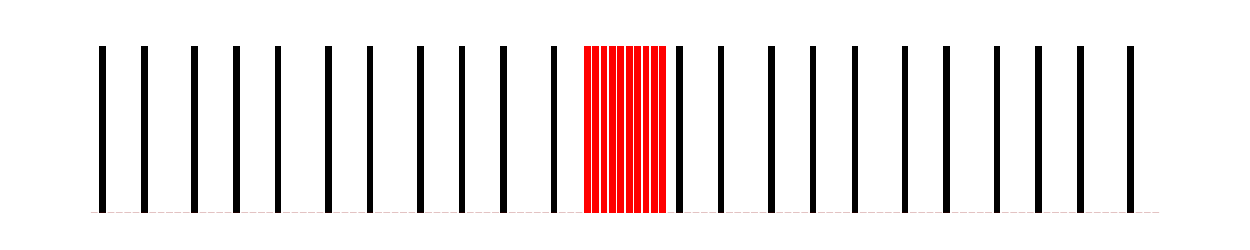}} 
     \subfigure[LOUPE]{\includegraphics[width=0.21\textwidth]{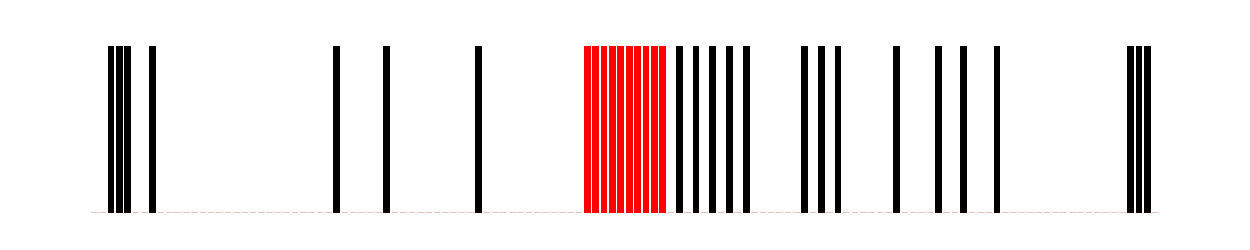}} 
     \subfigure[Sigmoid Policy]{\includegraphics[width=0.21\textwidth]{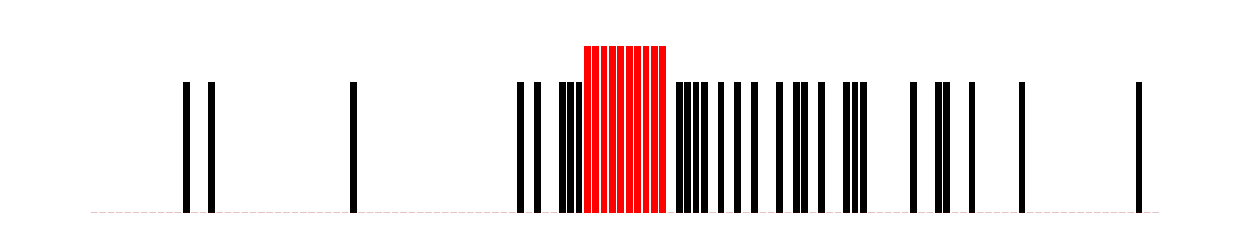}}
     \subfigure[Softplus Policy]{\includegraphics[width=0.21\textwidth]{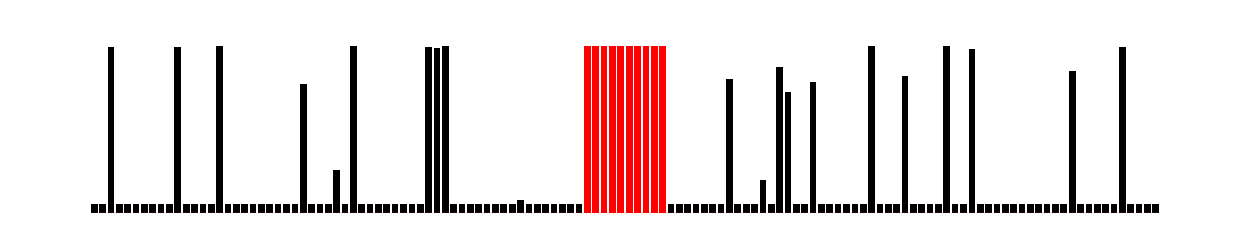}}
     \subfigure[Target]{\includegraphics[width=0.1\textwidth]{figures/target_1.pdf}}
     \subfigure[Equispaced]{\includegraphics[width=0.21\textwidth]{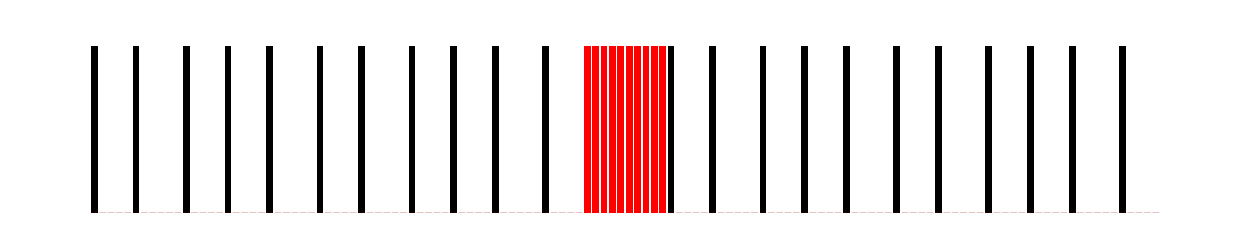}}
     \subfigure[LOUPE]{\includegraphics[width=0.21\textwidth]{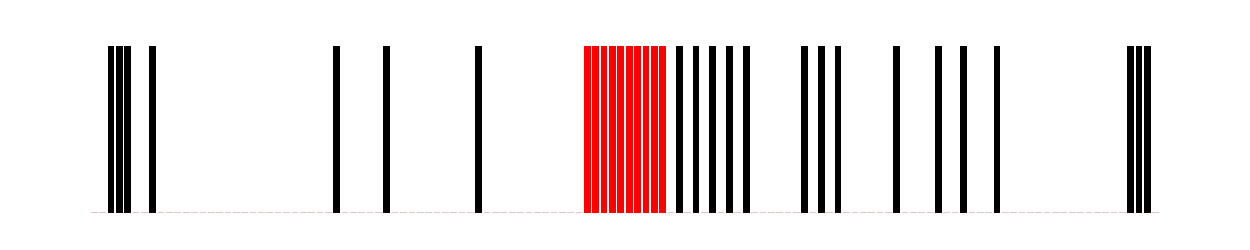}}
     \subfigure[Sigmoid Policy]{\includegraphics[width=0.21\textwidth]{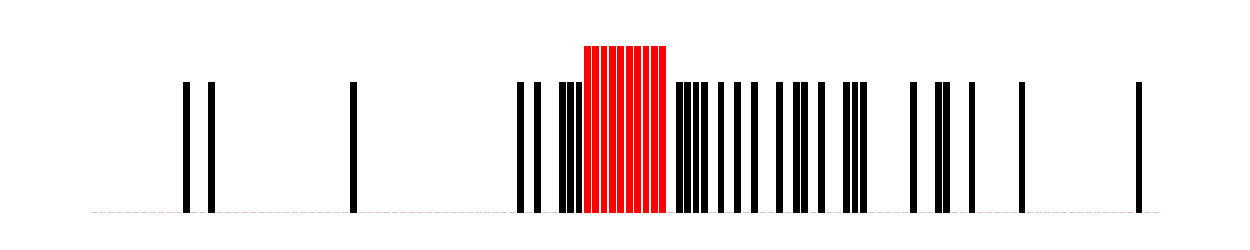}}
     \subfigure[Softplus Policy]{\includegraphics[width=0.21\textwidth]{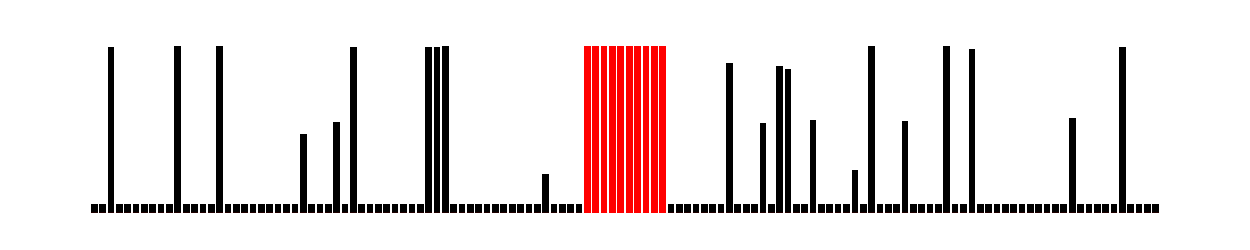}}
    \caption{Sampling strategies for the $4\times$ case. Every line corresponds to a possible $k$-space measurement. The height of a line denotes the probability of sampling that measurement. ACS measurements are shown in red and always have probability 1.}
    \label{fig:4x_policies}
\end{figure}

\begin{figure}[t]
     \centering
     \subfigure[Target]{\includegraphics[width=0.1\textwidth]{figures/target_0.pdf}} 
     \subfigure[Equispaced]{\includegraphics[width=0.21\textwidth]{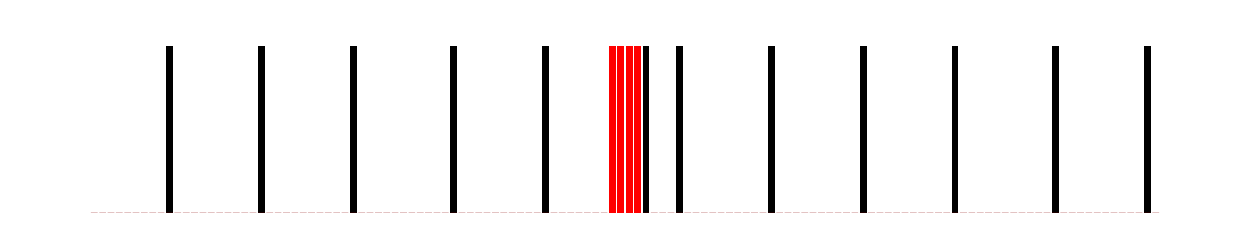}} 
     \subfigure[LOUPE]{\includegraphics[width=0.21\textwidth]{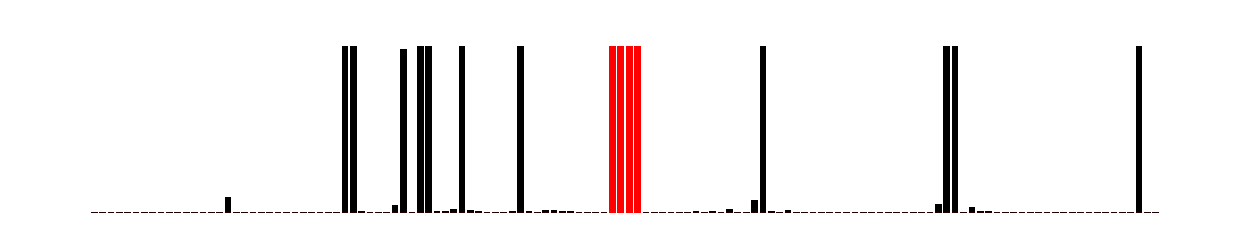}} 
     \subfigure[Sigmoid Policy]{\includegraphics[width=0.21\textwidth]{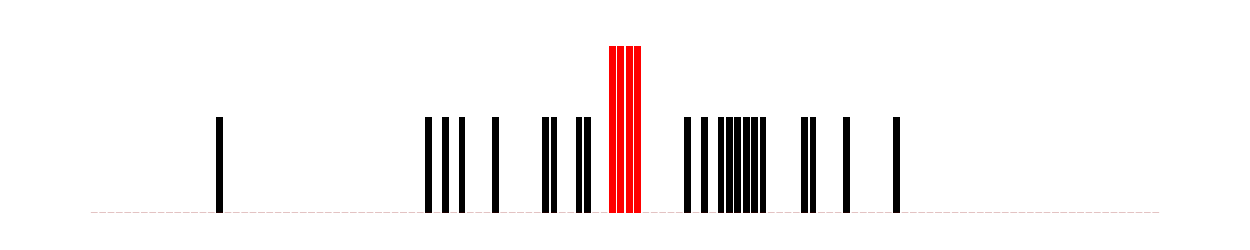}}
     \subfigure[Softplus Policy]{\includegraphics[width=0.21\textwidth]{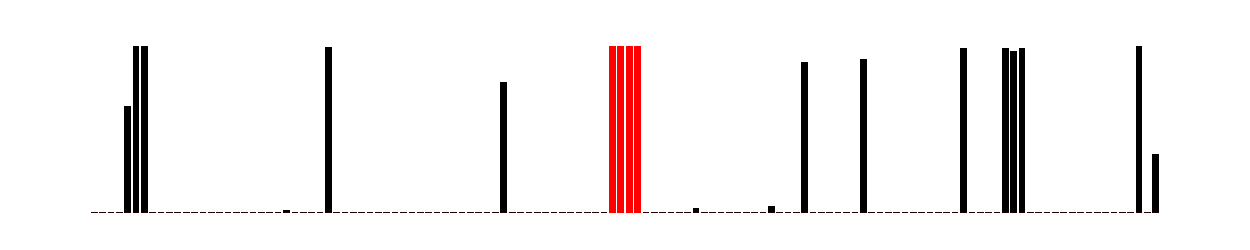}}
     \subfigure[Target]{\includegraphics[width=0.1\textwidth]{figures/target_1.pdf}}
     \subfigure[Equispaced]{\includegraphics[width=0.21\textwidth]{figures/8x_equidistant_policy_1.pdf}}
     \subfigure[LOUPE]{\includegraphics[width=0.21\textwidth]{figures/loupe_8x_policy_1.pdf}}
     \subfigure[Sigmoid Policy]{\includegraphics[width=0.21\textwidth]{figures/act_8x_policy_1.pdf}}
     \subfigure[Softplus Policy]{\includegraphics[width=0.21\textwidth]{figures/act_act_8x_policy_1.pdf}}
    \caption{Sampling strategies for the $8\times$ case. Every line corresponds to a possible $k$-space measurement. The height of a line denotes the probability of sampling that measurement. ACS measurements are shown in red and always have probability 1.}
    \label{fig:8x_policies}
\end{figure}

Figures~\ref{fig:4x_policies} and~\ref{fig:8x_policies} show sampling strategies (learned) by the various methods for respectively $4\times$ and $8\times$ acceleration. Note that softplus-based Policies assign nonzero probability to every action, while sigmoid-based Policies are more selective. This effect is also observed for LOUPE at $8\times$, although it vanishes at $4\times$ acceleration.

\subsubsection{Reconstructions} \label{app:recon}

\begin{figure}[t]
\small
     \centering
     \subfigure[Target]{\includegraphics[width=0.19\textwidth]{figures/target_0.pdf}} 
     \subfigure[Equispaced]{\includegraphics[width=0.19\textwidth]{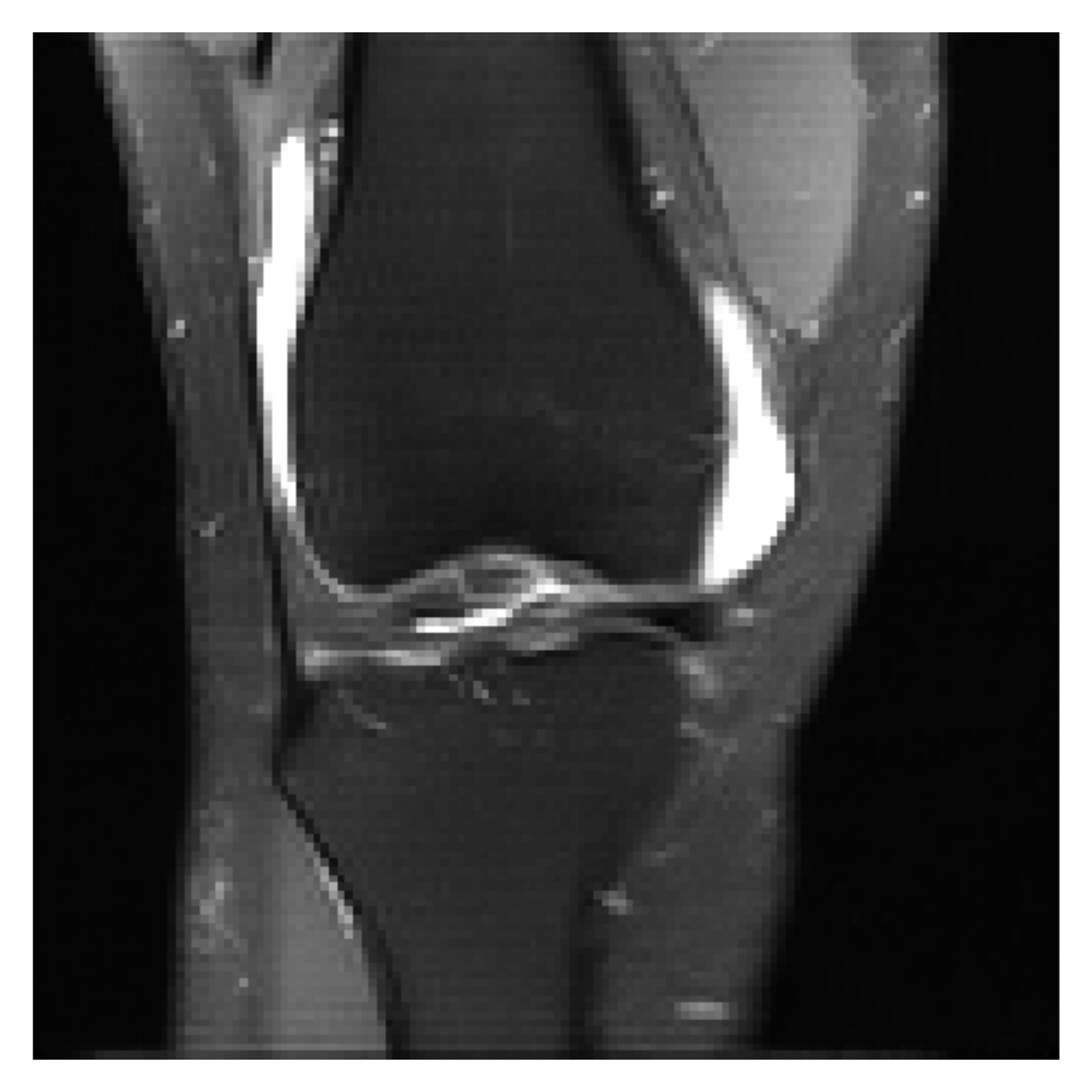}} 
     \subfigure[LOUPE]{\includegraphics[width=0.19\textwidth]{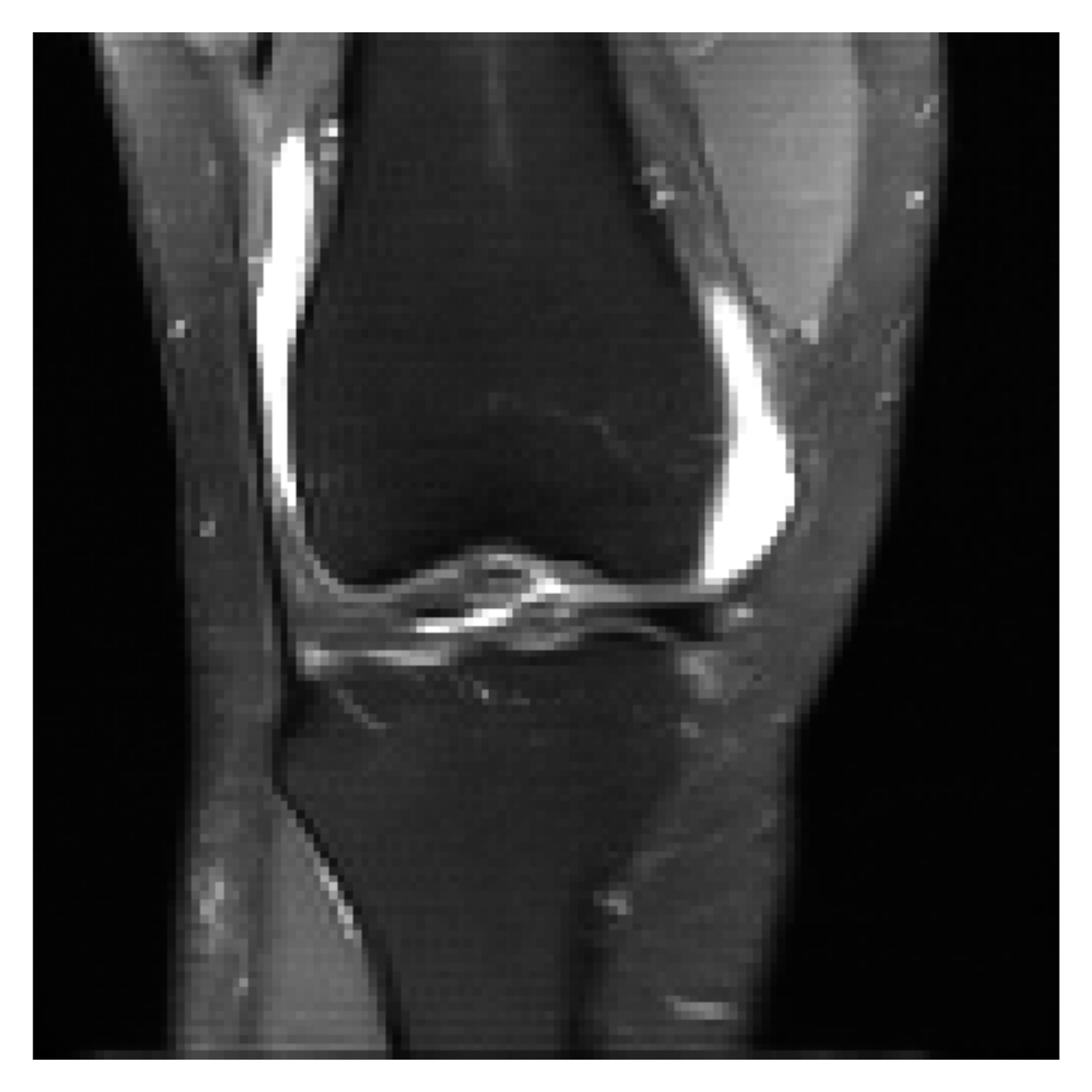}} 
     \subfigure[Sigmoid Policy]{\includegraphics[width=0.19\textwidth]{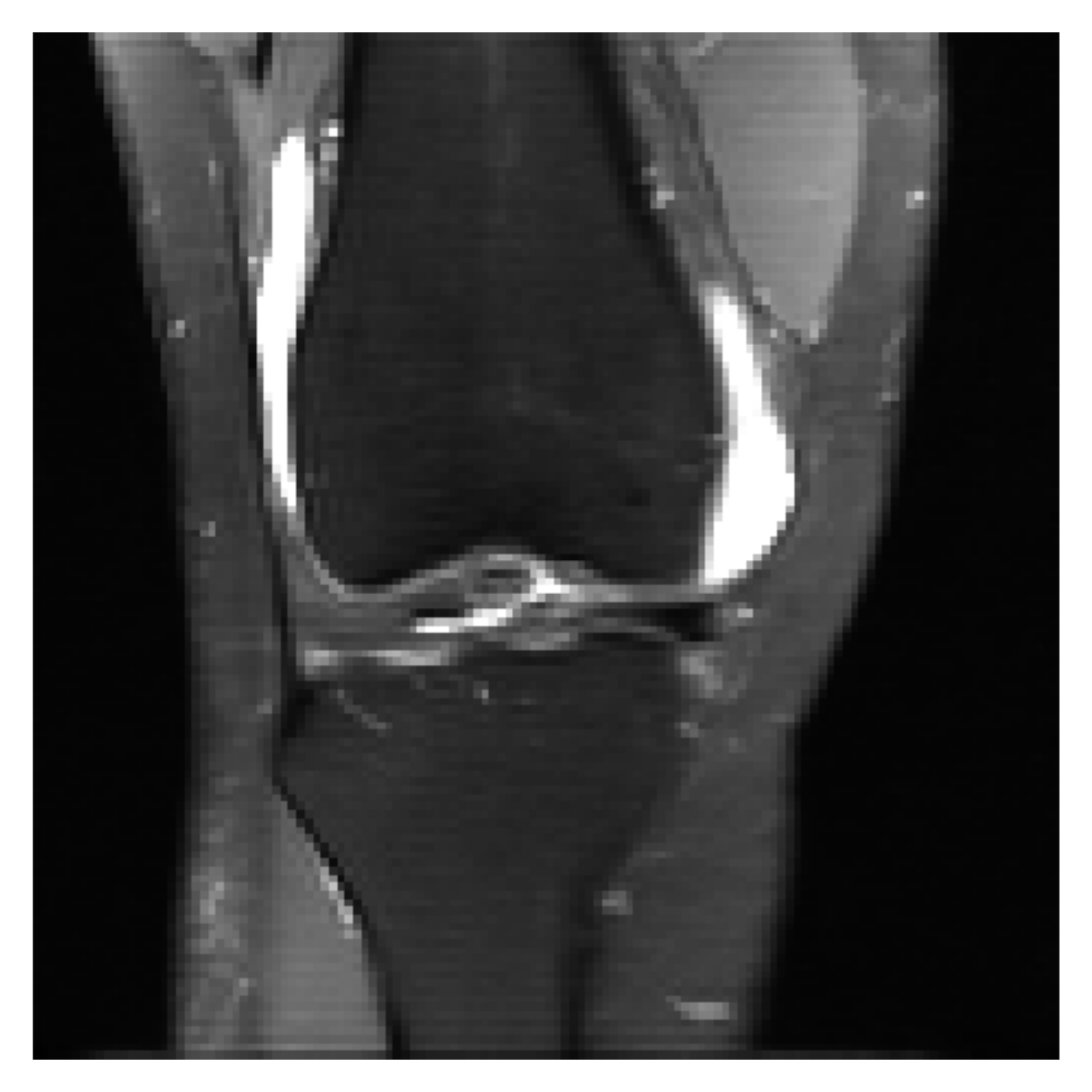}}
     \subfigure[Softplus Policy]{\includegraphics[width=0.19\textwidth]{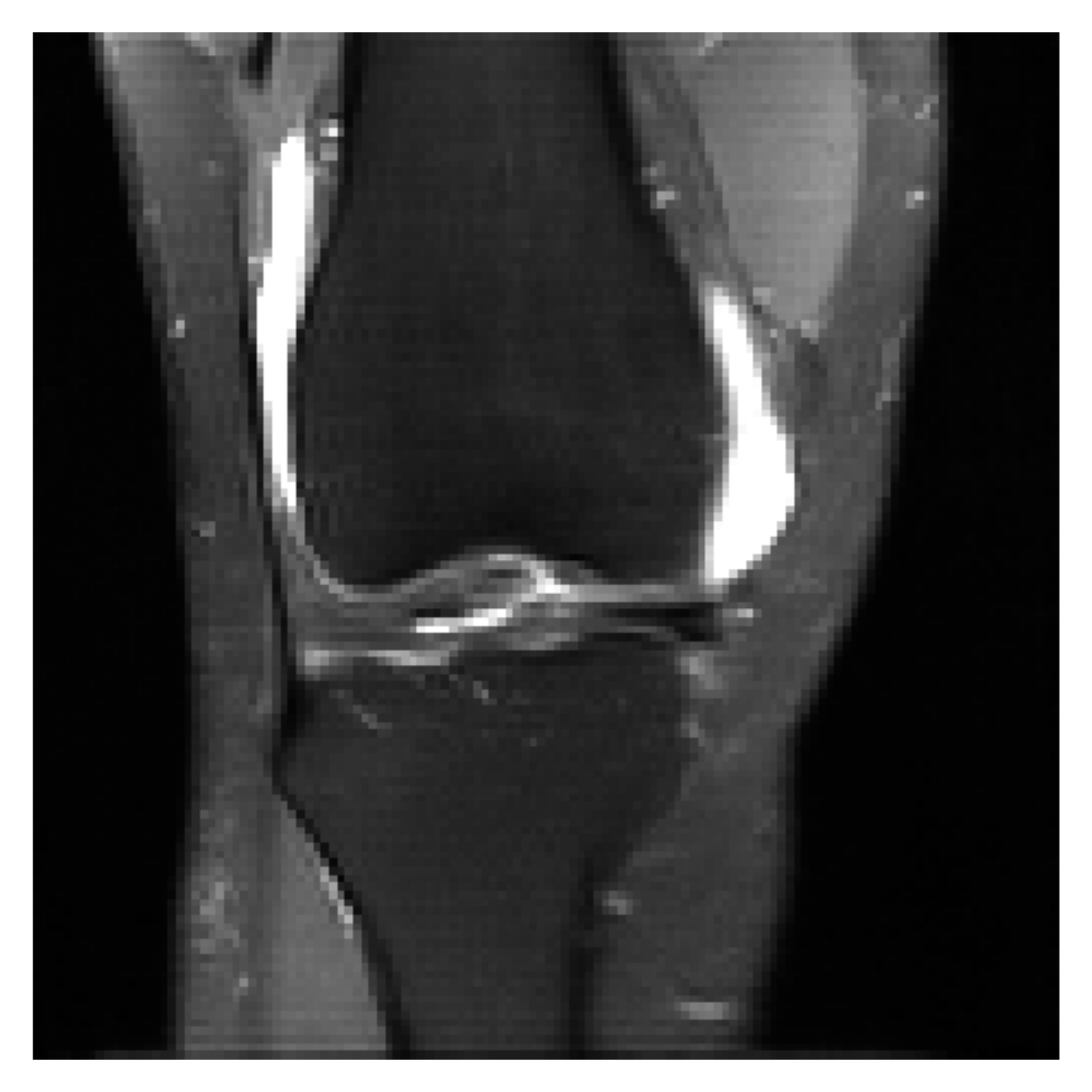}}
     \subfigure[Target]{\includegraphics[width=0.19\textwidth]{figures/target_1.pdf}}
     \subfigure[Equispaced]{\includegraphics[width=0.19\textwidth]{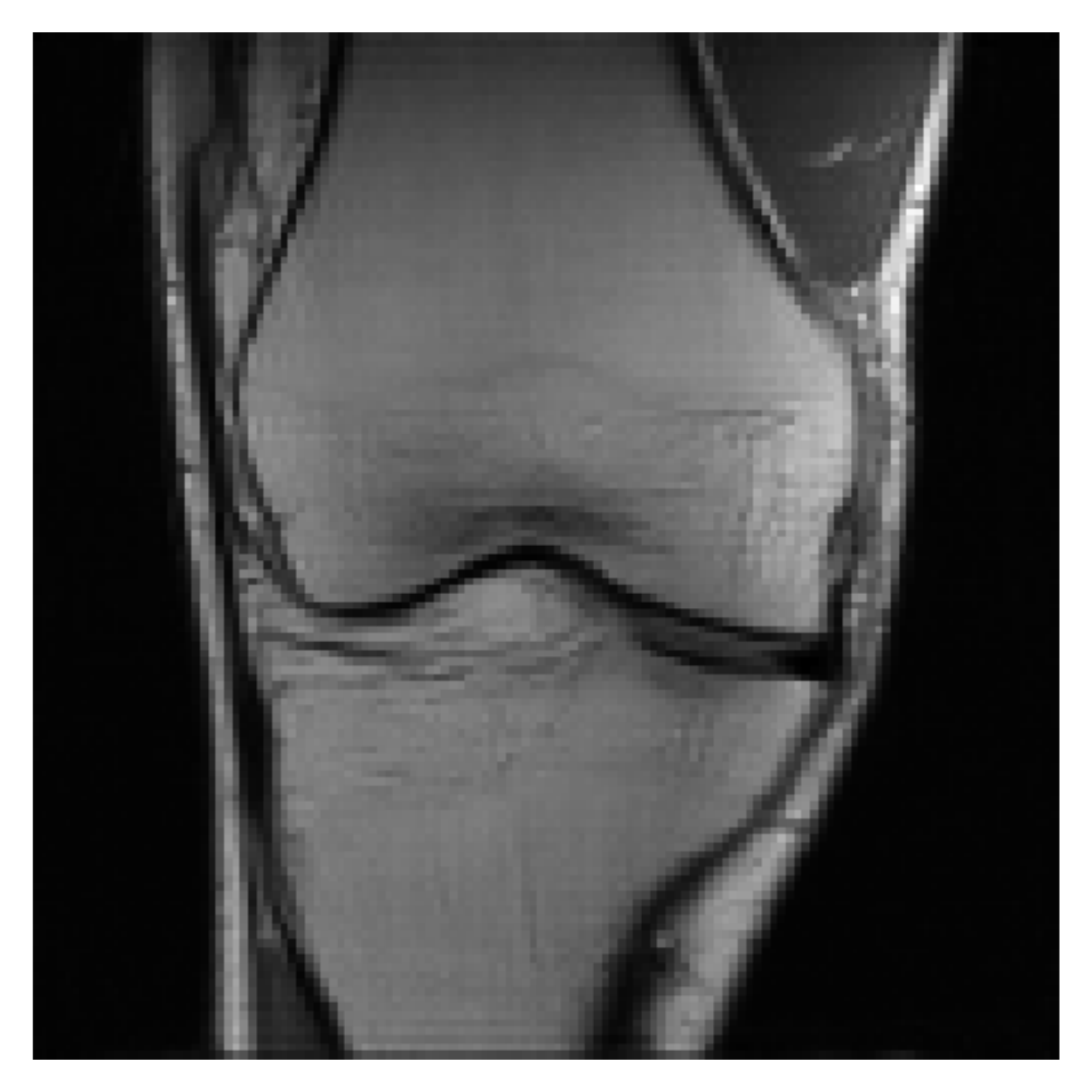}}
     \subfigure[LOUPE]{\includegraphics[width=0.19\textwidth]{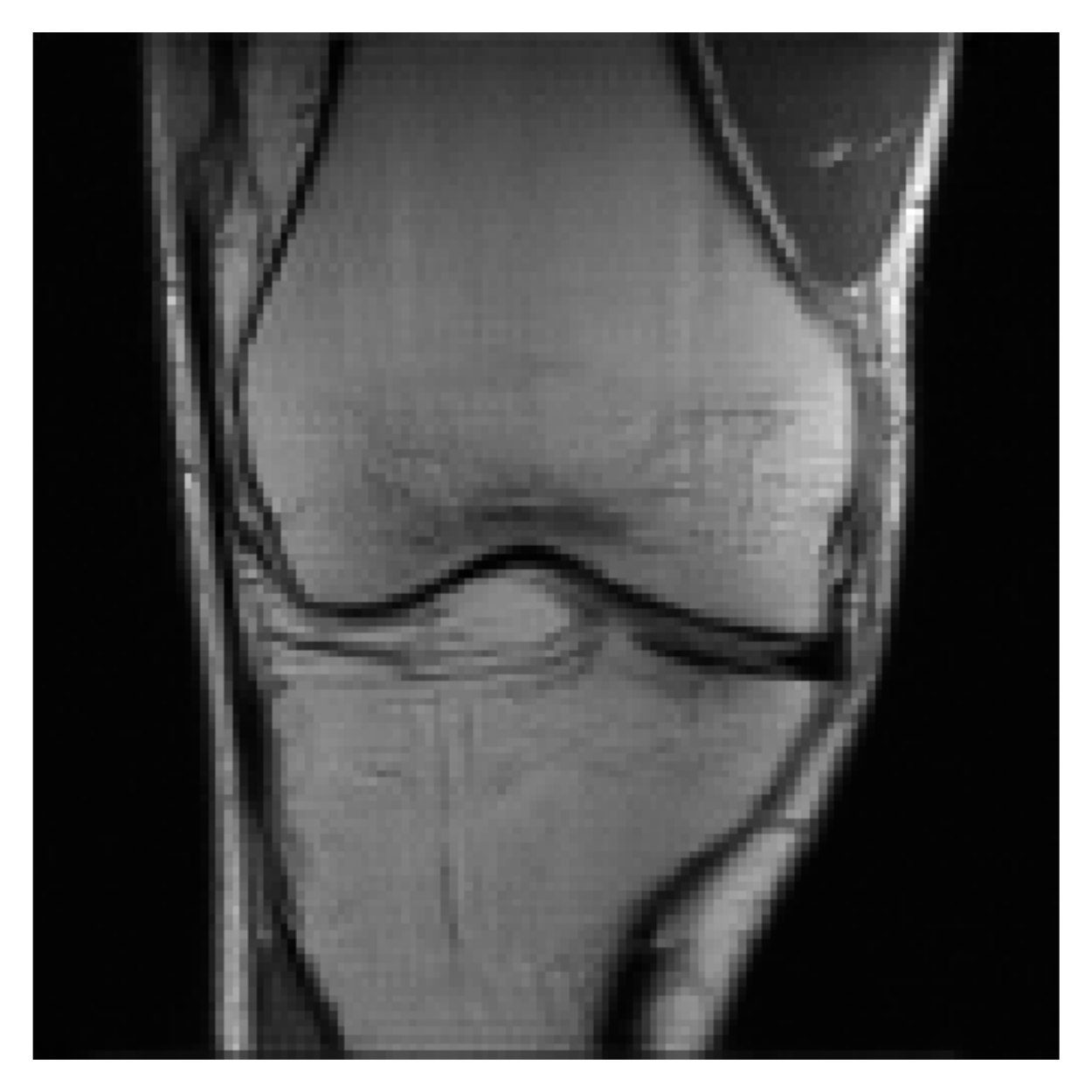}}
     \subfigure[Sigmoid Policy]{\includegraphics[width=0.19\textwidth]{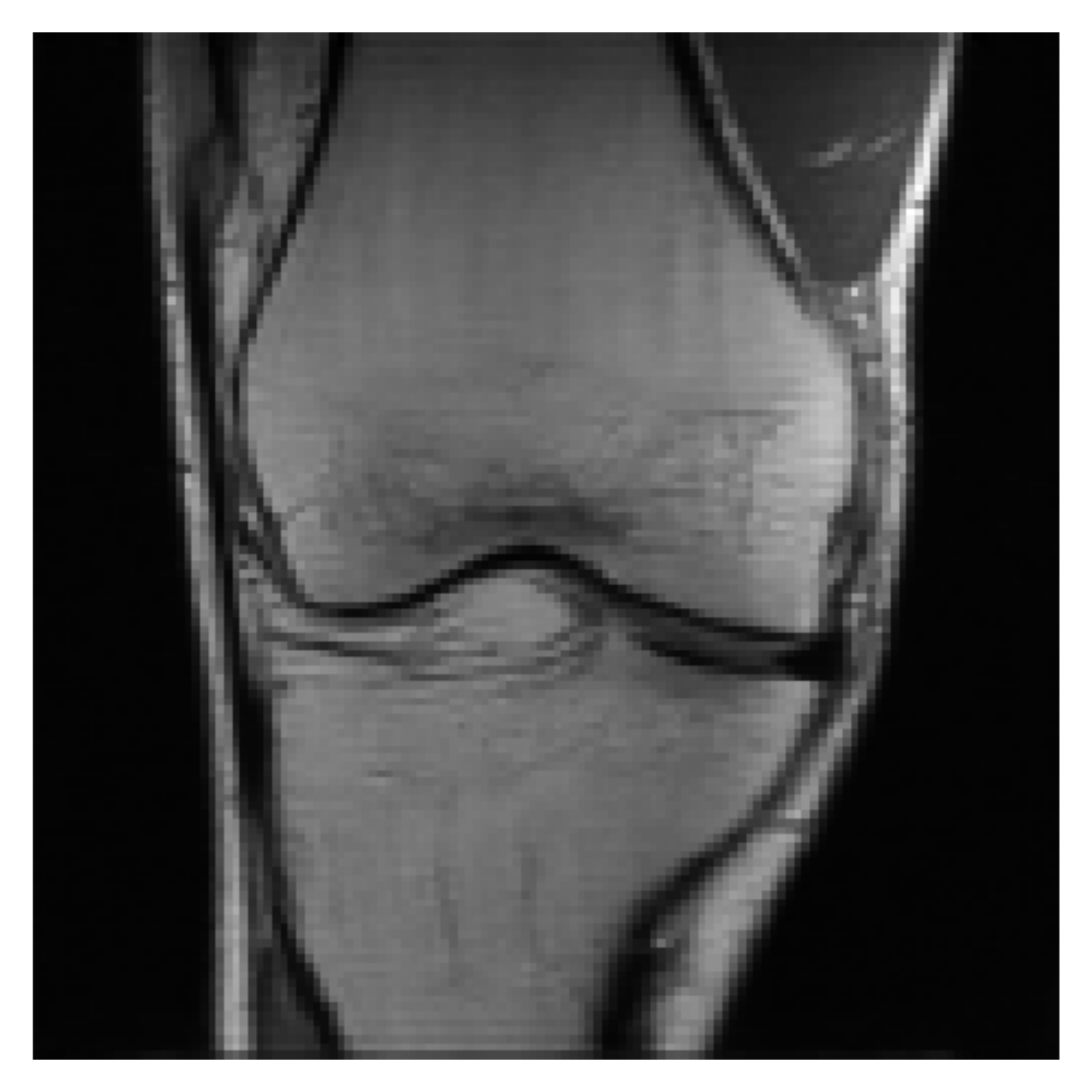}}
     \subfigure[Softplus Policy]{\includegraphics[width=0.19\textwidth]{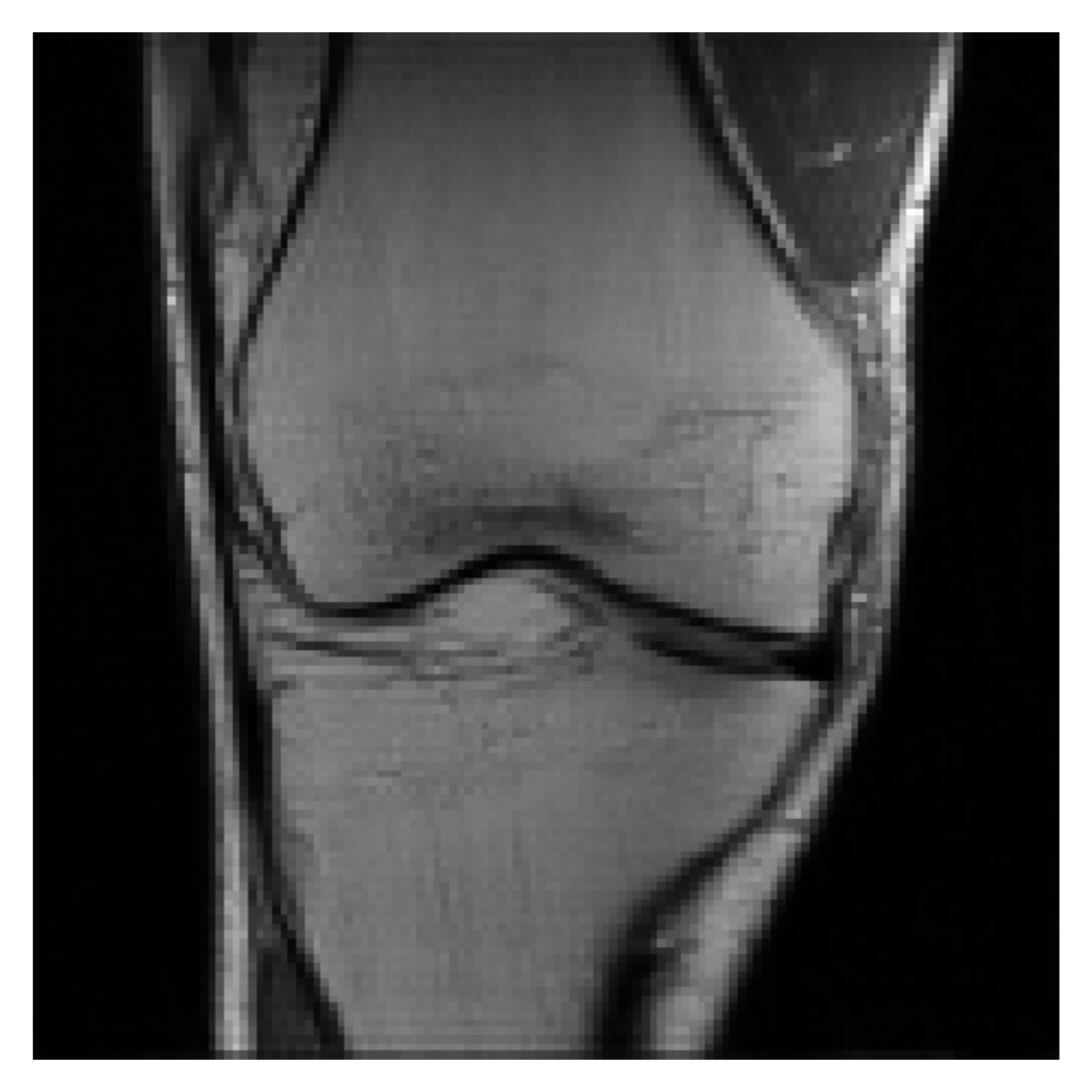}}
    \caption{Examples of $4\times$ reconstructions.}
    \label{fig:4x_recon}
\end{figure}

\begin{figure}[t]
\small
     \centering
     \subfigure[Target]{\includegraphics[width=0.19\textwidth]{figures/target_0.pdf}} 
     \subfigure[Equispaced]{\includegraphics[width=0.19\textwidth]{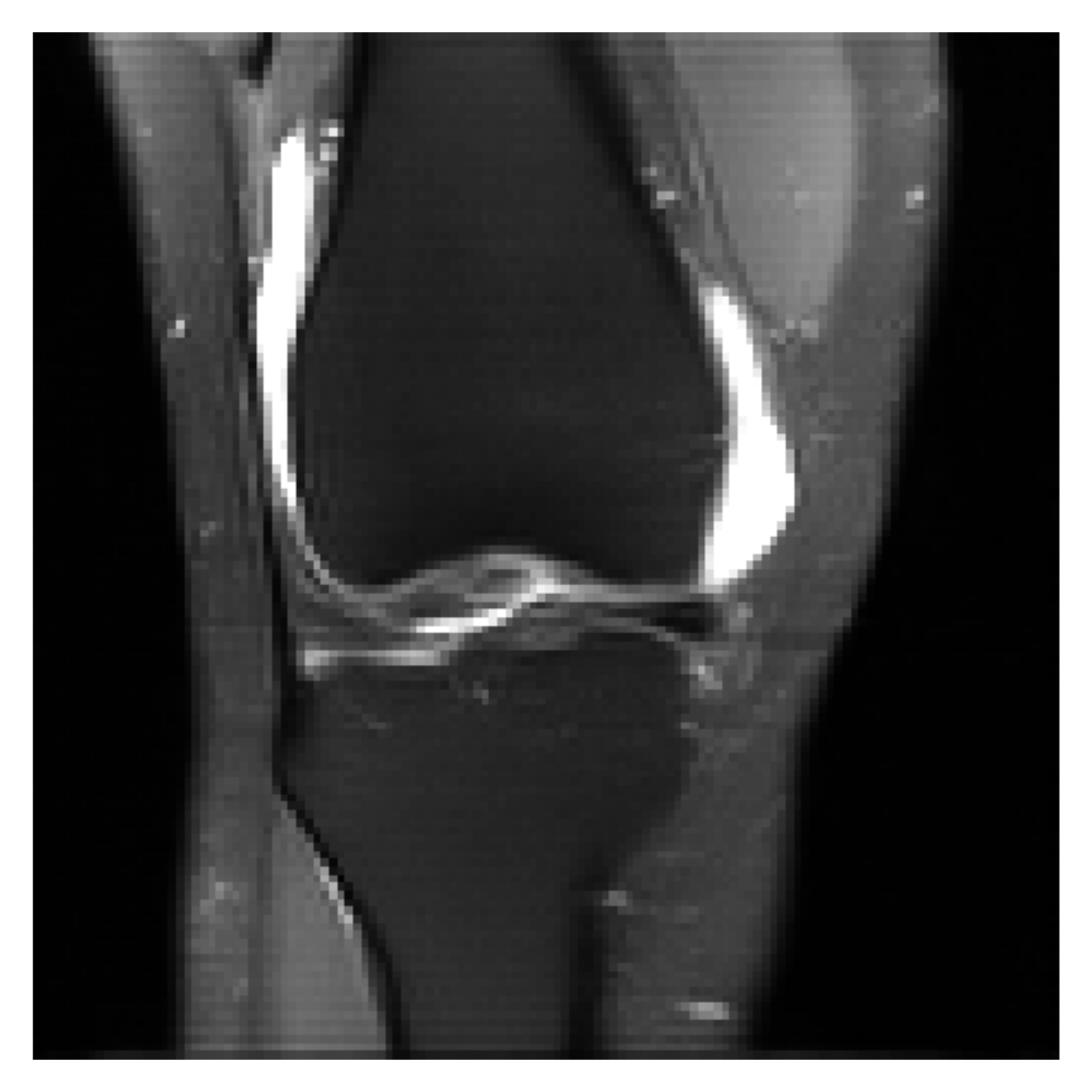}} 
     \subfigure[LOUPE]{\includegraphics[width=0.19\textwidth]{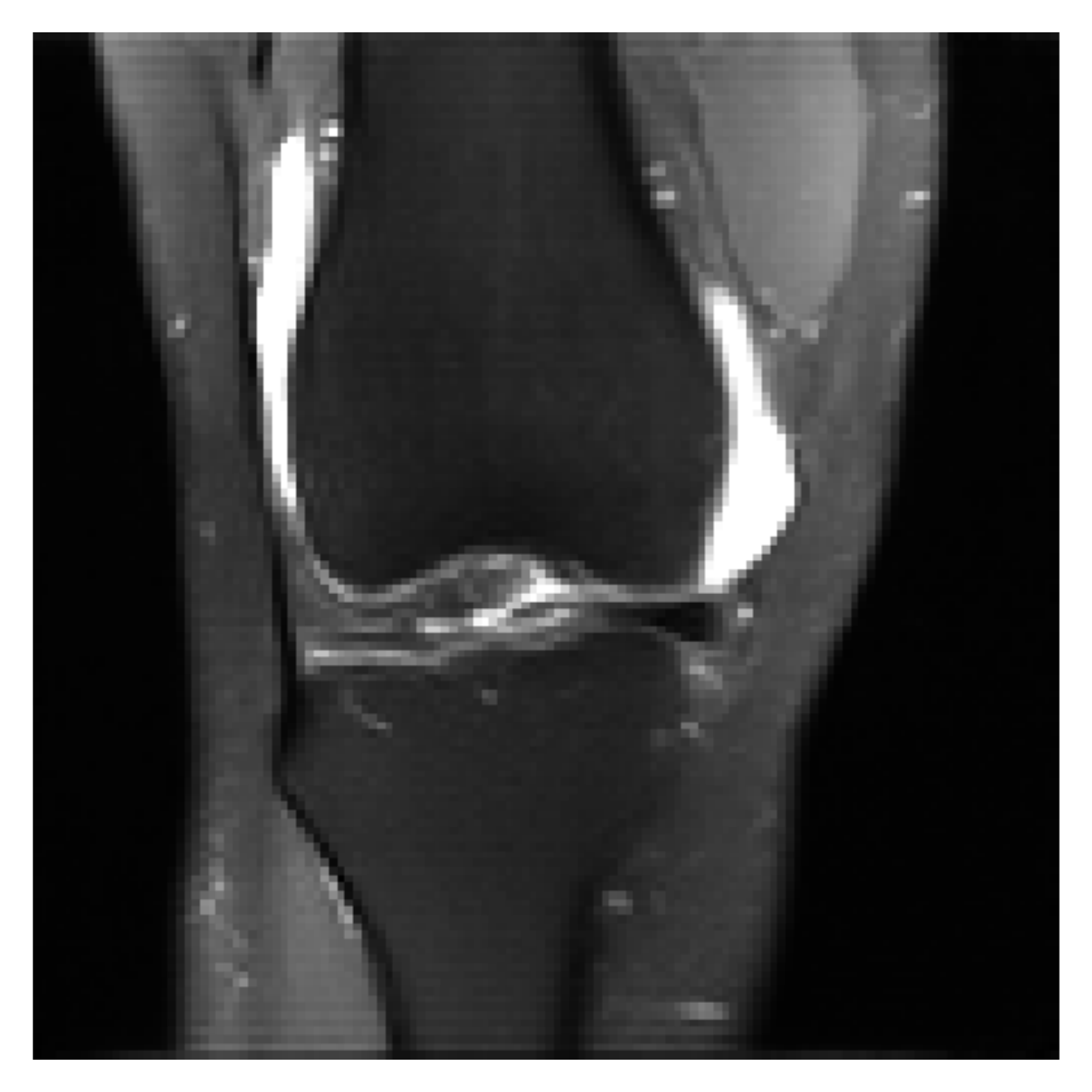}} 
     \subfigure[Sigmoid Policy]{\includegraphics[width=0.19\textwidth]{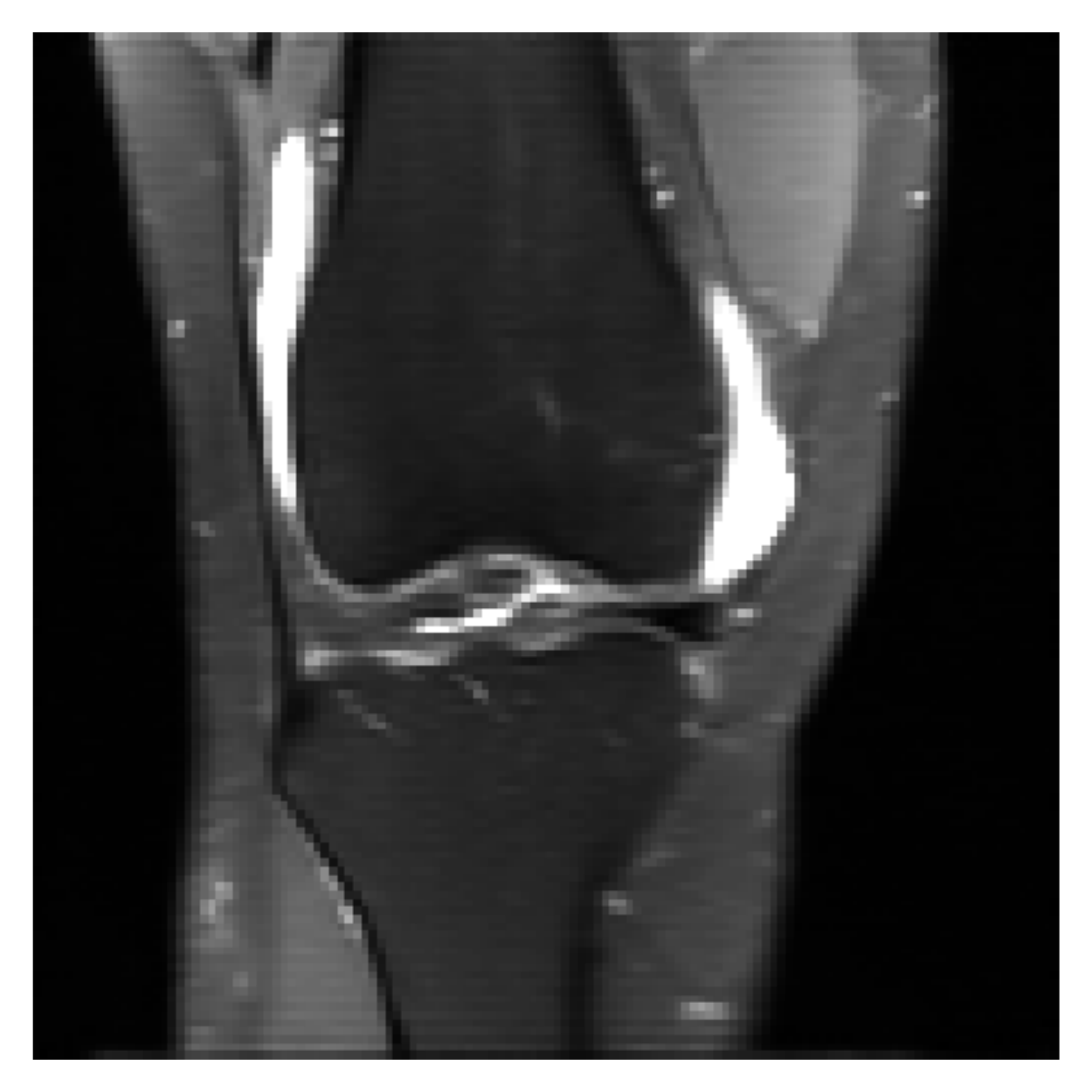}}
     \subfigure[Softplus Policy]{\includegraphics[width=0.19\textwidth]{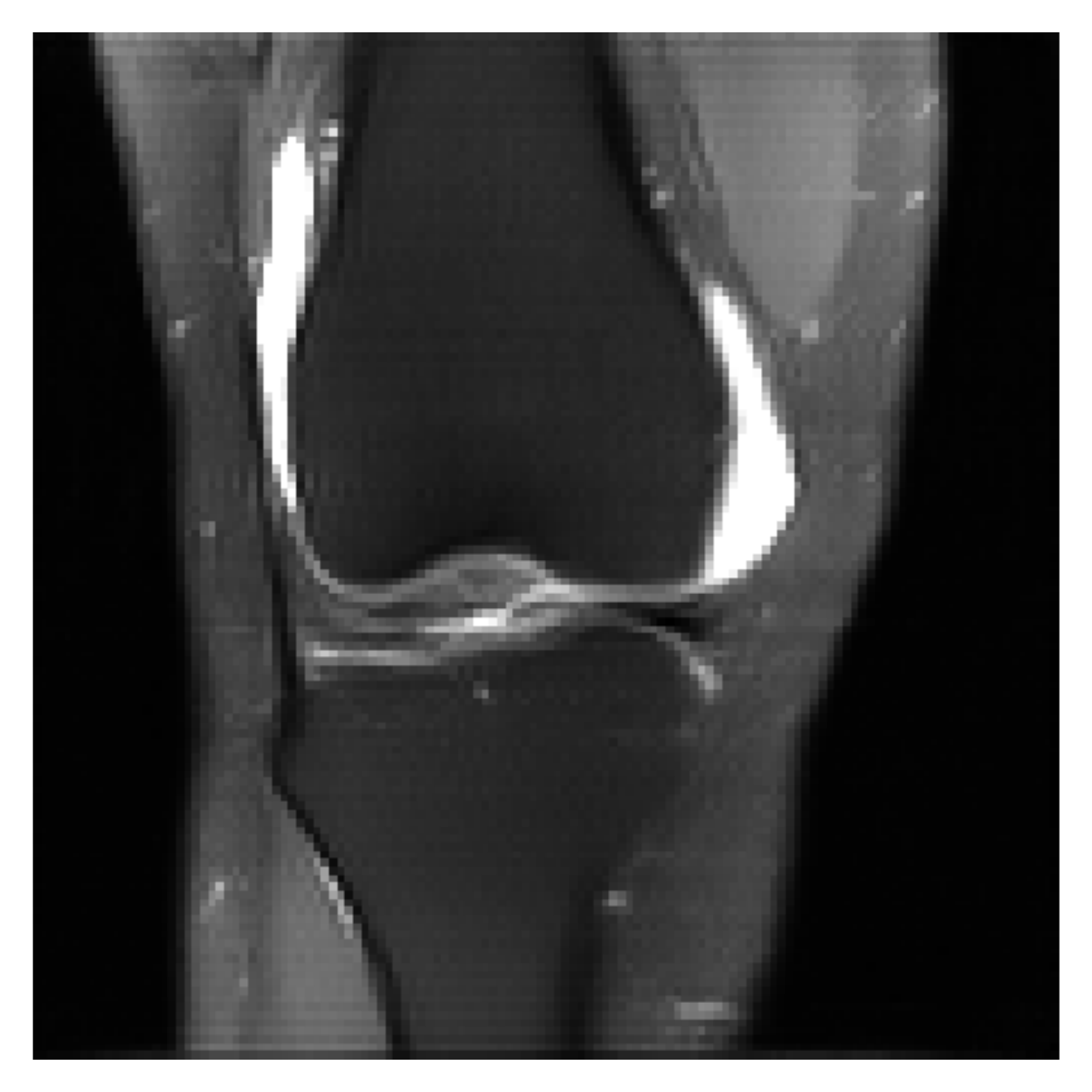}}
     \subfigure[Target]{\includegraphics[width=0.19\textwidth]{figures/target_1.pdf}}
     \subfigure[Equispaced]{\includegraphics[width=0.19\textwidth]{figures/non_act_8x_recon_1.pdf}}
     \subfigure[LOUPE]{\includegraphics[width=0.19\textwidth]{figures/loupe_8x_recon_1.pdf}}
     \subfigure[Sigmoid Policy]{\includegraphics[width=0.19\textwidth]{figures/act_8x_recon_1.pdf}}
     \subfigure[Softplus Policy]{\includegraphics[width=0.19\textwidth]{figures/act_act_8x_recon_1.pdf}}
    \caption{Examples of $8\times$ reconstructions.}
    \label{fig:8x_recon}
\end{figure}

Figures~\ref{fig:4x_recon} and~\ref{fig:8x_recon} show selected reconstructions by the various methods for respectively $4\times$ and $8\times$ acceleration. 

\subsubsection{Sensitivity maps} \label{app:sense}

\begin{figure}[t]
     \centering
     \subfigure[Equispaced]{\includegraphics[width=0.24\textwidth]{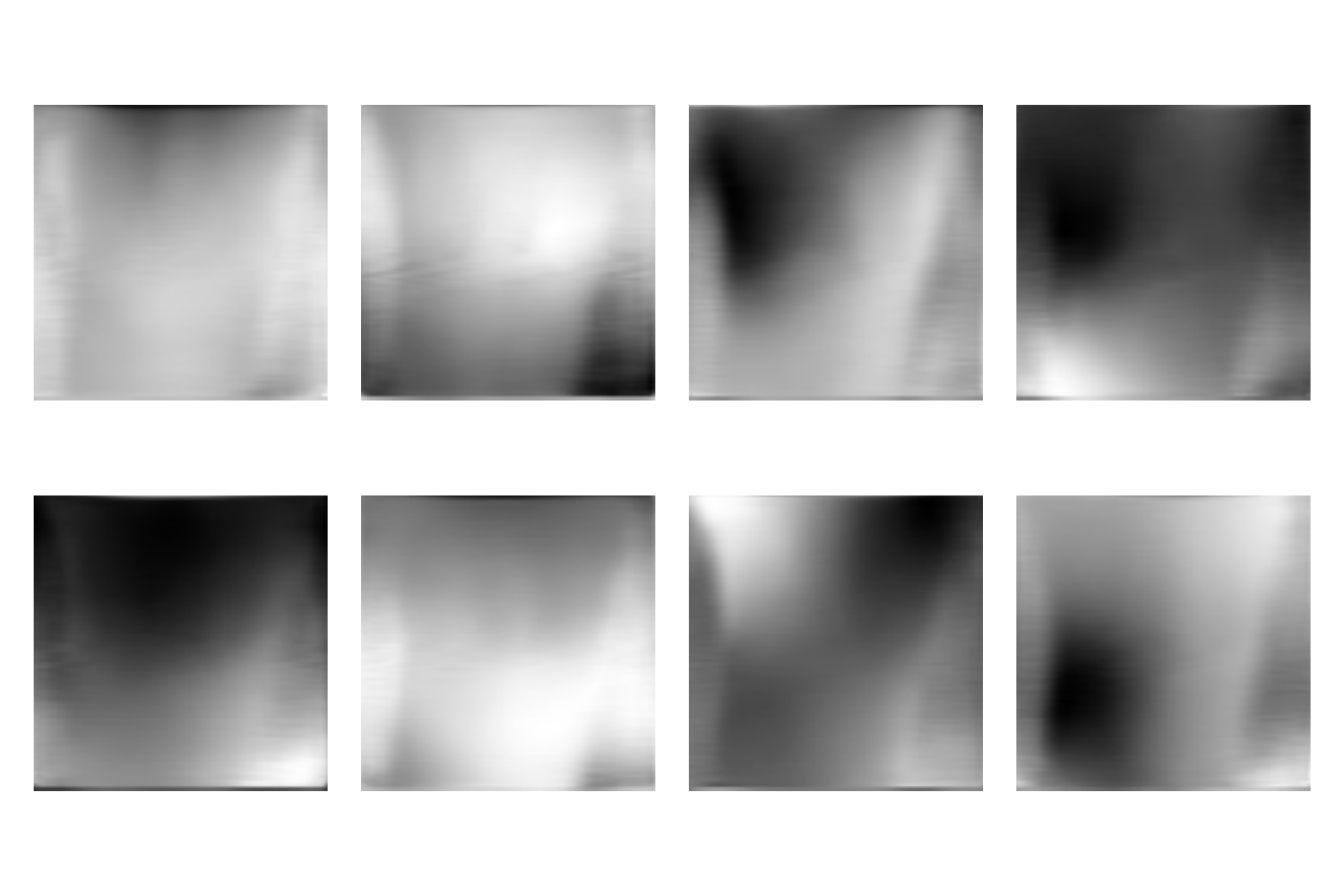}} 
     \subfigure[LOUPE]{\includegraphics[width=0.24\textwidth]{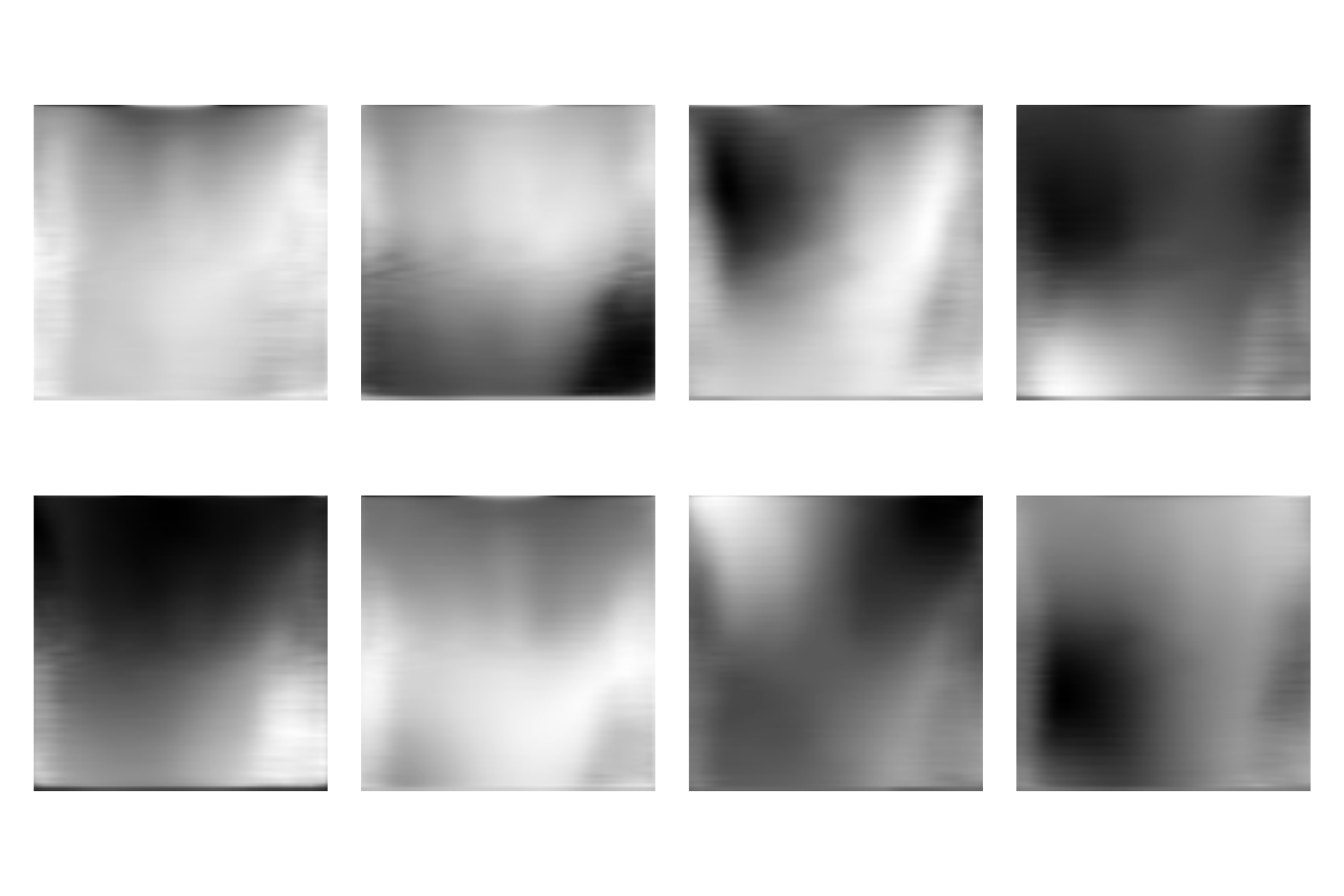}} 
     \subfigure[Sigmoid Policy]{\includegraphics[width=0.24\textwidth]{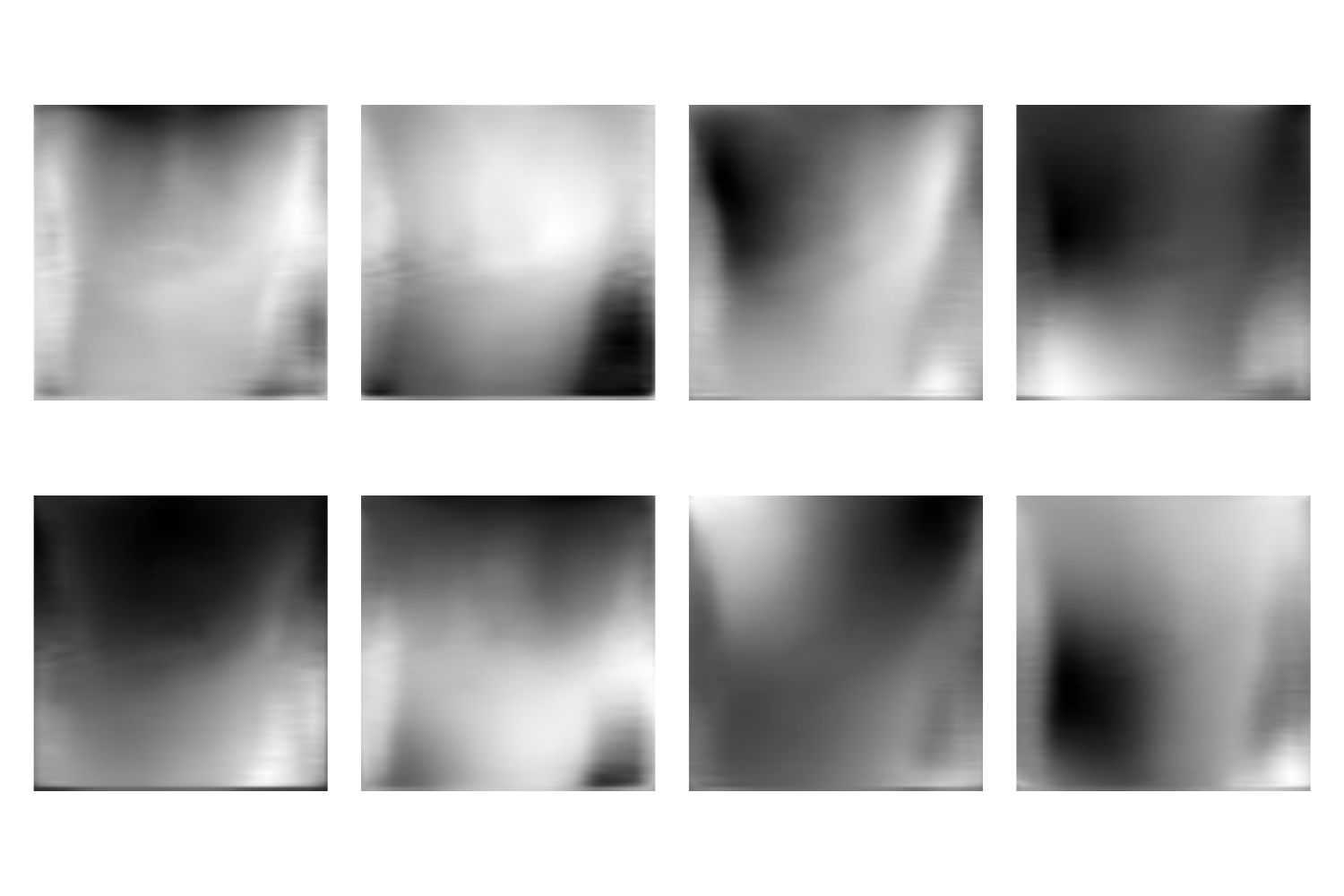}}
     \subfigure[Softplus Policy]{\includegraphics[width=0.24\textwidth]{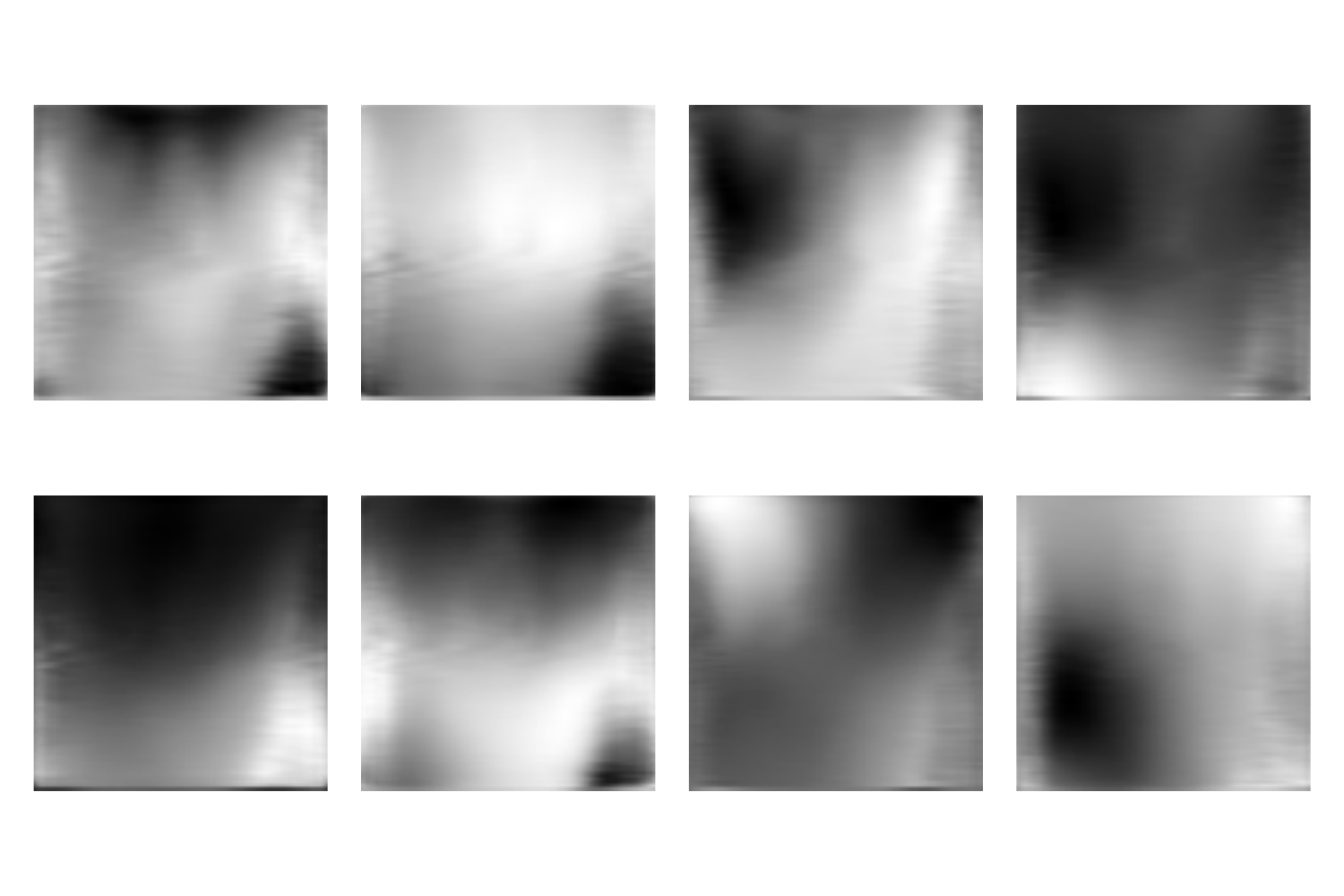}}
     \subfigure[Equispaced]{\includegraphics[width=0.24\textwidth]{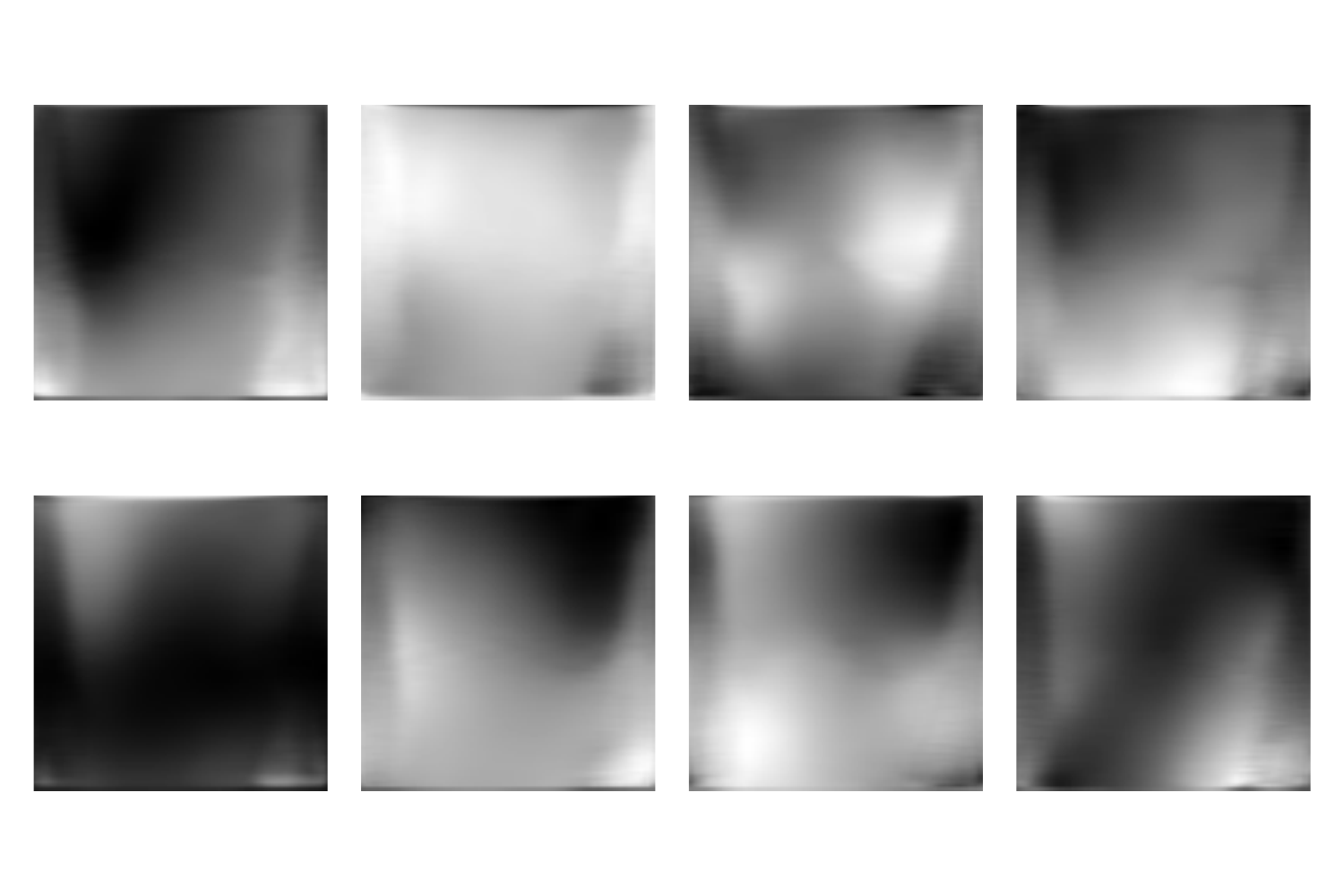}}
     \subfigure[LOUPE]{\includegraphics[width=0.24\textwidth]{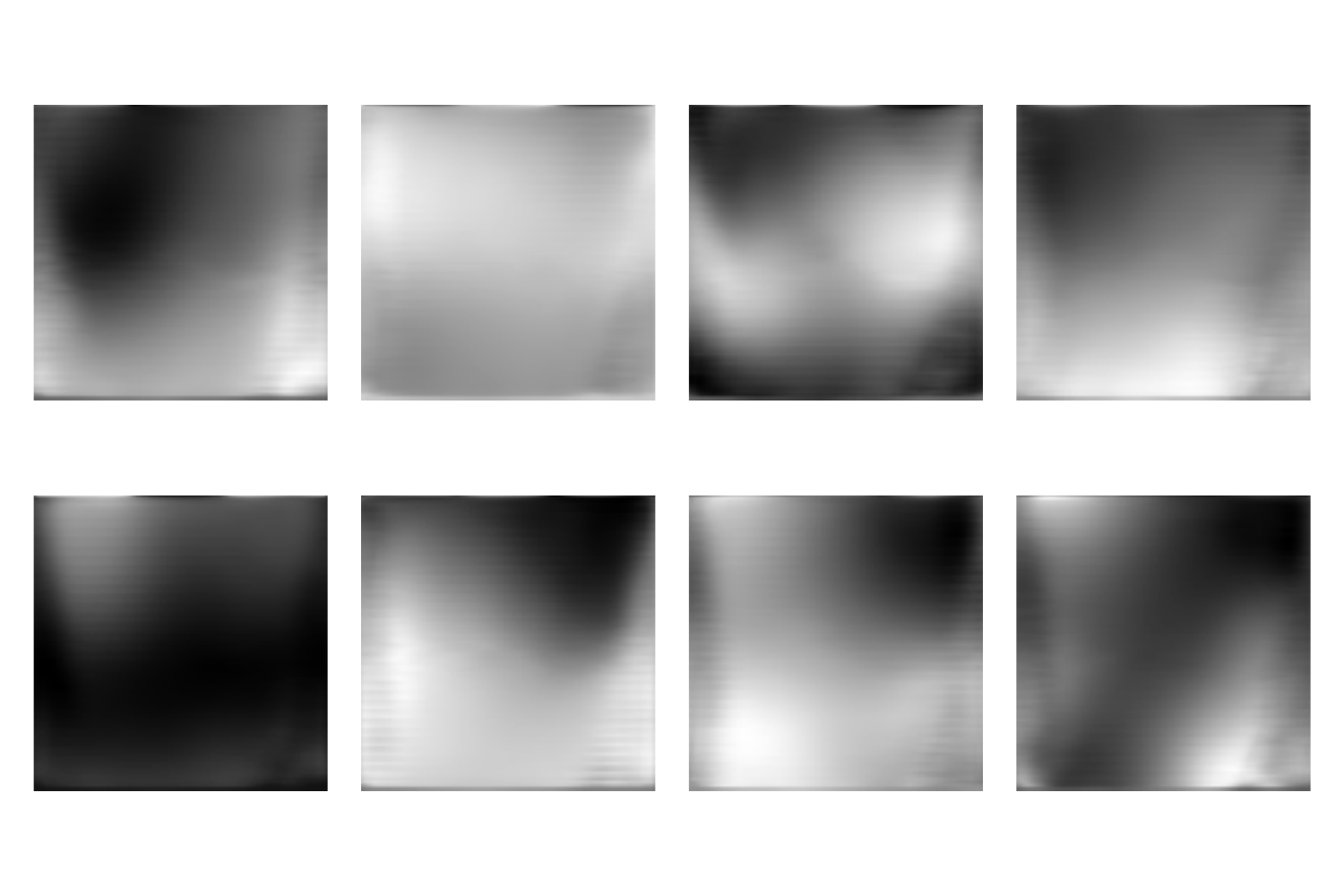}}
     \subfigure[Sigmoid Policy]{\includegraphics[width=0.24\textwidth]{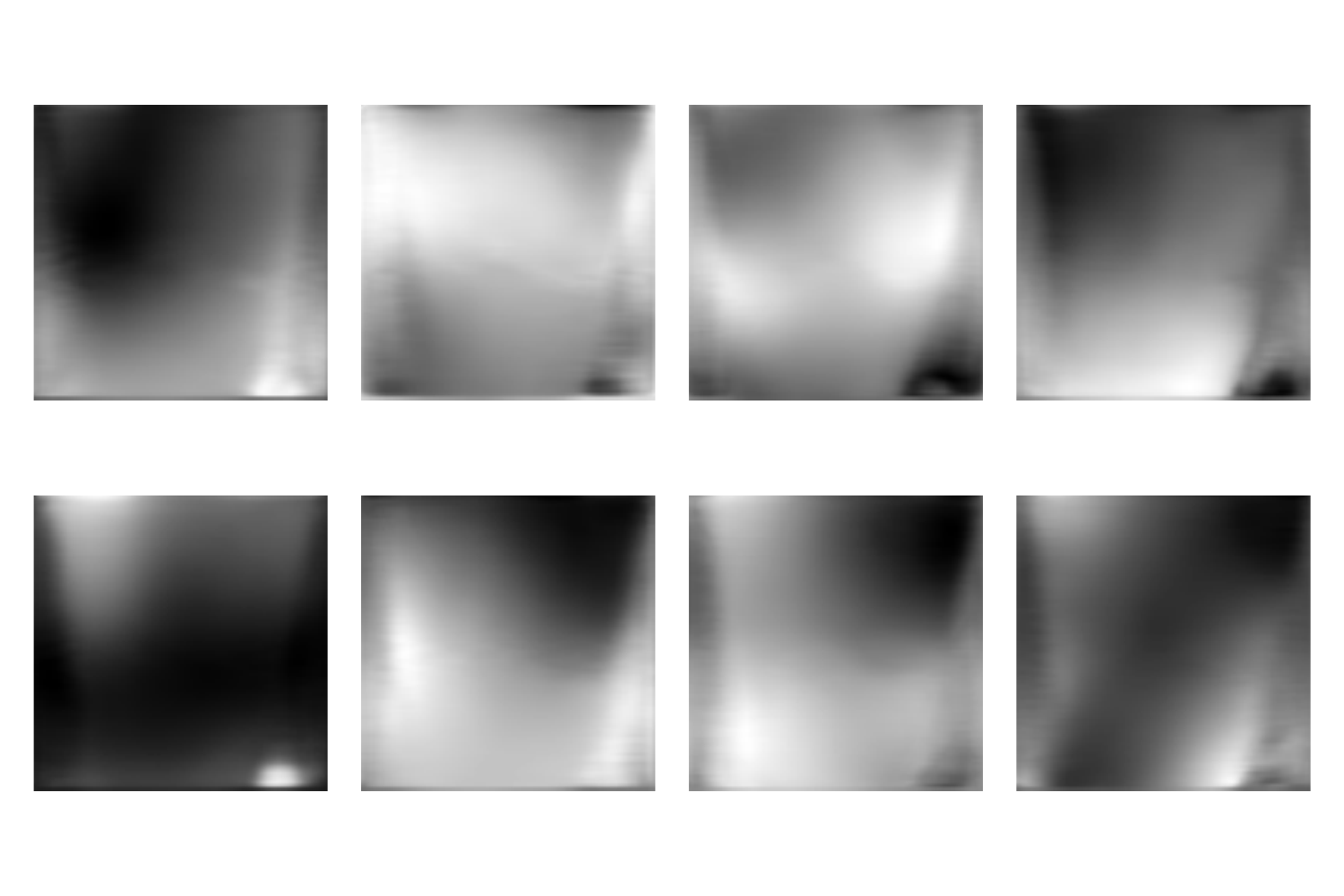}}
     \subfigure[Softplus Policy]{\includegraphics[width=0.24\textwidth]{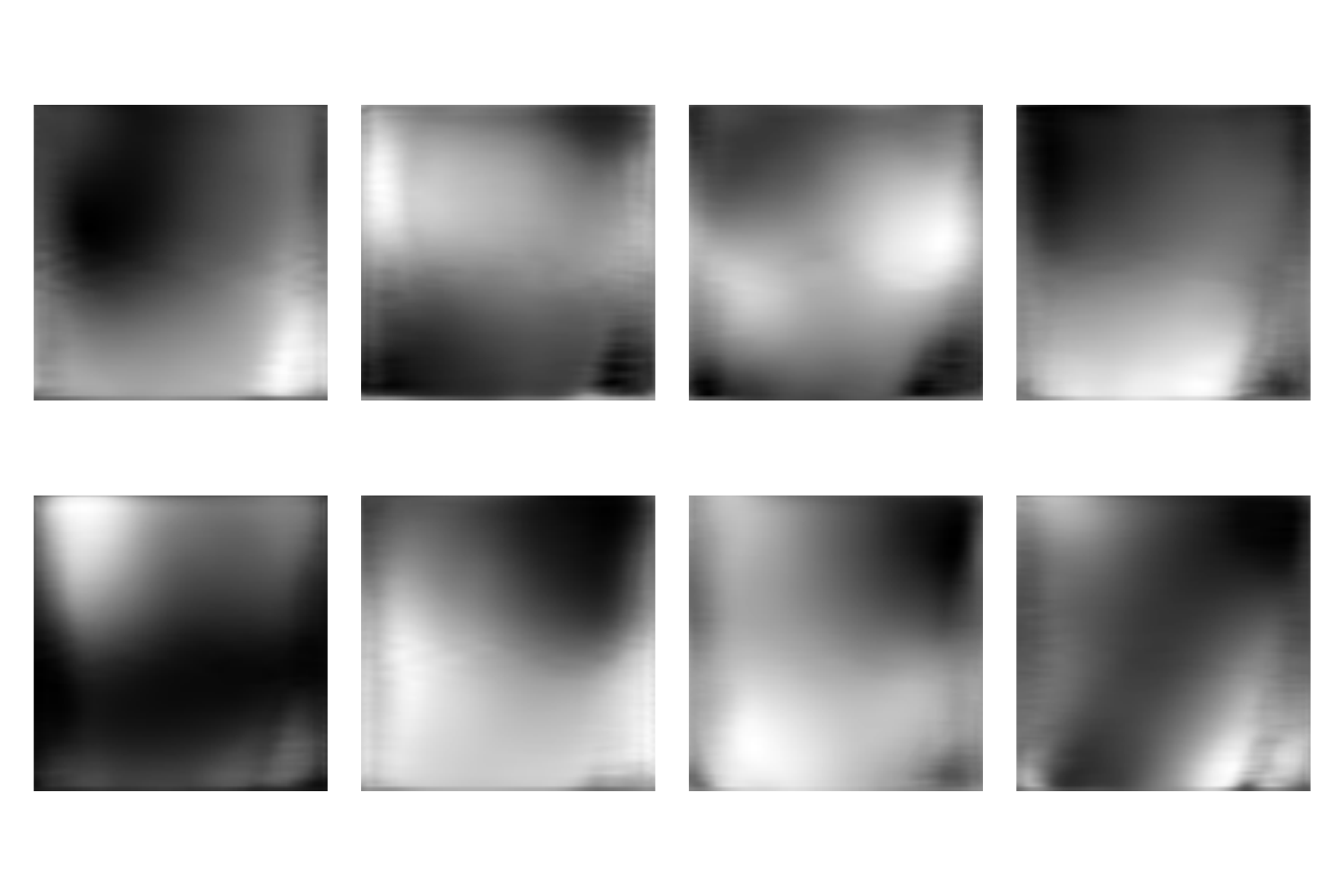}}
    \caption{Examples of $4\times$ sensitivity maps estimated by each model corresponding to the reconstructions displayed in Figure~\ref{fig:4x_recon}.}
    \label{fig:4x_sens_maps}
\end{figure}

\begin{figure}[t]
     \centering
     \subfigure[Equispaced]{\includegraphics[width=0.24\textwidth]{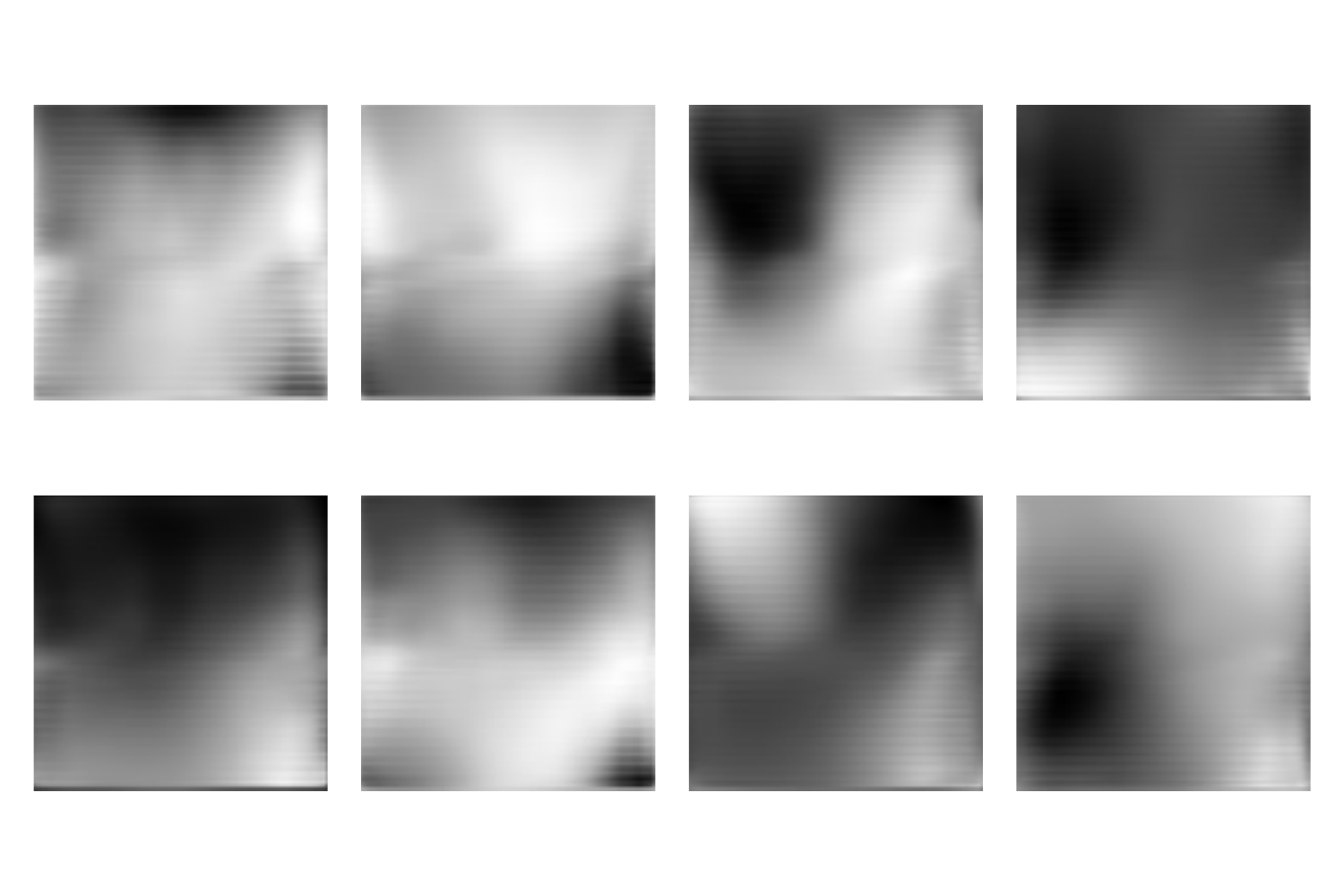}} 
     \subfigure[LOUPE]{\includegraphics[width=0.24\textwidth]{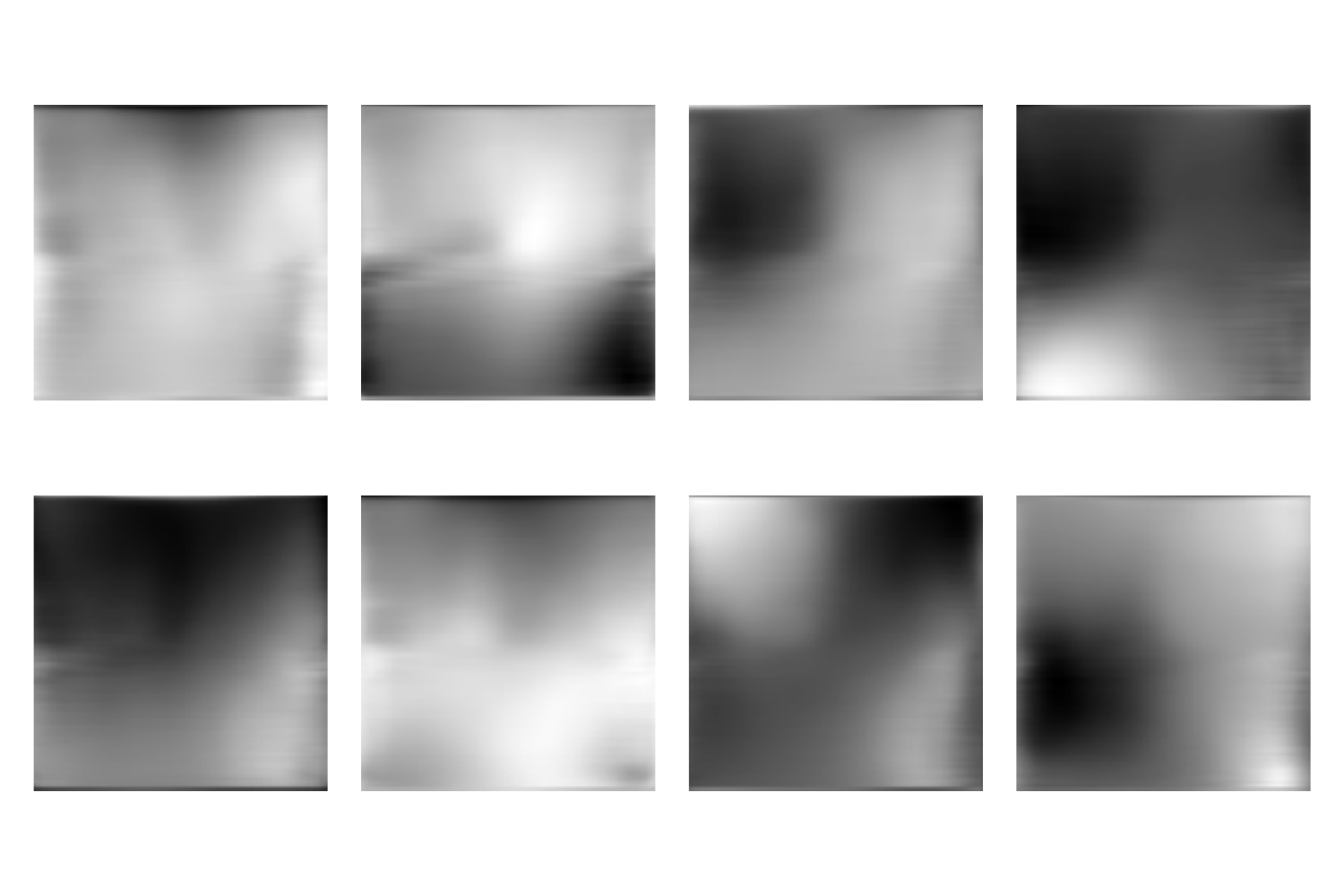}} 
     \subfigure[Sigmoid Policy]{\includegraphics[width=0.24\textwidth]{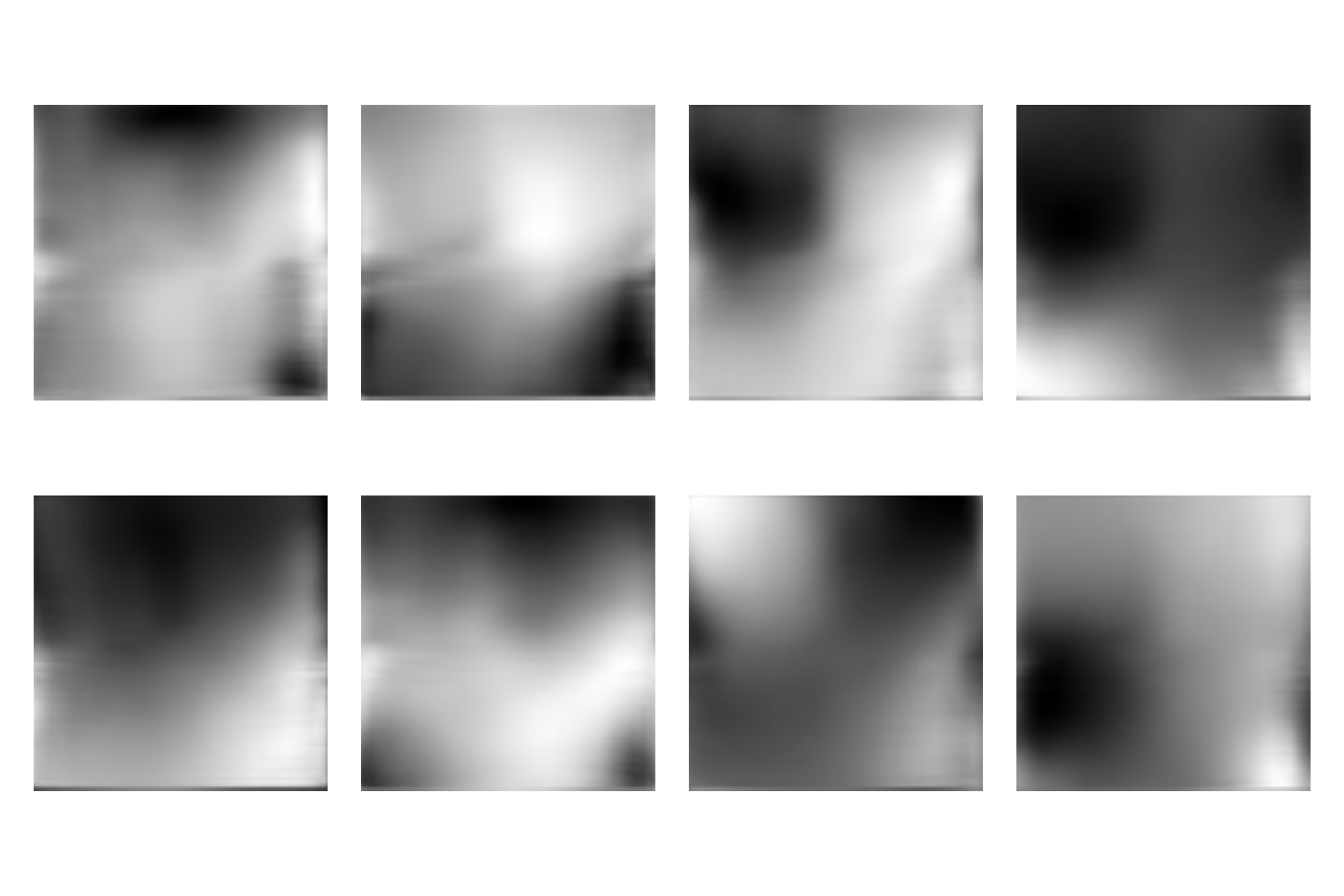}}
     \subfigure[Softplus Policy]{\includegraphics[width=0.24\textwidth]{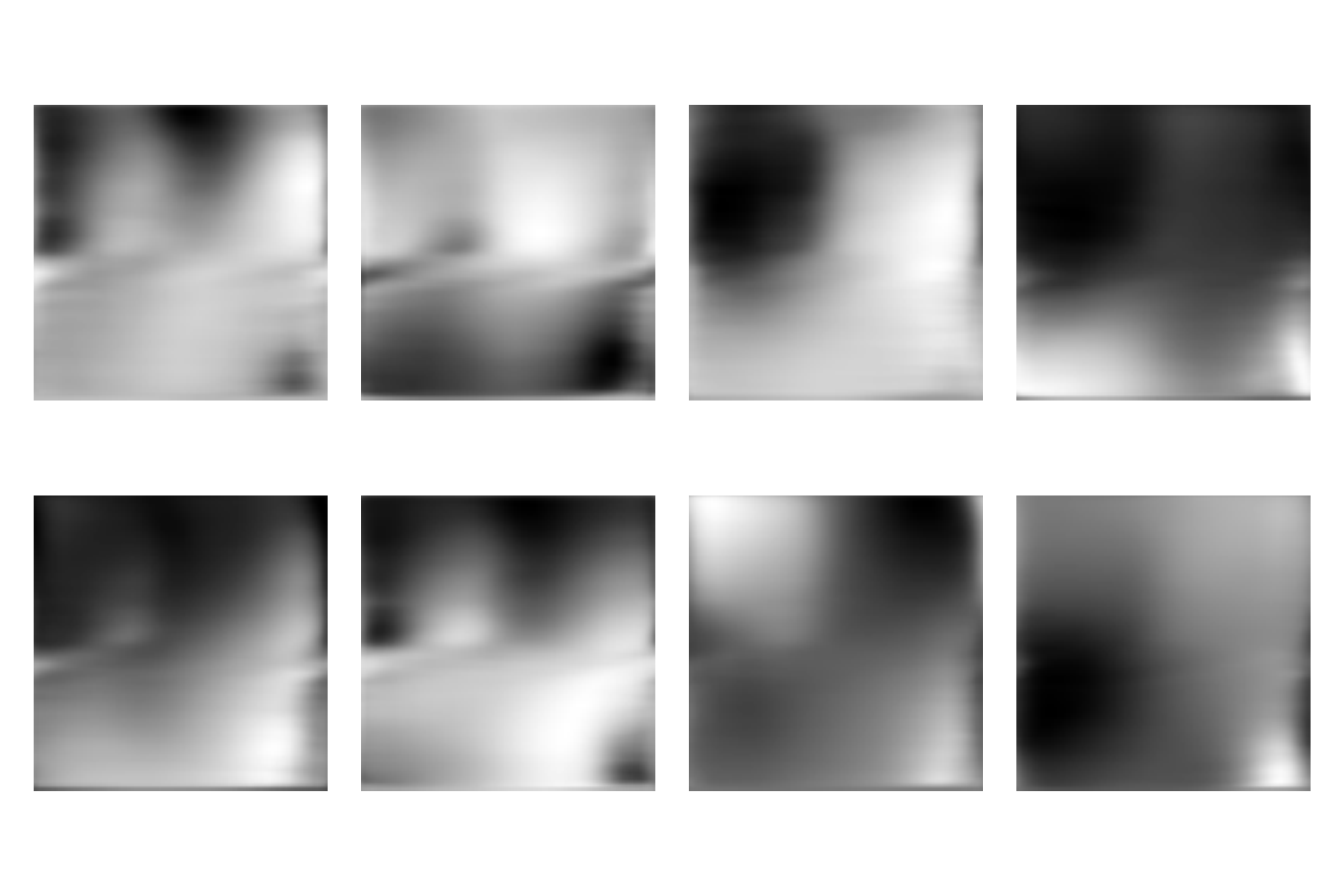}}
     \subfigure[Equispaced]{\includegraphics[width=0.24\textwidth]{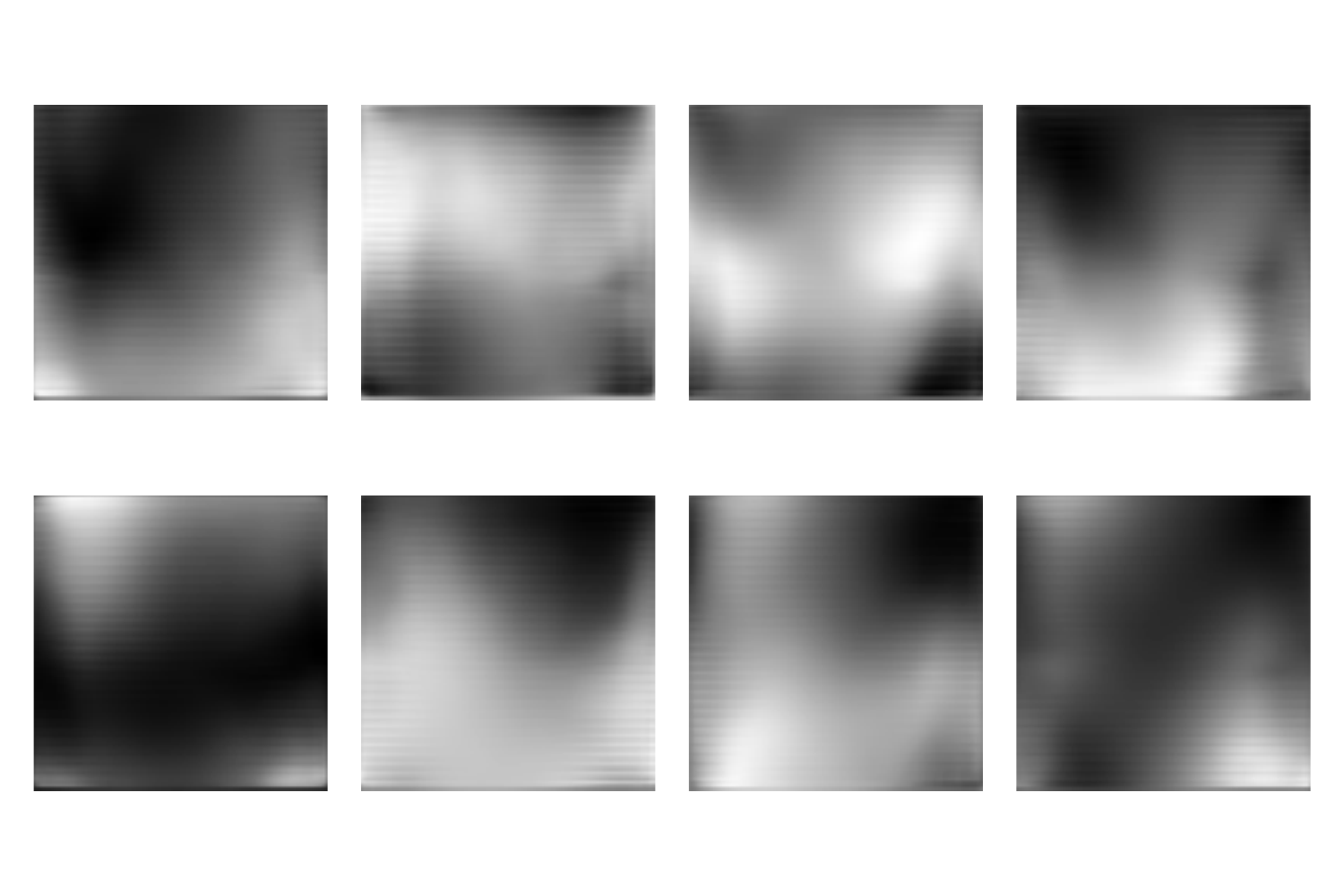}}
     \subfigure[LOUPE]{\includegraphics[width=0.24\textwidth]{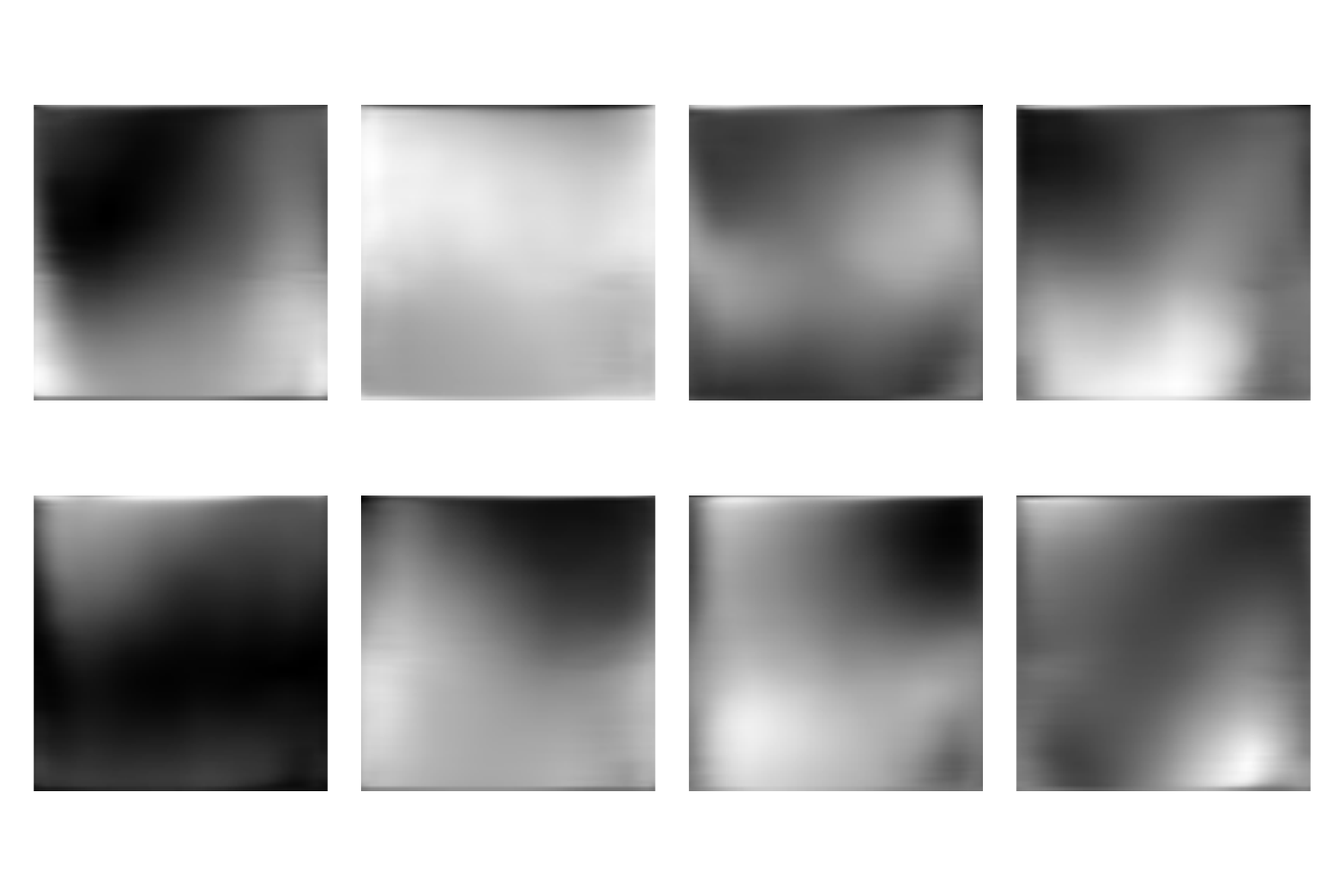}}
     \subfigure[Sigmoid Policy]{\includegraphics[width=0.24\textwidth]{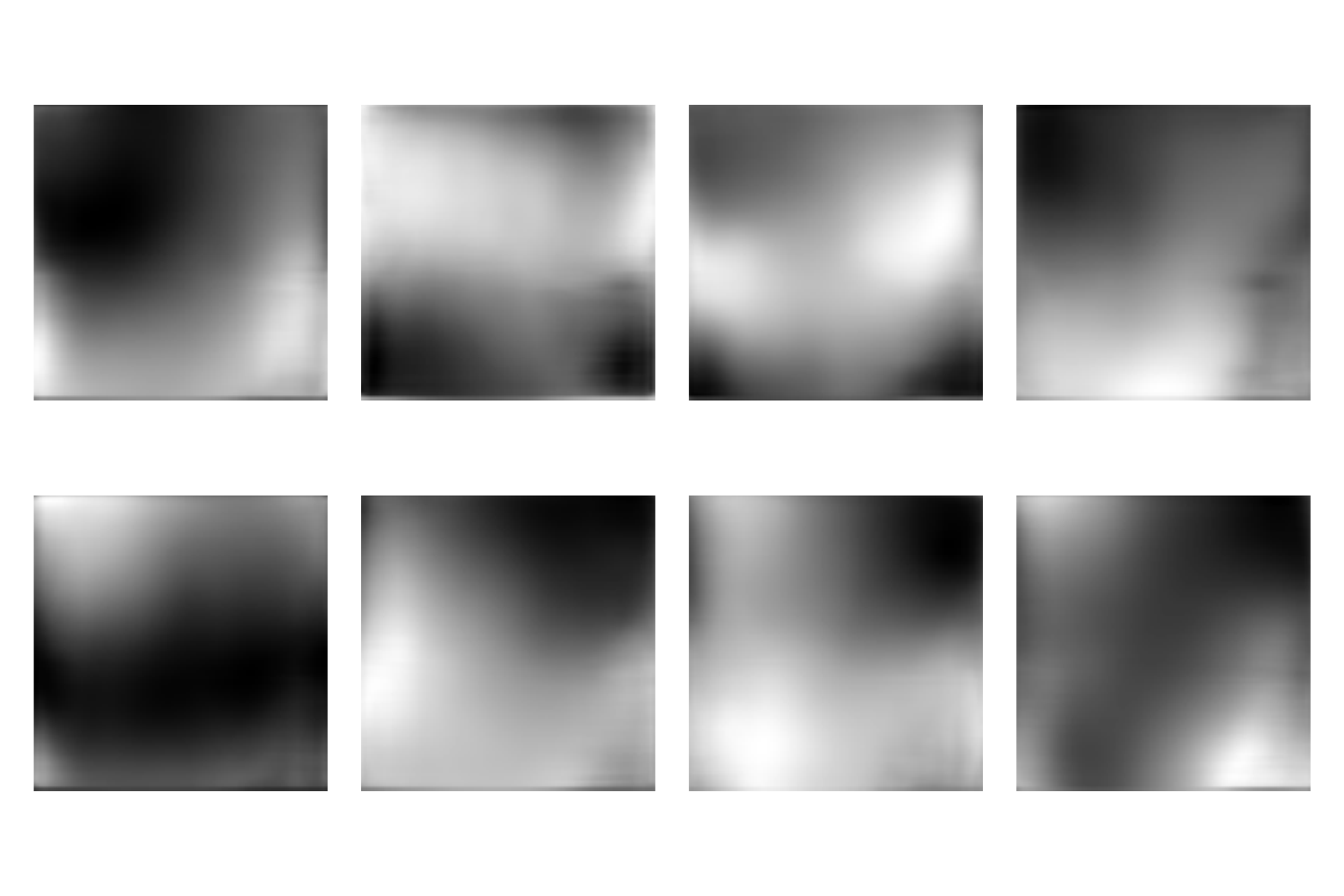}}
     \subfigure[Softplus Policy]{\includegraphics[width=0.24\textwidth]{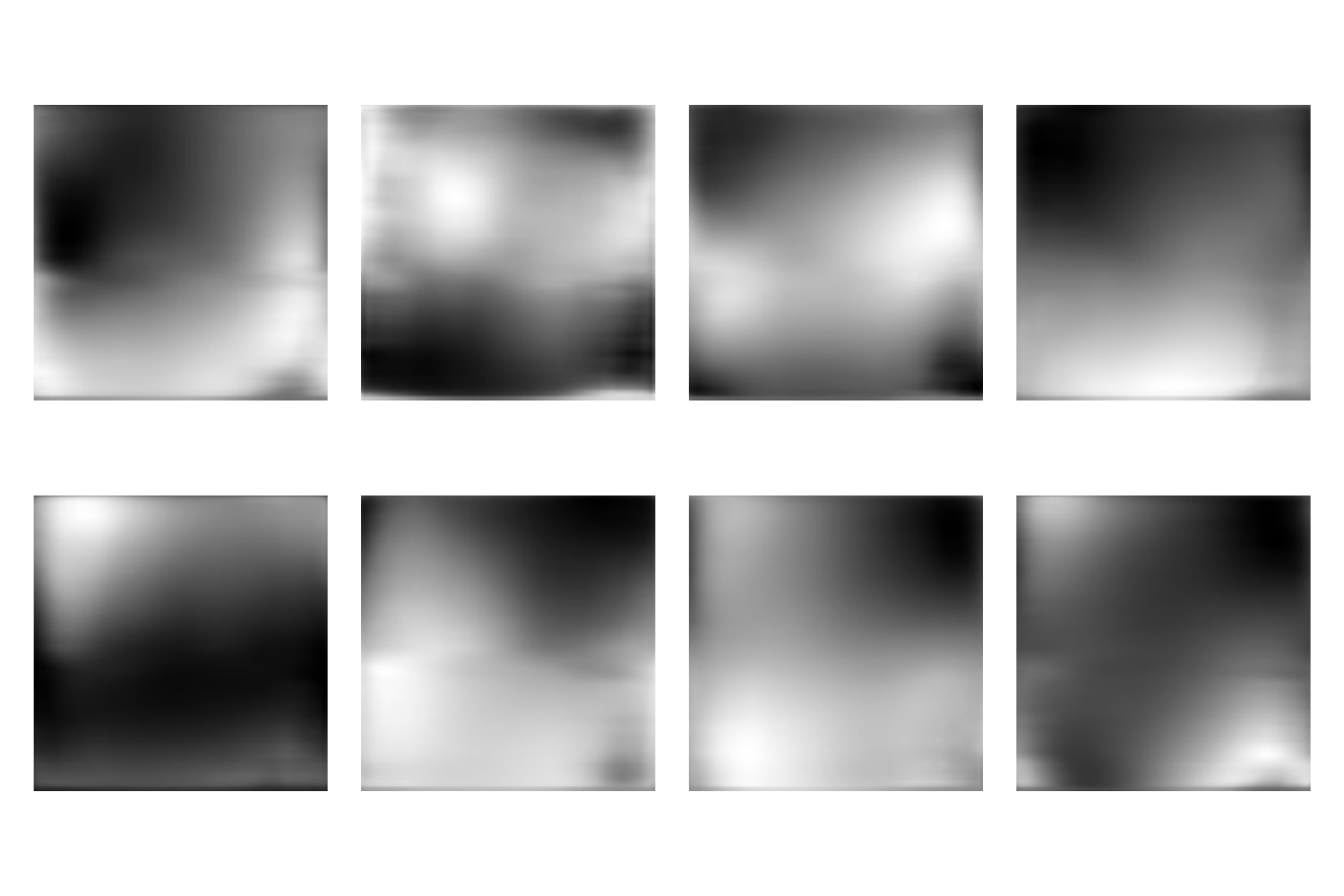}}
    \caption{Examples of $8\times$ sensitivity maps estimated by each model corresponding to the reconstructions displayed in Figure~\ref{fig:8x_recon}.}
    \label{fig:8x_sens_maps}
\end{figure}

Figures~\ref{fig:4x_sens_maps} and~\ref{fig:8x_sens_maps} show sensitivity maps learned by the various methods for respectively $4\times$ and $8\times$ acceleration. Further research is needed to explore the interaction between the sensitivity maps and the acquisition strategies.

\end{document}